\documentclass[a4paper, 11pt]{article}

\renewcommand{\baselinestretch}{1.2}
\usepackage{graphicx}
\usepackage{dcolumn}
\usepackage{bm}
\usepackage{multirow}

\usepackage{centernot}

\usepackage[normalem]{ulem}
\usepackage{cancel}

\usepackage{amsmath}
\usepackage{amsthm}
\usepackage{amstext}
\usepackage{amssymb}
\usepackage{mathrsfs}
\usepackage{amsfonts}
\usepackage{amsbsy} 
\usepackage{tensor}
\usepackage{physics}

\usepackage{latexsym}
\usepackage[american]{babel}
\usepackage{bbm}
\usepackage[nosort]{cite}

\usepackage{tocloft}
\usepackage[colorlinks=true, citecolor=blue!90!black, linkcolor=blue!90!black, linktocpage=true, urlcolor=red!70!black]{hyperref}

\newcommand*{\figref}[2][]{%
  \hyperref[{#2}]{%
    \ref*{#2}%
    \ifx\\#1\\%
    \else
      \,#1%
    \fi
  }%
}

\usepackage[all,matrix,cmtip]{xy}

\usepackage{csquotes}
\MakeOuterQuote{"}

\usepackage{color}
\definecolor{red}{rgb}{1,0,0}
\definecolor{blue}{rgb}{0,0,1}
\definecolor{dblue}{rgb}{0,0,0.4}
\definecolor{green}{rgb}{0,1,0}
\definecolor{black}{rgb}{0,0,0}
\definecolor{white}{rgb}{1,1,1}
\definecolor{niceBlue}{RGB}{20,10,237}

\usepackage[dvipsnames]{xcolor}

\definecolor{brn}{rgb}{.8,.4,.0}
\definecolor{redo}{rgb}{1,.5,.0}
\definecolor{ddgrn}{rgb}{0,0.4,0}
\definecolor{dgrn}{rgb}{0,0.55,0}
\definecolor{dbl}{rgb}{0,0,0.5}

\usepackage[bbgreekl]{mathbbol}

\newcommand{\Z}{\mathbb{Z}}
\newcommand{\C}{\mathbb{C}}
\newcommand{\R}{\mathbb{R}}

\newcommand{\p}[1]{\prime\,}

\renewcommand{\t}[1]{\widetilde{#1}}

\newcommand{\ii}{\hspace{1pt}\mathrm{i}\hspace{1pt}}
\newcommand{\ee}{\hspace{1pt}\mathrm{e}}
\renewcommand{\dd}{\hspace{1pt}\mathrm{d}}
\newcommand{\da}{\dagger}
\newcommand{\<}{\langle}
\renewcommand{\>}{\rangle}

\newcommand{\Rf}[1]{Ref.~\citenum{#1}}
\newcommand{\Rfs}[1]{Refs.~\citenum{#1}}

\newcommand{\pp}{\partial}

\newcommand{\ie}{{\it i.e.~}}

\newcommand{\bpm}{\begin{pmatrix}}
\newcommand{\epm}{\end{pmatrix}}
\newcommand{\bmm}{\begin{matrix}}
\newcommand{\emm}{\end{matrix}}

\newcommand{\cA}{\mathcal{A}} 
\newcommand{\cB}{ {\cal B} }
 
\newcommand{\cD}{ {\cal D} }

\newcommand{\cI}{\mathcal{I}}

\newcommand{\cL}{ {\cal L} }

\newcommand{\cO}{ {\cal O} } 
\newcommand{\cP}{ {\cal P} } 
\newcommand{\cQ}{\mathcal{Q}} 
 
\newcommand{\cS}{ {\cal S} }

\newcommand{\cX}{ {\cal X} } 
 
\newcommand{\cZ}{ {\cal Z} } 

\usepackage{euscript}

\newcommand{\al}{\alpha} 
\newcommand{\bt}{\beta} 
\newcommand{\del}{\delta} 
\newcommand{\Del}{\Delta}

\newcommand{\ga}{\gamma} 
\newcommand{\Ga}{\Gamma}

\newcommand{\La}{\Lambda} 
\newcommand{\om}{\omega}

\newcommand{\si}{\sigma} 
\newcommand{\Si}{\Sigma}

\renewcommand{\txt}[1]{\text{#1}}

\newcommand{\Hom}{\mathrm{Hom}}

\renewcommand{\mod}{\mathrm{mod}}

\newcommand{\Aut}{\mathrm{Aut}}

\def\wdg{{\mathchoice{\,{\scriptstyle\wedge}\,}{{\scriptstyle\wedge}}{{\scriptscriptstyle\wedge}}{{\scriptscriptstyle\wedge}}}} 

\renewcommand{\Vec}{\mathsf{Vec}}

\newcommand{\ssb}{\overset{\mathrm{SSB}}{\longrightarrow}}

\usepackage{tikz}
\usepackage{tikz-cd}
\usetikzlibrary{arrows}
\usetikzlibrary{intersections}
\usetikzlibrary{shapes.geometric}
\usetikzlibrary{decorations.pathmorphing, patterns,shapes}
\usetikzlibrary{decorations.markings}

\definecolor{Adef}{HTML}{d95f02}
\definecolor{Qdef}{HTML}{7570b3}
\definecolor{Wdef}{HTML}{1b9e77}
\definecolor{SymTFTcolor}{HTML}{CBD5E8}
\definecolor{latSymTFTcolor}{HTML}{7570B3}
\definecolor{reducedSymTFTcolor}{HTML}{FDCDAC}

\def\ie{\begin{equation}\begin{aligned}}
\def\fe{\end{aligned}\end{equation}}

\numberwithin{equation}{section}
\begin{document}

\thispagestyle{empty}
\fontsize{12pt}{20pt}
\hfill MIT-CTP/5884
\vspace{13mm}
\begin{center}
{\huge Spacetime symmetry-enriched SymTFT:
\\\vspace{6pt}
\LARGE from LSM anomalies to modulated symmetries and beyond}
\\[13mm]
{\large Salvatore D. Pace,
\"{O}mer M. Aksoy, 
Ho Tat Lam
}

\bigskip
{\it 
Department of Physics, Massachusetts Institute of Technology, Cambridge, MA 02139, USA \\
[.6em]	}

\bigskip
\today
\end{center}

\bigskip

\begin{abstract}
\noindent

We extend the Symmetry Topological Field Theory (SymTFT) framework beyond internal symmetries by including geometric data that encode spacetime symmetries. Concretely, we enrich the SymTFT of an internal symmetry by spacetime symmetries and study the resulting symmetry-enriched topological (SET) order, which captures the interplay between the spacetime and internal symmetries. We illustrate the framework by focusing on symmetries in ${1+1}$D. To this end, we first analyze how gapped boundaries of ${2+1}$D SETs affect the enriching symmetry, and apply this within the SymTFT framework to gauging and detecting anomalies of the ${1+1}$D symmetry, as well as to classifying ${1+1}$D symmetry-enriched phases. We then consider quantum spin chains and explicitly construct the SymTFTs for three prototypical spacetime symmetries: lattice translations, spatial reflections, and time reversal. For lattice translations, the interplay with internal symmetries is encoded in the SymTFT by translations permuting anyons, which causes the continuum description of the SymTFT to be a foliated field theory. Using this, we elucidate the relation between Lieb-Schultz-Mattis (LSM) anomalies and modulated symmetries and classify modulated symmetry-protected topological (SPT) phases. For reflection and time-reversal symmetries, the interplay can additionally be encoded by symmetry fractionalization data in the SymTFT, and we identify mixed anomalies and study gauging for such examples.

\end{abstract}

\vfill

\newpage

\pagenumbering{arabic}
\setcounter{page}{1}
\setcounter{footnote}{0}

{\renewcommand{\baselinestretch}{.88} \parskip=0pt
\setcounter{tocdepth}{3}
\tableofcontents}

\vspace{20pt}
\hrule width\textwidth height .8pt
\vspace{13pt}

\section{Introduction}

Symmetries have long provided a powerful perspective and an essential tool in theoretical physics. From the early triumphs of Noether’s theorem~\cite{Noether:1918zz} to the modern notion of generalized global symmetries initiated by~\cite{GW14125148} (see~\cite{M220403045, G230301817, S230518296, BH230600912, Bhardwaj:2023kri, LWW230709215, Shao:2023gho} for recent reviews), symmetry principles have yielded deep insights into a plethora of physics and phenomena, from phases of matter and their transitions, to dualities of quantum field theories. The power of symmetries lies in their ability to distill a system’s universal features while remaining agnostic to its potentially complicated details.

A powerful framework for separating the universal, symmetry-based properties of a theory from its theory-specific details is the symmetry topological field theory (SymTFT)~\cite{Witten:1998wy,GW14125148, KZ150201690, KZ170501087, FT180600008, KZ190504924, Pulmann:2019vrw, TW191202817, JW191213492, LB200304328, KZ200514178, GK200805960, AFM200808598, ABE211202092, CW220303596, CW220506244, MT220710712, FT220907471, Kaidi:2022cpf, KNZ230107112, ZC230401262, BS230517159, BS240106128, AB240110165, BZM240212347, ABD240214813, ABB240406601, WYP240419004, H240509611, BIT240509754, JC240702488, C240801490, Cordova:2024iti, Cvetic:2024dzu,  GarciaEtxebarria:2024jfv, Choi:2024tri, Bhardwaj:2024igy, VRS240906647, PLA240918113}.\footnote{Different communities---hep-th, cond-mat, and math-ph---independently initiated the developments over the past decade that led to the SymTFT. Depending on the community, the SymTFT is also called topological holography and Symmetry Topological Order (SymTO). In general, the TFT defining the SymTFT can have trivial topological order, potentially including non-trivial local operators~\cite{AGRS240200117,Y241114997,NSW241119683,LRR250321862, RR250514807} or by being an invertible theory~\cite{ABB240406601}. In this work, the SymTFTs considered always have non-trivial topological order. Related concepts to the SymTFT can be found in various foundational papers on quantum field theory, e.g., Refs.~\citeonline{Witten:1988hf} and~\citeonline{Elitzur:1989nr}. } The SymTFT is a symmetry-based bulk-boundary correspondence that relates a quantum theory with global symmetry to a topological theory in one higher dimension, as shown in Fig.~\ref{fig:TopHolo}. As we review in Appendix~\ref{SymTFTReview}, the SymTFT provides a physical realization/representation of the symmetry defects and charges of a (generalized) symmetry. It is useful for performing the ``calculus'' of symmetry defects and deducing symmetry-allowed phenomena, thereby providing conceptual clarity and finding new applications of symmetries. Among its various applications, those we consider in this paper are its applications to diagnosing 't Hooft anomalies~\cite{TW191202817, Kaidi:2022cpf, KNZ230107112, ZC230401262, CHZ230811706, AB240110165, PR240504619, PR250313633} and to classifying phases of matter~\cite{TW191202817, KZ200514178, CW220506244, MT220710712, CW221214432, BBP231217322, BPS240300905, BPS240805266, ACS240805585, BS241019040, BST250220440, BGH250312699, AW250321764}.

Since its inception, SymTFT has been undergoing rapid development and has been formulated for discrete and continuous internal (generalized) symmetries. Despite this impressive progress, a significant gap remains: spacetime symmetries remain largely outside the standard SymTFT framework. Spacetime symmetries, such as time-reversal, spatial reflections, lattice or continuum translations, and boosts, are ubiquitous in realistic systems and underpin a wide range of physical phenomena. The most straightforward generalization of SymTFT for spacetime symmetries would involve dynamically gauging spacetime symmetries. However, unlike their internal counterparts, spacetime symmetries fail to admit conventional background gauge fields and symmetry defects. The difficulties of formulating such a gauge theory leave a systematic understanding of SymTFT for spacetime symmetries still elusive.

\begin{figure}
\centering
\begin{tikzpicture}[scale=1, baseline={([yshift=-.5ex]current bounding box.center)},thick]
\fill[SymTFTcolor] (0,0) rectangle (4,2);
\draw[black, line width=4pt] (0, 0) -- (4,0);
\draw[black, line width=4pt] (0, 2) -- (4,2);
\node[black, left] at (0, 0) {\Large $\mathfrak{B}^{\text{phys}}_{\mathfrak{T}^\cS}$};
\node[black, left] at (0, 2) {\Large $\mathfrak{B}^{\text{sym}}_{\cS}$};
\node[black, left] at (0, 1) {\Large $\mathfrak{Z}(\cS)$};
\node[black] at (2, 1) {\normalsize ${X_{d+1}\times[0,1]}$};
\draw[-{>[scale=1.5]}, line width=1.75pt] (5,1) -- node[yshift=-15] {Compactification} node[yshift=15] {Interval} (8,1);
\draw[black, line width=4pt] (9.75, 0) -- (13.75,0);
\node[black, left] at (9.75, 0) {\Large $\mathfrak{T}^\cS$};
\node[black, above] at (11.75, 0) {\normalsize $X_{d+1}$};
\end{tikzpicture}
\caption{The $\cS$-symmetric theory $\mathfrak{T}^\cS$ on ${(d+1)}$D spacetime ${X_{d+1}}$ is related to the ${((d+1)+1)}$D SymTFT $\mathfrak{Z}(\cS)$ of $\cS$ by the interval compactification of the slab theory ${(\mathfrak{B}^{\text{sym}}_\cS, \mathfrak{Z}(\cS), \mathfrak{B}^{\text{phys}}_{\mathfrak{T}^\cS})}$ on spacetime ${X_{d+1}\times [0,1]}$. The $\mathfrak{B}^{\text{sym}}_\cS$ boundary realizes the $\cS$ symmetry defects and the boundary $\mathfrak{B}^{\text{phys}}_{\mathfrak{T}^\cS}$ encodes the dynamics of $\mathfrak{T}^\cS$. The interval compactification is an exact relation due to the topological properties of $\mathfrak{Z}(\cS)$. More specifically, the interval compactification induces an isomorphism from the ``sandwich'' ${(\mathfrak{B}^{\text{sym}}_\cS, \mathfrak{Z}(\cS), \mathfrak{B}^{\text{phys}}_{\mathfrak{T}^\cS})}$ to $\mathfrak{T}^\cS$.}
\label{fig:TopHolo}
\end{figure}
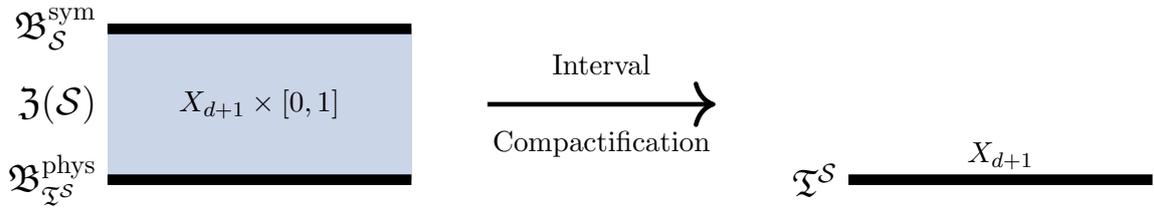

One approach to formulating a SymTFT for spacetime symmetries is to replace the spacetime symmetry with an effective internal symmetry and consider the SymTFT for that effective internal symmetry. This approach closely follows the philosophy that lies at the heart of the crystalline equivalence principle~\cite{TE161200846, MB200510265, ZYQ201215657, D210202941, MCB221002452}, which states that the classification of \textit{some} invertible phases protected by crystalline symmetries is in one-to-one correspondence with the classification of invertible phases protected by an internal symmetry. However, significant conceptual differences exist between internal and spacetime symmetries that would not be captured by simply replacing the spacetime symmetry with an effective internal symmetry. For example, the difference between their defects (see~\cite{CS221112543, SSS240112281} for a detailed survey in the context of one-dimensional lattice translations, and~\cite{PR171111044} for discussion on the fractonic nature unique to crystalline defects in higher dimensions). Furthermore, crystalline equivalence principle type arguments can fail for phases failing to meet the ``liquid'' property introduced in~\cite{TE161200846}. These include, for example, subsystem symmetry protected topological (SPT) phases~\cite{YDB180302369, DYB180504097, DWY180805300, D181202721, DSW191001630, Burnell:2021reh, ZCB221015596, PLS250518119 }. Hence, it is desirable to have a construction of the SymTFT that treats spacetime symmetries honestly, which we will do in this work.

\subsection{Summary}\label{SummarySection}

In this paper, we extend the SymTFT framework to spacetime symmetries using a symmetry-enriched topological (SET) order as the SymTFT. The intrinsic topological order of the SET realizes the SymTFT of an internal symmetry, and its enriching symmetry is the spacetime symmetry. Different symmetry enrichments correspond to different interplays between the internal and spacetime symmetries (e.g., mixed anomalies and group extensions). From a field theory point of view, our construction yields the SymTFT of an internal symmetry coupled to background fields corresponding to geometric structures of spacetime (e.g., Stiefel–Whitney classes and foliation structures). Among other things, we apply this spacetime symmetry-enriched SymTFT to classifying spacetime symmetry-enriched phases of matter and diagnosing Lieb-Schultz-Mattis (LSM) anomalies.

We start in Section~\ref{generalSection} discussing how SETs can be used as SymTFTs and why they naturally arise for symmetries that cannot be straightforwardly gauged (i.e., spacetime symmetries). Throughout the section, and this section alone, we will assume that the enriching symmetry has well-defined topological defects. By doing so, we can use aspects of the well-known framework for bosonic SETs~\cite{BBC14104540} while developing helpful intuition for later when the enriching symmetry is a spacetime symmetry.

Gapped boundaries play a central role in the SymTFT framework, and in Section~\ref{gappedBdySETsec}, we examine how a gapped boundary modifies an SET's enriching symmetry $Q$ in ${2+1}$D. We assume $Q$ is a finite invertible symmetry and that the topological order is bosonic. 
A gapped boundary of an SET is obtained by condensing anyons that form a Lagrangian condensable algebra $\cL$. 
If there is a non-trivial action of $Q$ on the anyons, denoted by $\rho$, the presence of a gapped boundary can explicitly break $Q$ down to a subgroup ${Q_\cL = \{q\in Q \mid \rho_q(\cL)\cong \cL\}}$.
The remaining $Q_\cL$ symmetry operators, however, do not necessarily form the group $Q_\cL$: the gapped boundary can cause $Q_\cL$ to be extended by a group $A_\cL$ describing operators on the boundary. 
These $A_\cL$ operators themselves are symmetries, implemented by uncondensed anyon lines on the boundary. We show that the enriching symmetry $Q_\cL$ and boundary symmetry $A_\cL$ form the total symmetry group $G_\cL$ described by the group extension (see Fig.~\ref{fig:SETDirbdy})
\begin{equation}\label{summaryExt}
1 \to A_\cL \to G_\cL \to Q_\cL \to 1.
\end{equation}
The action of $Q_\cL$ on $A_\cL$ in $G_\cL$ is inherited from the action $\rho$ of $Q$ on the anyons. We further show that the extension class in~\eqref{summaryExt} is non-trivial if and only if there is an anyon in $\cL$ with fractional $Q_\cL$ symmetry charge. In this paper, we do not investigate the interplay with possible non-invertible symmetries on the boundary, instead focusing on examples with invertible symmetries and Abelian $A_\cL$ and $Q_\cL$ for simplicity.

Having established how $Q$ is affected by a gapped boundary and how it interplays with $A_\cL$, we then apply the symmetry-enriched SymTFT to study discrete gauging and to classify $Q$-enriched gapped phases. 

As in the standard SymTFT framework, discrete gauging is implemented by changing the symmetry boundary (the top boundary in Fig~\ref{fig:TopHolo}). For a symmetry boundary with Lagrangian algebra $\cL$, the group $G_\cL$ describes the symmetry captured by the symmetry-enriched SymTFT. This includes the symmetry encoded by symmetry-enrichment and the symmetry encoded by the symmetry boundary. Changing the symmetry boundary amounts to discrete gauging only the $A_\cL$ sub-symmetry, but not the full $G_\cL$ symmetry. Changing the symmetry boundary generally changes the groups $A_\cL$ and $Q_\cL$ as well as their extension~\eqref{summaryExt}. These changes capture the properties of the dual symmetry arising from gauging, which includes new interplays involving $Q$ (e.g., extensions and anomalies).

$Q$-enriched phases are phases where ${Q_\cL\cong Q}$ is a subgroup of $G_\mathcal{L}$ and is not spontaneously broken. We show how such phases are classified using a sandwich of the $Q$-enriched SymTFT, similar to that shown in Fig.~\ref{fig:TopHolo}. As explained in Section~\ref{gappedSymPhasesSec}, here, the symmetry and physical boundaries of the SymTFT are restricted to $Q$-symmetric Lagrangian algebras. A $Q$-symmetric Lagrangian algebra $\cL$ satisfies ${\rho_q(\cL)\cong \cL}$ for all ${q\in Q}$ and carries no fractional $Q$ symmetry charge. These Lagrangian algebras correspond to boundaries that do not affect the enriching symmetry. That is, they are ones for which ${Q_\cL\cong Q}$ and the extension class of~\eqref{summaryExt} is trivial. There can, however, still be an action of $Q$ on $A_\cL$, making $G_\cL \cong A_\cL \rtimes Q$. A powerful application of this classification is diagnosing 't Hooft anomalies of the total symmetry. Indeed, if there are no $Q$-enriched SPTs, there will be no $G_\cL$ SPTs and hence an 't Hooft anomaly for $G_\cL$.

An example of a symmetry-enriched SymTFT we consider throughout Section~\ref{generalSection} is $\Z_2$ topological order in ${2+1}$D enriched by a $\Z_2$ symmetry. For this SET, the enriching $\Z_2$ symmetry's action $\rho$ is trivial and the $e$ anyons carry fractional $\Z_2$ charge (see Appendix~\ref{Z2 enriched TC App} for a commuting projector model realizing this SET). We show that the symmetry-enriched SymTFT with Lagrangian algebra ${\cL_m = 1\oplus m}$ describes an anomalous ${\Z_2\!\times \Z^{(e)}_2}$ symmetry. This is anomalous because $\cL_m$ is the only $\Z_2$-symmetric Lagrangian algebra, so the only $\Z_2$-enriched gapped phase has $\Z_2^{(e)}$ spontaneously broken and there are no ${\Z_2\!\times \Z^{(e)}_2}$ SPTs. The other Lagrangian algebra is ${\cL_e = 1\oplus e}$, which is not $\Z_2$-symmetric since $e$ carries fractional $\Z_2$ charge. Due to the symmetry fractionalization, the symmetry-enriched SymTFT with this symmetry boundary describes a $\Z_4$ symmetry, which extends $\Z_2$ by the boundary $\Z_2^{(m)}$ symmetry. Changing from the $\cL_e$ boundary to the $\cL_m$ boundary amounts to gauging the $\Z_2^{(e)}$ sub-symmetry of the anomalous ${\Z_2\!\times \Z^{(e)}_2}$ symmetry and, as expected, leads to a dual $\Z_4$ symmetry.

After Section~\ref{generalSection}, the remainder of the paper focuses on explicit models of spacetime symmetry-enriched SymTFTs. We concentrate on spacetime symmetries ubiquitous to $1+1$D quantum spin chains: lattice translations in Section~\ref{translationSec}, spatial reflections in Section~\ref{sec reflection SymTFT}, and time reversal in Section~\ref{sec TRS SymTFT}. We present examples of the symmetry-enriched SymTFTs with these spacetime symmetries from both the Euclidean field theory and quantum lattice model perspectives, often times providing derivations from one to the other.

Lattice translations of a one-dimensional spin chain in the translation-enriched SymTFT become discrete spatial translations in the direction parallel to the boundary in Fig.~\ref{fig:TopHolo}. One-dimensional lattice translations in $2+1$D SETs cannot undergo symmetry fractionalization. Instead, their interplay with anyons is only through non-trivial anyon automorphisms, which makes the anyons position-dependent excitations~\cite{PW220407111}. When this occurs, the symmetry-enriched SymTFT is not an honest-to-goodness TFT. Instead, it is a foliated field theory that couples to a foliation structure of spacetime that causes the SymTFT to be topological only in the interval compactification direction.\footnote{Strictly speaking, the SymTFT does not need to be topological in all directions. It only needs to be topological in the direction parallel to the interval compactification direction to induce the isomorphism in Fig.~\ref{fig:TopHolo}.} Throughout Section~\ref{translationSec}, we show how this translation-enriched SymTFT can be applied to modulated symmetries\footnote{Modulated symmetries are internal symmetries that act in a non-uniform, spatially modulated way. They are generalizations of, for example, dipole symmetries. We refer the reader to \Rf{PDL240612962} for an introduction and recent survey of modulated symmetries.} and LSM anomalies involving lattice translations. A useful application is in classifying modulated SPTs, which we do for specific modulated symmetries in examples and later present a general classification in Section~\ref{classModSPTsSec}.

We consider dipole and exponential symmetries as examples in Sections~\ref{sec:exponential sym} and~\ref{ZNdipEx}, respectively, and general LSM anomalies of translation and finite Abelian symmetries in Section~\ref{LSMwithTranslationSec}. In both of these, we consider lattice and field theory perspectives. The translation-enriched SymTFT makes the relation between LSM anomalies and modulated symmetries transparent. In particular, we show in Section~\ref{LSMtoModsymbyGauging} that there is always a discrete gauging that relates LSM anomalies between lattice translations and finite Abelian group symmetries to modulated symmetries.

As an example, we consider a translation-enriched SymTFT whose underlying topological order is ${\Z_2\times\Z_2}$ topological order that is enriched by lattice translations $T_x$ in the $x$ direction. The anyons of the topological order are generated by $e_1$, $e_2$, $m_1$, and $m_2$, with $e_i$ and $m_i$ having mutual braiding by the phase $\pi$. The $T_x$ translation symmetry enriches this topological order by transforming the anyons as
\begin{equation}\label{TxactionSum}
T_x(e_1) = e_1e_2,
\qquad
T_x(e_2) = e_2,
\qquad
T_x(m_1) = m_1,
\qquad
T_x(m_2) = m_1 m_2.
\end{equation}
This action of $T_x$ on the anyons is the only non-trivial data specifying the symmetry enrichment. Because of the enrichment by $T_x$, this SymTFT is only topological in the $y$ direction.

This simple example is discussed in both Sections~\ref{ZNdipEx} and~\ref{LSMwithTranslationSec}, and we present a quantum code and Euclidean field theory description of it. The stabilizers of the former are
\begin{align}
A_{\bm{r}}
=
\begin{tikzpicture}[scale = 0.5, baseline = {([yshift=-.5ex]current bounding box.center)}]
\draw[line width=0.015in, gray] (4.5,0) -- (13.5, 0);
\draw[line width=0.015in, gray] (7.5,-3) -- (7.5, 3);
\draw[line width=0.015in, gray] (10.5,-3) -- (10.5, 3);
\node at (6, .45) {\normalsize $\textcolor[HTML]{cb181d}{X}$};
\node at (12, .45) {\normalsize $\textcolor[HTML]{cb181d}{X}$};
\node at (10.5, 1.5) {\normalsize $\textcolor[HTML]{cb181d}{X}$};
\node at (10.5, -1.5) {\normalsize $\textcolor[HTML]{cb181d}{X}$};
\fill[black] (10.5,0) circle (6pt);
\node at (10, -.5) {\normalsize $r$};
\end{tikzpicture}~,
\hspace{40pt}
B_{\bm{r}}=\begin{tikzpicture}[scale = 0.5, baseline={([yshift=-.5ex]current bounding box.center)}]
\draw[line width=0.015in, gray] (-3, 0) -- (-3, 3) -- (3, 3) -- (3, 0) -- cycle;
\draw[line width=0.015in, gray] (0,0) -- (0, 3);
\node at (-1.5, 0.4) {\normalsize $\textcolor[HTML]{2171b5}{Z}$};
\node at (-1.5, 3.45) {\normalsize $\textcolor[HTML]{2171b5}{Z}$};
\node at (-3, 1.5) {\normalsize $\textcolor[HTML]{2171b5}{Z}$};
\node at (3, 1.5) {\normalsize $\textcolor[HTML]{2171b5}{Z}$};
\fill[black] (-3,0) circle (6pt);
\node at (-3.5, -.5) {\normalsize $r$};
\end{tikzpicture},
\end{align}
where $X$ and $Z$ are Pauli matrices transforming qubits that reside on the edges of the square lattice. The continuum Euclidean action of the latter is the foliated field theory
\begin{equation}
S[e] = -\frac{2\ii}{2\pi} \int \left(\t{a} \wdg \dd b -\t{b}\wdg\dd a - \t{a}\wdg \t{b} \wdg e\right),
\end{equation}
where ${e = \La \dd x}$ is a background foliation field with $\La^{-1}$ a necessary UV cutoff (i.e., a lattice spacing). We write down the logical operators/topological defect lines explicitly, showing they satisfy~\eqref{TxactionSum}. There are six Lagrangian condensable algebras, four of which are translation-symmetric while the other two are not. The translation-symmetric Lagrangian algebras are
\begin{align}
\cL_{1} &= 1\oplus e_1 \oplus e_2\oplus e_1e_2,
\qquad\hspace{10pt}
\cL_{2} = 1\oplus m_1 \oplus m_2\oplus m_1m_2,
\\
\cL_{3} &= 1\oplus m_1 \oplus e_2\oplus m_1e_2,
\qquad
\cL_{4} = 1\oplus m_1e_2 \oplus e_1m_2\oplus e_1e_2m_1m_2.
\end{align}
The Lagrangian algebras
\begin{align}
\cL_{5} = 1\oplus e_1 \oplus m_2\oplus e_1m_2,
\qquad
\cL_{6} = 1\oplus e_1e_2 \oplus m_1m_2\oplus e_1e_2m_1m_2,
\end{align}
are not translation-symmetric because they satisfy ${T_x(\cL_{5})\cong \cL_6}$ and ${T_x(\cL_6)\cong \cL_{5}}$. When specifying symmetry boundaries, $\cL_{1,2,4}$ correspond to $\Z_2$ dipole symmetries, $\cL_3$ to a ${\Z_2\times\Z_2}$ symmetry with an LSM anomaly involving translations, and $\cL_{5,6}$ to a ${\Z_2\times\Z_2}$ symmetry with a non-invertible translation. The $\cL_1$ boundary, for example, has a dipole symmetry because its symmetry defects $m_1$ and $m_2$ transform under $T_x$ as a monopole and dipole, respectively. On the other hand, the $\cL_3$ boundary describes a ${\Z_2\times\Z_2}$ symmetry. It is not modulated because its symmetry defects $e_1$ and $m_2$ do not transform under $T_x$ when $\cL_3$ is condensed, and it has an LSM anomaly because $\cL_3$ has no translation-symmetric magnetic Lagrangian algebras. Lastly, the $\cL_5$ and $\cL_6$ boundaries describe non-invertible translations implemented by first discrete gauging to change $\cL_5$ or $\cL_6$ to, for example, $\cL_2$, then performing $T_x$ (a symmetry of $\cL_2$), and then discrete gauging to return to $\cL_5$ or $\cL_6$.

In Sections~\ref{sec reflection SymTFT} and~\ref{sec TRS SymTFT}, we consider SymTFTs enriched by spatial reflections and time reversal, respectively. Unlike one-dimensional translations, these spacetime symmetries can exhibit symmetry fractionalization. For simplicity, we focus on  $\Z_2$ topological orders in ${2+1}$D enriched by these two symmetries. In both cases, the reflection and time-reversal symmetries enrich the topological order by acting trivially on $m$ anyons while fractionalizing onto the $e$ anyon. These examples closely resemble the internal $\Z_2$ symmetry fractionalization example discussed above. Namely, the $e$ condensed boundary is a symmetry boundary where the reflection and time reversal are extended by a $\Z_2$ symmetry operator, giving rise to a $\Z_4$ reflection/time reversal symmetry. The $m$ condensed boundary encodes a mixed anomaly between an internal $\Z_2$ symmetry and the reflection/time reversal symmetries. In both of these Sections, we present lattice models of these symmetry-enriched SymTFTs as well as quantum spin chain models realizing these symmetries.

For example, in Section~\ref{sec reflection SymTFT}, the stabilizer code Hamiltonian we consider can be recast as the toric code, whose stabilizers
\begin{align}\label{SumSecTC}
A_{\bm{r}}
=
\begin{tikzpicture}[scale = 0.5, baseline = {([yshift=-.5ex]current bounding box.center)}]
\draw[line width=0.015in, gray] (7.5,0) -- (13.5, 0);
\draw[line width=0.015in, gray] (10.5,-3) -- (10.5, 3);
\node at (9, .45) {\normalsize $\textcolor[HTML]{cb181d}{X}$};
\node at (12, .45) {\normalsize $\textcolor[HTML]{cb181d}{X}$};
\node at (10.5, 1.5) {\normalsize $\textcolor[HTML]{cb181d}{X}$};
\node at (10.5, -1.5) {\normalsize $\textcolor[HTML]{cb181d}{X}$};
\fill[black] (10.5,0) circle (6pt);
\node at (10, -.5) {\normalsize $r$};
\end{tikzpicture}~,
\hspace{40pt}
B_{\bm{r}}=\begin{tikzpicture}[scale = 0.5, baseline={([yshift=-.5ex]current bounding box.center)}]
\draw[line width=0.015in, gray] (-3, 0) -- (-3, 3) -- (0, 3) -- (0, 0) -- cycle;
\node at (-1.5, 0.4) {\normalsize $\textcolor[HTML]{2171b5}{Z}$};
\node at (-1.5, 3.45) {\normalsize $\textcolor[HTML]{2171b5}{Z}$};
\node at (-3, 1.5) {\normalsize $\textcolor[HTML]{2171b5}{Z}$};
\node at (0, 1.5) {\normalsize $\textcolor[HTML]{2171b5}{Z}$};
\fill[black] (-3,0) circle (6pt);
\node at (-3.5, -.5) {\normalsize $r$};
\end{tikzpicture},
\end{align}
The $e$ and $m$ anyons correspond to a violations of ${A_{\bm{r}} = 1}$ and ${B_{\bm{r}} = 1}$, respectively.
The spatial reflection symmetry operator we study, which commutes with the toric code Hamiltonian, is\footnote{We derive the stabilizer code and reflection operator in Section~\ref{reflectionEnrichedSymTFTStabCode} by gauging a $\Z_4$ reflection operator in a paramagnet Hamiltonian. The resulting stabilizer code~\eqref{eq:hamitlonian_reflection_SymTFT} and reflection operator~\eqref{eq:reflection_symTFT} appear different from~\eqref{SumSecTC} and~\eqref{SumSecReflec}, respectively, but are equivalent. The operators in Section~\ref{reflectionEnrichedSymTFTStabCode} become the simplified versions presented here by (1) enforcing the stabilizer ${Z_{\bm{r}} = 1}$, (2) shifting the site qubits acted by Pauli operators ${(\t{X}_{\bm{r}},\t{Z}_{\bm{r}})}$ to neighboring horizontal links and dropping the tildes: ${(\t{X}_{\bm{r}},\t{Z}_{\bm{r}})\mapsto (X_{\bm{r},x},Z_{\bm{r},x})}$, and (3) shifting each qubit on a plaquette rightward to the nearest vertical link: ${(X_{p},Z_{p})\mapsto (X_{\bm{r},y},Z_{\bm{r},y})}$.}
\begin{equation}\label{SumSecReflec}
    U_R = R\prod_{\bm{r}}X_{\bm{r},x},
\end{equation}
where $R$ is the lattice reflection operator about the center of the links ${\<(0,r_y),(1,r_y)\>}$ and $X_{\bm{r},x}$ acts on the qubit at the link ${\<(r_x,r_y),(r_x+1,r_y)\>}$. This reflection operator satisfies ${U_R^2 = 1}$ and generates a $\Z_2^R$ reflection symmetry. The $e$ anyons of the toric code carry fractional $U_R$ symmetry charge. Indeed, the string operator $Z_{(0,0),x}$, which creates a pair of $e$ anyons about the $U_R$ reflection center, carries $-1$ symmetry charge under $U_R$. Therefore, for an $R$ symmetric configuration of $e$ anyons, each $e$ anyon carries fractional $U_R$ symmetry charge. For the smooth boundary condition---the $m$ condensing boundary---of the toric code, $U_R$ operator is unaffected and acts as a $\Z_2^R$ operator which locally anti-commutes with the boundary ${W_e = \prod_{\text{boundary}} Z}$ logical operator signaling an LSM anomaly. The smooth boundary, therefore, describes an LSM anomalous ${\Z_2^R\times \Z_2}$ symmetry. On the other hand, for the rough boundary---the $e$ condensing boundary---the $U_R$ operator becomes a $\Z_4^R$ operator 
\begin{equation}
U_R = R\prod_{\bm{r}}X_{\bm{r},x}\prod_{\bm{\t{r}}\in\text{boundary}}X^{\t{r}_x}_{\bm{\t{r}},y}.
\end{equation}
Here, the $U_R$ symmetry operator is dressed by a modulated boundary operator so that it commutes with the boundary stabilizers.
The new reflection symmetry operator satisfies ${U_R^2 = W_m}$, where $W_m$ is the boundary ${W_m= \prod_{\text{boundary}} X}$ logical operator. Therefore, the rough boundary describes a $\Z_4^R$ reflection symmetry. Thus, gauging the $\Z_2$ symmetry of the LSM anomalous ${\Z_2^R\times \Z_2}$ symmetry leads to a dual $\Z_4^R$ reflection symmetry.

\section{Symmetry enriched SymTFT}\label{generalSection}

In this paper, we extend the SymTFT framework to incorporate spacetime symmetries and their possible non-trivial interplays with internal symmetries. Such interplays could arise from 't Hooft/LSM anomalies or non-trivial symmetry extensions. To systematically capture such symmetries using the SymTFT framework, we will consider SymTFTs of internal symmetries enriched by spacetime symmetries. The interplay between internal and spacetime symmetries is encoded in the spacetime symmetry enrichment of the SymTFT. An example of such a spacetime symmetry-enriched SymTFT was considered in~\cite{PLA240918113}. 

To motivate why spacetime symmetry-enriched SymTFTs provide a natural description, let us first recall a common construction of SymTFTs. Typically, a SymTFT is constructed by gauging the symmetry of an SPT phase. For simplicity, in this paper we will specialize to ${1+1}$D symmetry described by a finite group \(G\). 
In this setting, the SymTFT is a Dijkgraaf-Witten (DW) theory, which can be constructed by starting with the \(2+1\)D \(G\)-SPT associated with the 't Hooft anomaly ${\om\in H^3(BG,U(1))}$ of \(G\) and subsequently gauging $G$.

Suppose the total symmetry group $G$ has a normal subgroup $N$, and let $Q$ denote the quotient group $G/N$. 
When $G$ consists of only internal symmetries, there are two equivalent ways to construct the SymTFT for $G$. One approach is to start with the $G$-SPT and gauge $G$. Alternatively, one can first gauge $N$, arriving at a $Q$-enriched $N$-DW theory, and then gauge $Q$.\footnote{When $N$ is not a normal subgroup of $G$, the quotient ${Q = G/N}$ does not define an ordinary group symmetry but rather a non-invertible ${G/N}$ coset symmetry~\cite{HKZ240520401,HKZ250300105}. Gauging this non-invertible symmetry in the $N$ Dijkgraaf–Witten theory produces the $G$ DW theory.} This process can be summarized in the following commutative diagram:\footnote{ Mathematically, the symmetry defects of the $2+1$D \( G \)-SPT are described by the fusion 2-category \( 2\text{-}\Vec_G \). Meanwhile, the symmetry defects of the \( G \)-DW theory (modulo condensation defects) are captured by the center \( \mathcal{Z}(\Vec_G^\omega) \) of \( \Vec_G^\omega \). When \( \omega \) is trivial, the symmetry defects of the \( Q \)-enriched \( N \)-DW theory are given by the relative center \( \mathcal{Z}_{\Vec_G}(\Vec_N) \), which forms a \( Q \)-crossed braided extension of \( \mathcal{Z}(\Vec_N) \)~\cite{Gelaki:2009blp, BJL181100434}. Equivariantizing the \( Q \)-crossed braided fusion category amounts to gauging \( Q \)~\cite{BBC14104540}. In this case, the equivariantization of \( \mathcal{Z}_{\Vec_G}(\Vec_N) \) yields \( \mathcal{Z}(\Vec_G) \), as expected from gauging \( Q \). For non-trivial \( \omega \), the symmetry defects are described by a different \( Q \)-crossed braided fusion category, whose equivariantization results in \( \mathcal{Z}(\Vec^\omega_G) \).}
\begin{equation}\label{diagramgauging}
\begin{tikzpicture}[ baseline = {([yshift=-.5ex]current bounding box.center)}, scale=2]
\node (SPT) at (0, 0) {\normalsize \( G \)-SPT};
\node (TO) at (2.0, 0) {\normalsize \( G \)-DW Theory};
\node (SET) at (1.0, -1.2) {\normalsize \( Q \)-enriched \( N \)-DW Theory};

\draw[-{>[scale=1.2]}] (SPT) -- node[below left] { \( N \)} (SET);
\draw[-{>[scale=1.2]}] (SET) -- node[below right] { \( Q \)} (TO);
\draw[-{>[scale=1.2]}] (SPT) -- node[above] {\( G \)} (TO);
\end{tikzpicture},
\end{equation}
where each arrow is labeled by the symmetry being gauged. The interplay between \( N \) and \( Q \)  is encoded in the symmetry enrichment data of \( Q \) on the \( N \)-DW theory. The \( G \)-DW theory is deducible from the \( Q \)-enriched \( N \)-DW theory~\cite{BN12066625, BBC14104540}. Thus, while the \( G \)-DW theory serves as the SymTFT for the \( G \) symmetry, the \( Q \)-enriched \( N \)-DW theory provides an alternative perspective for understanding various aspects of the \( G \) symmetry. While we specialized to DW theories enriched by invertible symmetries, this perspective
applies to general SymTFTs with non-invertible symmetry enrichment. 

We remark that symmetry-enriched SymTFTs have also been studied in the context of non-invertible duality symmetries~\cite{ZC230401262, CHZ230811706, Antinucci:2023ezl, LSZ250107787}. These duality symmetries are constructed by half-gauging some symmetry \( \mathcal{S} \)~\cite{CCH211101139,CCH220409025}. The SymTFT for the full symmetry including both $\cS$ and the duality symmetry can be obtained by gauging an anyon-automorphism symmetry of the \( \mathcal{S} \) SymTFT. Therefore, by analyzing the anyon-automorphism symmetry enrichment on the \( \mathcal{S} \) SymTFT, one can learn about the non-invertible duality symmetry.

In this paper, we are interested in symmetries $G$ that involve
both internal and spacetime symmetries with a non-trivial interplay. Specifically, we consider scenarios where the quotient group ${Q\cong G/N}$
represents a spacetime symmetry while the normal subgroup \( N \) represents an internal symmetry. In this case, it is not clear how to gauge the spacetime symmetry \( Q \), which obstructs the direct construction of the SymTFT for the full symmetry \( G \). Nevertheless, the \( Q \)-enriched \( N \)-DW theory can still be constructed by starting with the \( G \)-SPT and gauging \( N \). 

In the remainder of this section, we will explore the general structure of \( Q \)-symmetry-enriched SymTFTs, treating \( Q \) as if it were an internal symmetry to develop intuition. Such an internal symmetry could describe a spacetime symmetry through the crystalline equivalence principle~\cite{TE161200846, MB200510265, ZYQ201215657, D210202941, MCB221002452}. In the subsequent sections, we will study explicit examples of spacetime symmetry-enriched SymTFTs, both on the lattice and in the continuum.

\subsection{SymTFT frameworks with symmetry enrichment}

Two fruitful applications of the SymTFT frameworks are discrete gauging and classifying phases, which are reviewed in Appendix~\ref{SymTFTReview}. The former is implemented by changing the gapped (topological) symmetry boundary, while the latter uses topological interfaces like those in Fig.~\ref{fig:TopHoloClass}. In this section, we will discuss how these applications generalize for ${2+1}$D symmetry-enriched SymTFTs. Similar to SymTFTs without symmetry enrichment, gapped boundaries will play a central role in these applications of the symmetry-enriched SymTFT. However, gapped boundaries of SETs have received considerably less attention than those of topological orders (see~\cite{BJL181100434, CW200202984, K220308156, S240810832, PLA240918113, LSZ250107787} for some previous discussion on SET boundaries). Therefore, before presenting these generalizations, we will first discuss some general aspects of gapped boundaries for SETs.

\subsubsection{An aside: gapped boundaries of SETs}\label{gappedBdySETsec}

This subsection will discuss gapped boundaries of ${2+1}$D $Q$-enriched bosonic topological orders, where $Q$ is a discrete group. A systematic study of the boundaries of SETs lies outside the scope of this paper. Instead, we will highlight a few properties relevant to our present discussion. Recall that the symmetry enrichment data consists of the action of $Q$ on the anyons of the topological order, which we denote by $\rho$,\footnote{The action $\rho$ can change the anyon labels and can act non-trivially on the anyons fusion/splitting spaces.} and the symmetry fractionalization class given by ${[\eta] \in H^2(BQ, A^\rho)}$,\footnote{Symmetry fractionalization can be understood as the dressing of trivalent junctions of $Q$ symmetry defects by Abelian anyons, which defines an $A$ 1-form symmetry. The 2-cocycle $\eta(q_1, q_2)$ specifies which Abelian anyon decorates the junction formed by fusing $q_1$ and $q_2$ symmetry defects.} where $A$ is the group of Abelian anyons~\cite{BBC14104540}.\footnote{We assume the Postnikov class ${H^3(BQ,A^\rho)}$ vanishes. A non-trivial Postnikov class forces $Q$ to spontaneously break in the bulk, which is not allowed in the symmetry-enriched SymTFT framework we are exploring.} There is also the ${H^3(BQ,\mathrm{U}(1))}$ class describing whether $Q$ is realized anomalously on boundaries, which we always assume is trivial.

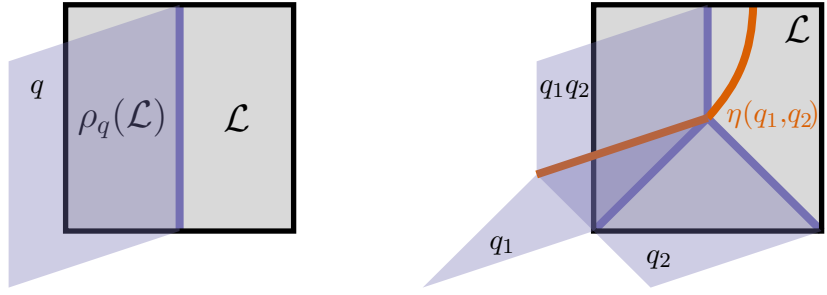
\begin{figure}
\centering
\begin{tikzpicture}[scale=.75, baseline={([yshift=-.5ex]current bounding box.center)},thick]
\fill[black, opacity=0.15] (0,0) rectangle (4,4);
\node[black] at (3, 2) {\Large $\cL$};
\node[black] at (1, 2) {\Large $\rho_q(\cL)$};
\draw[Qdef, line width=3pt] (2, 0) -- (2,4);
\draw[black, line width=2pt] (0,0) rectangle (4,4);
\fill[Qdef, opacity = .35] (-1,3) -- (2,4) -- (2,0) -- (-1,-1) -- cycle;
\node[black] at (-.5, 2.5) {\normalsize $q$};
\end{tikzpicture}
\hspace{40pt}
\begin{tikzpicture}[scale=.75, baseline={([yshift=-.5ex]current bounding box.center)},thick]
\fill[black, opacity=0.15] (0,0) rectangle (4,4);
\node[black, below left] at (4, 4) {\Large $\cL$};
\draw[Qdef, line width=3pt] (2, 2) -- (2,4);
\draw[Qdef, line width=3pt] (0, 0) -- (2,2);
\draw[Qdef, line width=3pt] (2, 2) -- (4,0);
\draw[Adef, line width=3pt] (2, 2) to[bend right=20] (2.8,4);
\draw[black, line width=2pt] (0,0) rectangle (4,4);
\fill[Qdef, opacity = .35] (-3,-1) -- (0,0) -- (2,2) -- (-1,1) -- cycle;
\draw[Adef, line width=3pt] (-1, 1) -- (2,2);
\fill[Qdef, opacity = .35] (1,-1) -- (4,0) -- (2,2) -- (-1,1) -- cycle;
\fill[Qdef, opacity = .35] (-1,3) -- (2,4) -- (2,2) -- (-1,1) -- cycle;
\node[black] at (-1.6, -.25) {\normalsize $q_1$};
\node[black] at (1.15, -.5) {\normalsize $q_2$};
\node[black] at (-.5, 2.5) {\normalsize $q_1q_2$};
\node[Adef] at (3.15, 2.05) {\normalsize $\eta(q_1,\!q_2\!)$};
\end{tikzpicture}
\vspace{10pt}
\caption{
Consider a gapped ${1+1}$D boundary (shown in gray) of a ${2+1}$D $Q$-enriched topological order. When $Q$ symmetry defects (shown in purple) end on the boundary, they have two distinct types of interplay. (Left) When the $Q$-action $\rho$ on anyons is non-trivial, it induces a $Q$ action on the boundary Lagrangian algebra $\cL$, denoted by ${\rho_q(\cL)}$. (Right) When the symmetry fractionalization class $[\eta]$ is non-trivial, a trivalent junction of $Q$ symmetry defects can end on the boundary, sourcing a $\eta(q_1,q_2)$ topological defect line on the boundary.
}
\label{fig:SETDirbdy}
\end{figure}

Consider a gapped boundary of the SET whose condensed anyons are described by the Lagrangian algebra $\cL$ of the underlying topological order. When $\rho$ is non-trivial, this boundary typically breaks the bulk symmetry $Q$ explicitly down to a subgroup ${Q_\cL}$. Indeed, the $Q$-action $\rho$ on the anyons induces an action on a Lagrangian algebra ${\cL \to \rho_q(\cL)}$. This transforms the anyons of ${\cL = \bigoplus_a n_a \, a}$ to the anyons
\begin{equation}
\rho_q(\cL) = \bigoplus_a n_a \, \rho_q(a),
\end{equation}
and also acts on the fusion space of the Lagrangian algebra, i.e., on the junctions of the anyons in $\cL$.
This implies that, as illustrated on the left panel of Fig.~\ref{fig:SETDirbdy}, a $Q$ symmetry defect can act on the boundary and transform $\cL$ to $\rho_q(\cL)$. When ${\rho_q(\cL) \cong \cL}$ for all ${q\in Q}$, the Lagrangian algebra $\cL$ is said to be $Q$-stable, and the enriching bulk symmetry ${Q_\cL \cong Q}$. However, if $\cL$ is not $Q$-stable, there is a ${q\in Q}$ such that ${\rho_q(\cL)\neq \cL}$, which causes $Q$ to be explicitly broken down to
\begin{equation}
Q_\cL = \{q\in Q\,\mid\, \rho_q(\cL) \cong \cL\}.
\end{equation}
From a Hamiltonian perspective, this means that the symmetry operator $U_q$ changes the boundary conditions from $\cL$ to ${\rho_q(\cL)}$, and does not commute with the Hamiltonian near the boundary when ${\rho_q(\cL)\not\cong\cL}$.

The gapped boundary also has its own boundary symmetries, generated by the uncondensed anyons on the boundary. The uncondensed Abelian anyons on the boundary form the group ${A_\cL}$. We denote by $G_\cL$ the group formed by all of the invertible symmetry operators, both the $Q_\cL$ operators that act non-trivially in the bulk and the $A_\cL$ operators that act non-trivially only on the boundary. It is described by the group extension\footnote{In general, the full symmetry can be a non-invertible symmetry due to the non-Abelian anyons. By restricting to the invertible symmetry $A_\cL$, we focus on its invertible sub-symmetry.}
\begin{equation}\label{extensionGcL}
1 \to A_\cL \to G_\cL \to Q_\cL \to 1.
\end{equation}
The action of $Q_\cL$ on $A_\cL$ in~\eqref{extensionGcL} is inherited from the action $\rho$. The extension class of~\eqref{extensionGcL} is non-trivial if there exists a non-trivial ${\eta(q_1,q_2)}\not\in\cL$ for ${q_1,q_2\in Q_\cL}$. Indeed, if such an anyon exists, the $Q_\cL$ symmetry would be enlarged by a group extension on the boundary, as shown by the right panel of Fig.~\ref{fig:SETDirbdy}. From a Hamiltonian perspective, this means that the symmetry operators $U_q$ satisfy 
\begin{equation}
U_{q_1}\times U_{q_2} = U_{q_1q_2}\times W_{\eta(q_1,q_2)}~,
\end{equation}
where $W_{\eta(q_1,q_2)}$ is the string operator for the anyon $\eta(q_1,q_2)$ acting along the gapped boundary. When the symmetry-enrichment is trivial, i.e., when $\rho$ is the identity automorphism and $[\eta]$ is trivial, ${Q_\cL \cong Q}$ and ${G_\cL \cong A_\cL \times Q}$.

An important class of gapped boundaries are those whose Lagrangian condensable algebra is $Q$-symmetric~\cite{BJL181100434, CW200202984, K220308156}. A $Q$-symmetric condensable algebra $\cA$ is a condensable algebra that:\footnote{A condensable algebra satisfying these two conditions is a $Q$-equivariant algebra~\cite{BJL181100434}} 
\begin{enumerate}
\item is $Q$-stable, i.e., ${\rho_q(\cA)\cong\cA}$\,;
\item  no anyons in $\cA$ carry fractional $Q$ charge.
\end{enumerate}
We note that a boundary whose condensable algebra has anyons carrying fractional symmetry charge generally causes the symmetry to spontaneously break on the boundary.

Notably, a gapped boundary with a $Q$-symmetric condensable algebra $\cL$  neither explicitly breaks nor extends $Q$. First, it does not explicitly break $Q$ since $\cL$ is $Q$-stable. Second, perhaps less obviously, it does not extend $Q$, i.e., the sequence~\eqref{extensionGcL} 
splits.
This follows from $\cL$ having to condense all ${\eta(q_1,q_2)}$ anyons to satisfy property 2 of being $Q$-symmetric. Indeed, because $\cL$ is Lagrangian, every anyon not in $\cL$ must braid non-trivially with at least one anyon in $\cL$.\footnote{This can be proven using that a Lagrangian algebra ${\cL = \bigoplus_a n_a \, a}$ satisfies ${\sum_a S_{ba} n_{a}  = n_b}$, where $S$ is the topological $S$ matrix~\cite{KKO210713091}. For any anyon $b$ not in $\cL$, ${n_b = 0}$ and this equation becomes ${\sum_a S_{ba} n_{a} = 0}$. Because ${S_{b1} = d_b/\cD>0}$, where $\cD$ is the total quantum dimension, ${\sum_{a\neq 1}  S_{ba} n_{a}< 0}$. Therefore, there must be some anyon $a$ with ${n_a \neq 0}$ that has ${S_{ba} \neq 0}$ and braids non-trivially with $b$. We thank Carolyn Zhang for informing us of this proof.} Since a $Q$-symmetric $\cL$ does not include anyons with fractional $Q$ charge, the $\eta(q_1,q_2)$ anyons do not braid with any anyons in $\cL$, and therefore they must be condensed since $\cL$ is Lagrangian. Because every $\eta(q_1,q_2)$ anyon is condensed, the boundary extension mechanism never occurs.\footnote{Using the same reasoning, we find that any boundary for which the extension class of~\eqref{extensionGcL} is non-trivial must condense anyons carrying fractional symmetry charge and spontaneously break the enriching symmetry.} Importantly, not every $Q$-enriched topological order admits a $Q$-symmetric Lagrangian algebra.\footnote{For example, $2+1$D $\Z_2$ topological order, i.e., $2+1$D $\Z_2$ toric code, enriched by a $\Z_2$ symmetry that exchanges $e$ and $m$ anyons does not have a $\Z_2$-symmetric Lagrangian algebra.}

Let us consider an example where a $\Z_2$ symmetry enriches a $2+1$D $\Z_2$ topological order, i.e., the $\Z_2$ toric code. The anyons are denoted by $\{1, e, m, f\}$, where $e$ and $m$ are bosons that braid non-trivially with each other, and ${f = e \times m}$ is a fermion. There are two Lagrangian condensable algebras: 
\begin{equation} 
\cL_e = 1 \oplus e, \qquad \cL_m = 1 \oplus m.
\end{equation}
Suppose the $\Z_2$ action $\rho$ on the anyons is trivial, but the symmetry fractionalization class is non-trivial. In particular, we consider the fractionalization pattern ${\eta(-1,-1) = m}$, such that the $e$ anyon carries a fractional $\Z_2$ charge (see Appendix~\ref{Z2 enriched TC App} for an exactly solvable lattice model realizing this SET). In this case, the Lagrangian algebra $\cL_m$ is $\Z_2$-symmetric. In contrast, the Lagrangian algebra $\cL_e$ is $\Z_2$-stable but not $\Z_2$-symmetric since the $e$ anyon carries a fractional charge. In the presence of an $e$-condensed boundary, the $\Z_2$ symmetry operator squares to an $m$ string operator on the boundary. This extends the ${Q_\cL = \Z_2}$ bulk symmetry to ${G_\cL = \Z_4}$ on the boundary. 

In general, there can be two gapped boundaries that share the same Lagrangian algebra but differ by the spontaneous breaking of $Q_\cL$. We will always work with the gapped boundary for a given $\cL$ that minimally spontaneously breaks $Q_\cL$ and has the fewest local topological operators.

\subsubsection{Discrete gauging with the quiche}

Having discussed some aspects of gapped boundaries of SETs, we now turn to applying the symmetry-enriched SymTFT to discrete gauging. As reviewed in 
Appendix~\ref{SymTFTReview}, discrete gauging for the standard SymTFT is performed by changing the symmetry boundary. For the $Q$-enriched SymTFT, suppose the total $G$-SymTFT can be constructed by gauging $Q$ in a $Q$-enriched $N$-SymTFT (i.e., the situation shown in~\eqref{diagramgauging}). $N$-SymTFT is nothing but a $N$ gauge theory (i.e., the ground state subspace of Kitaev's $N$ quantum double model). As mentioned, the non-trivial enrichment of $Q$ will encode the interplay between the $Q$ and $N$ symmetries that defines the total symmetry $G$. The corresponding ``quiche'' for this symmetry-enriched SymTFT is
\begin{equation*}
\begin{tikzpicture}[scale=.75, baseline={([yshift=-.5ex]current bounding box.center)},thick]
\fill[SymTFTcolor] (-1.5,0) rectangle (7,2);
\draw[black, line width=4pt] (-1.5, 2) -- (7,2);
\node[black, left] at (-1.5, 2) {\Large $\mathfrak{B}^{\text{sym}}_{G}$};
\node[black] at (2.75, 1) {\Large $Q\text{-enriched $N$ gauge theory}$};
\end{tikzpicture}.
\end{equation*}
The gapped boundary $\mathfrak{B}^{\text{sym}}_{G}$ has the electric Lagrangian algebra $\cL_e$ of $N$ gauge theory condensed, and hosts topological defects describing the ${A_{\cL_e}\cong N}$ symmetry.
In the presence of this gapped boundary, the enriching symmetry $Q$ becomes $Q_{\cL_e}$, and the symmetry-enriched quiche describes the total symmetry ${G_{\cL_e} \cong G}$ through the group extension~\eqref{extensionGcL}.

Changing the Lagrangian algebra specifying the boundary condition from $\cL_e$ to $\cL'$ leads to a new symmetry boundary condition. It follows from the SymTFT without enrichment that this change corresponds to gauging the $N$ sub-symmetry of $G$. When $N$ is Abelian, the resulting quiche is 
\begin{equation*}
\begin{tikzpicture}[scale=.75, baseline={([yshift=-.5ex]current bounding box.center)},thick]
\fill[SymTFTcolor] (-1.5,0) rectangle (7,2);
\draw[black, line width=4pt] (-1.5, 2) -- (7,2);
\node[black, left] at (-1.5, 2) {\Large $\mathfrak{B}^{\text{sym}}_{{G^\vee}}$};
\node[black] at (2.75, 1) {\Large $Q\text{-enriched $N$ gauge theory}$};
\end{tikzpicture},
\end{equation*}
which has an enriching $Q_{\cL'}$ $0$-form symmetry in its bulk and a $A_{\cL'}\cong N^\vee$ symmetry on its boundary.
This quiche now describes the total symmetry ${G_{\cL'}\cong G^\vee}$ formed by the enriching $Q_{\cL'}$ and boundary ${A_{\cL'}\cong N^\vee}$ symmetries. 
The interplay between $Q_{\cL'}$ and $N^\vee$ is generally different from the interplay between $Q_{\cL_e}$ and $N$, which captures the effect of $Q$ on the dual symmetry $N^\vee$ arising from gauging $N$. While the $Q$-enriched SymTFT can be used to gauge the $N$ sub-symmetry of $G$, it cannot gauge a sub-symmetry of $G$ involving the $Q$ symmetry.

Let us contextualize this general discussion to the previously considered example of a $\Z_2$-enriched $\Z_2$ topological order (we will consider more examples throughout the paper). The two Lagrangian algebras $\cL_m$ and $\cL_e$ both describe ${\Z_2}$ symmetries on the top boundary (i.e., ${A_{\cL_e} = A_{\cL_m} = \Z_2}$). When the top boundary has $\cL_m$ condensed, the $\Z_2$-enriched symmetry is unaffected, and the total symmetry described by the quiche is ${Q\times A_{\cL_m}= \Z_2\times\Z_2}$. When the top boundary has $\cL_e$ condensed, the enriched symmetry is extended by $A_{\cL_{e}}$, and the quiche describes a $\Z_4$ symmetry. Therefore, gauging the ${A_{\cL_{m}} = \Z_2}$ sub-symmetry of ${G_{\cL_m} = Q\times A_{\cL_m} = \Z_2\times\Z_2}$ leads to a dual ${G_{\cL_e} = \Z_4}$ symmetry. This suggests that the quiche with the $\cL_m$ boundary describes an anomalous ${\Z_2\times\Z_2}$ symmetry~\cite{BT170402330}, which we will confirm in the next subsection. 

This example is a particular instance of something more general. Namely, starting with a symmetry boundary specified by a $Q$-symmetric Lagrangian algebra and changing boundary conditions to one whose Lagrangian algebra contains anyons with fractional symmetry charges will result in an extension of $Q$ via the mechanism described in the previous section.

\subsubsection{Gapped phases with the sandwich}\label{gappedSymPhasesSec}

We now discuss how symmetry-enriched SymTFTs can be used to classify gapped phases. For simplicity, we will assume the entire symmetry group $G$ is formed by a normal, Abelian subgroup $N$ of $G$ and quotient group ${Q\cong G/N}$, and consider the SymTFT for $N$ enriched by $Q$. The generalization to non-Abelian $N$ is straightforward.

\begin{figure}
\centering
\begin{tikzpicture}[scale=.75, baseline={([yshift=-.5ex]current bounding box.center)},thick]
\fill[SymTFTcolor] (0,0) rectangle (4,1.5);
\fill[reducedSymTFTcolor] (0,-1.5) rectangle (4,0);
\draw[black, line width=4pt] (0, 0) -- (4,0);
\draw[black, line width=4pt] (0, 1.5) -- (4,1.5);
\node[black, left] at (0, 0) {\Large $\mathcal{I}_{\cA}$};
\node[black, left] at (0, 1.5) {\Large $\mathfrak{B}^{\text{sym}}_{\cS}$};
\node[black] at (2, 0.75) {\Large $\mathfrak{Z}(\cS)$};
\node[black] at (2, -0.75) {\Large $\mathfrak{Z}(\cS)/\cA$};
\end{tikzpicture}
\caption{
The classification of quantum phases characterized by a symmetry $\cS$ using the SymTFT $\mathfrak{Z}(\cS)$ is based on interfaces $\cI_\cA$ with condensable algebras $\cA$ of the SymTFT. See Appendix~\ref{SymTFTReview} for an introduction.
}
\label{fig:TopHoloClass}
\end{figure}

Like for SymTFTs without symmetry enrichment, we consider the sandwich configuration shown in Fig.~\ref{fig:TopHoloClass}, where the ${1+1}$D symmetry is determined by the gapped top boundary and different gapped bottom boundaries correspond to different gapped phases. Since the top and bottom boundaries are gapped, they have corresponding Lagrangian algebras. While the Lagrangian algebra of the top boundary is generally unconstrained, not every Lagrangian algebra can be condensed on the bottom boundary. Indeed, recall that the enriching symmetry can be modified by a gapped boundary whose Lagrangian algebra is not $Q$-symmetric. Since the total symmetry $G$ should not depend on the bottom boundary, the Lagrangian algebra for the bottom boundary must not change the enriching symmetry. This is always satisfied, for instance, if the Lagrangian algebra for the bottom boundary is $Q$-symmetric. However, depending on the top boundary, a non $Q$-symmetric Lagrangian algebra can also leave the enriching symmetry invariant.\footnote{For example, an extension of $Q$ due to a non $Q$-symmetric Lagrangian algebra condensed on the bottom boundary can be trivialized by a top boundary with appropriate condensed Lagrangian algebra.}

In this paper, we will consider symmetry-enriched SymTFT sandwiches whose top and bottom boundaries have $Q$-symmetric Lagrangian algebras condensed.\footnote{Going beyond $Q$-symmetric Lagrangian algebras is subtle. For one, the typical separation of kinematics and dynamics provided by the SymTFT's top and bottom boundaries does not apply. For instance, the top boundary can spontaneously break the enriching symmetry by condensing anyons with fractional charge. Furthermore, it is unclear whether anomalies can be detected using non $Q$-symmetric Lagrangian algebras. In particular, the symmetry boundary could describe an anomaly-free symmetry but spontaneously break the enriching symmetry, thereby causing the sandwich never to have a unique ground state.} 
In this case, the enriching $Q$ symmetry is not affected by either gapped boundary, and the SymTFT describes the semidirect product symmetry group ${G= N \rtimes Q}$.
These symmetric sandwiches of the symmetry-enriched SymTFT can realize all $Q$-enriched gapped phases of the total symmetry $G$, which are phases where ${Q\subset G}$ is not spontaneously broken.
This construction is the natural extension of the Lagrangian algebra approach to classifying gapped phases, which is reviewed in Appendix \ref{SymTFTReview} for the case without symmetry enrichment.
While it cannot realize all gapped phases of $G$, it can still detect anomalies of $G$, even those that depend on $Q$. Indeed, if there are no $Q$-enriched gapped phases that are SPTs, then there are no $G$ SPTs and $G$ is anomalous.\footnote{The anomalies of $G$ captured here are ones that involve the subsymmetry $N$. Self-anomalies of $Q$, on the other hand, would not be detected using only the data provided. They are reflected in the defectification class $H^3(BQ,\mathrm{U}(1))$ of the SET, which we have assumed to be trivial.} Mixed anomalies involving $Q$ and $N$ generally manifest by the Lagrangian algebras that would have corresponded to SPTs not being $Q$-symmetric. Importantly, this approach applies to spacetime symmetries, which is the paper's focus, in which case these mixed anomalies are the 
so-called LSM anomalies.

To consider an example, let us return to the $\Z_2$-enriched $\Z_2$ topological order discussed previously. The only symmetric Lagrangian algebra is the magnetic Lagrangian algebra ${\cL_m = 1 \oplus m}$. When $\cL_m$ is condensed on the top boundary and bottom boundaries, the sandwich describes a ${\Z_2 \times \Z_2}$ symmetry with the spontaneous symmetry breaking (SSB) pattern ${\Z_2 \times \Z_2\to \Z_2}$ (where the unbroken symmetry is the enriching $\Z_2$ symmetry). Since $\cL_m$ is the only $\Z_2$-symmetric Lagrangian algebra, there are no other $\Z_2$-symmetric ${\Z_2\times\Z_2}$ gapped phases and, therefore, no SPTs. We conclude that $\cL_m$ describes a ${\Z_2 \times \Z_2}$ symmetry that is anomalous. This is consistent with how changing $\cL_m$ to $\cL_e$ in the quiche causes ${\Z_2 \times \Z_2}$ to become $\Z_4$: the anomaly leads to an extension, as expected. 

This example is a special case of a more general property. Namely, if the symmetry fractionalization class $[\eta]$ is non-trivial, then every symmetric Lagrangian algebra $\cL$  corresponds to an anomalous symmetry. Indeed, every $Q$-symmetric $\cL$ contains $\eta(q_1, q_2)$, resulting in a nontrivial overlap among all $Q$-symmetric $\cL$ and therefore spontaneous symmetry breaking phases.

\subsection{Symmetry enrichment from symmetry interplays}

Having discussed how the quiche and sandwich frameworks generalize to symmetry enriched SymTFTs, we now explore the possible forms of symmetry enrichment that can occur in these theories.

\subsubsection{Group extensions}\label{extensionsSubSec}

Consider an anomaly-free 0-form symmetry in ${(d+1)}$-dimensional spacetime described by a finite group $G$. Suppose $G$ has an Abelian normal subgroup $A$. Then, $G$ can be formulated as a group extension:
\begin{equation} \label{SESsec2}
\mathbf{1} \to A \to G \to Q \to \mathbf{1},
\end{equation}
where ${Q \cong G/A}$ is the quotient group. As reviewed in Appendix~\ref{grpExtApp}, this group extension is characterized by two key pieces of data.
Firstly, denoting the automorphism group of $A$ by $\Aut(A)$, there exists a group homomorphism ${\rho: Q \to \Aut(A)}$ that describes the action of $Q$ on $A$. Secondly, for a given $\rho$, inequivalent extensions are labeled by the cohomology classes ${[c] \in H^2(BQ, A^\rho)}$, where $BQ$ is the classifying space of $Q$, and $A^\rho$ is a $Q$-module with underlying Abelian group $A$ and $Q$-action given by $\rho$. The elements of $G$ can be represented by pairs ${(a, q) \in A \times Q}$, with group multiplication given by
\begin{equation} \label{multiplicationSec2}
(a_1, q_1) \cdot (a_2, q_2) = \big(a_1 + \rho_{q_1}(a_2) + c(q_1, q_2), q_1 q_2\big),
\end{equation}
where $c$ is a representative 2-cocycle of the cohomology class $[c]$.

We graphically represent the symmetry defects of $G$ by black lines and label them by the group elements of $G$ as ${(a,q)}$. A general $G$ symmetry defect can be decomposed into an $A$ and a $Q$ symmetry defect, which we represent by red and blue lines, respectively. The decomposition follows the group multiplication law~\eqref{multiplicationSec2}, which are summarized by the following fusion diagrams:
\begin{equation}
\begin{tikzpicture}[baseline={([yshift=-.5ex]current bounding box.center)},>=Triangle, thick]
\draw[Adef, line width=2pt] (3, 0) -- (2,1);
\draw[Adef, ->, line width=2pt,] (3, 0) -- (2.4, .6);
\node[Adef, above] at (2, 1) {$a$};
%
\draw[line width=2pt, Qdef] (3, 0) -- (4,1);
\draw[->, line width=2pt, Qdef] (3, 0) -- (3.6, 0.6);
\node[Qdef, above] at (4, 1) {$q$};
\draw[black, line width=2pt] (3, -1) -- (3,0);
\draw[->, line width=2pt, black] (3, -1) -- (3, -.3);
\node[black, below] at (3, -1) {$(a,q)$};
\end{tikzpicture}
\hspace{40pt}
\begin{tikzpicture}[baseline={([yshift=-.5ex]current bounding box.center)},>=Triangle, thick]
\draw[Qdef, line width=2pt] (3, 0) -- (2,1);
\draw[Qdef, ->, line width=2pt,] (3, 0) -- (2.4, .6);
\node[Qdef, above] at (2, 1) {$q$};
%
\draw[line width=2pt, Adef] (3, 0) -- (4,1);
\draw[->, line width=2pt, Adef] (3, 0) -- (3.6, 0.6);
\node[Adef, above] at (4, 1) {$\rho_{q^{-1}}(a)$};
\draw[black, line width=2pt] (3, -1) -- (3,0);
\draw[->, line width=2pt, black] (3, -1) -- (3, -.3);
\node[black, below] at (3, -1) {$(a,q)$};
\end{tikzpicture}.
\end{equation}
Using these fusion rules, we arrive at the following interplays between the $a$ and $q$ symmetry defects:
\begin{equation}
\begin{tikzpicture}[baseline={([yshift=-.5ex]current bounding box.center)}, >=Triangle, thick]
\draw[Adef, line width=2pt] (4, -1) -- (2,1);
\draw[Adef, ->, line width=2pt,] (4, -1) -- (3.3, -.3);
\draw[Adef, ->, line width=2pt,] (4, -1) -- (2.4, .6);
\node[Adef, above] at (2, 1) {$\rho_q(a)$};
\node[Adef, below] at (4, -1) {$a$};
%
\draw[line width=2pt, Qdef] (2, -1) -- (4,1);
\draw[->, line width=2pt, Qdef] (2, -1) -- (2.7, -0.3);
\draw[->, line width=2pt, Qdef] (2, -1) -- (3.6, 0.6);
\node[Qdef, above] at (4, 1) {$q$};
\node[Qdef, below] at (2, -1) {$q$};
\end{tikzpicture}
\hspace{50pt}
\begin{tikzpicture}[baseline={([yshift=-.5ex]current bounding box.center)},>=Triangle, thick]
\draw[line width=2pt, Adef] (7, 0) -- (6,1);
\draw[->, line width=2pt, Adef] (7, 0) -- (6.4, 0.6);
\node[Adef, above] at (6, 1) {$c(q_1,q_2)$};
\draw[Qdef, line width=2pt] (6, -1) -- (7,0);
\draw[Qdef, ->, line width=2pt,] (6, -1) -- (6.7, -.3);
\node[Qdef, below] at (6, -1) {$q_1$};
%
\draw[Qdef, line width=2pt] (8, -1) -- (7,0);
\draw[Qdef, ->, line width=2pt,] (8, -1) -- (7.3, -.3);
\node[Qdef, below] at (8, -1) {$q_2$};
%
\draw[Qdef, line width=2pt] (7, 0) -- (8,1);
\draw[Qdef, ->, line width=2pt,] (7, 0) -- (7.6, .6);
\node[Qdef, above] at (8, 1) {$q_1 q_2$};
\fill[Adef] (7,0) circle (4pt);
\end{tikzpicture}.
\end{equation}
The diagram on the left shows how $Q$ symmetry defects act on $A$ symmetry defects via the group homomorphism $\rho$. The diagram on the right shows how $A$ symmetry defects can terminate at trivalent junctions of $Q$ symmetry defects, where we highlight the boundary of the $A$ symmetry defects with red dots. By Poincar{\'e} duality, this configuration implies that on a triangulated spacetime, the background gauge fields $\cQ$ and $\cA$ for $Q$ and $A$, respectively, satisfy the following conditions on each $2$-simplex of ordered vertices ${(ijk)}$:
\begin{equation}
\cQ_{ij}\cQ_{jk}=\cQ_{ik},\hspace{40pt} \cA_{ij} + \rho_{\cQ_{ij}}(\cA_{jk}) - \cA_{ik} = c(\cQ_{ij}, \cQ_{jk}).
\end{equation}
These conditions can be compactly written as ${\dd \cQ = 1}$ and ${\dd_\rho \cA = \cQ^*c}$, where $\cQ^*c$ denotes the pullback of the 2-cocycle $c$ by $\cQ$.

We now proceed to construct the SymTFT for the $A$ sub-symmetry, while carefully tracking its interplay with the $Q$ symmetry. This is achieved by gauging the $A$ sub-symmetry of a trivial $G$-SPT in ${(d+2)}$ dimensions. It leads to an $A$ gauge theory enriched by the $Q$ symmetry. From a Euclidean field theory perspective, the symmetry enrichment is encoded in the coupling between the $Q$ background field and the $A$ SymTFT, which gives rise to the following partition function for the SymTFT:
\begin{equation}\label{SymTFTsec2.1}
\cZ[\cQ] = \sum_{\cA,\cB} \exp[\ii \int_X\,\cB\cup (\dd_\rho \cA-\cQ^*c)].
\end{equation}
Here, the sum is over all 1-cochains ${\cA\in C^1(X_{d+2 },A)}$ and $d$-cochains ${\cB\in C^d(X_{d+2 },A^\vee)}$. When the background gauge field $\cQ$ is turned off, this reduces to an $A$ gauge theory, that is, the $A$ SymTFT. The coupling to $\cQ$ implies that the $Q$ symmetry defects have a non-trivial interplay with the electric and magnetic defects of $A$ SymTFT. In what follows, we will show that a non-trivial $\rho$ results in a non-trivial action of $Q$ on both the electric and magnetic defects, whereas a non-trivial $[c]$ induces fractional $Q$ symmetry charges on the electric defects. We will follow the analysis in \cite{BT170402330, T171209542,HT190411550}.

The $A$ SymTFT has a ${A^\vee = \Hom(A,U(1))}$ $d$-form symmetry, which is dual to the $A$ 0-form symmetry that were gauged. The symmetry defects of this $A^\vee$ $d$-form symmetry are the Wilson lines of $A$. These Wilson lines are labeled by the group elements $\chi \in A^\vee$, which are in one-to-one correspondence with the irreducible representations of $A$. They take the form
\ie
{W_\chi(C) = \ee^{\ii \oint_C \chi({\cA})}}~,
\fe
where $C$ is a 1-cycle in spacetime. These Wilson lines are the electric defects of the $A$ SymTFT.  Other than these electric defects, the SymTFT also has codimension-2 magnetic defects, which are the boundaries of the codimension-1 $A$ symmetry defects that were gauged. Since these $A$ symmetry defects become trivial after gauging, their boundaries become genuine codimension-2 defects. These magnetic defects are labeled by the group elements $a\in A$ and take the form
\ie
V_a(M)=\ee^{\ii \oint_M a(\cB)}
\fe
where $M$ is a $d$-cycle in spacetime. They generate an $A$ 1-form symmetry.

Before gauging $A$, the Wilson lines $W_\chi$ have non-trivial interactions with the $G$ symmetry defects. Representing the Wilson lines as green lines, we have the following relations:
\begin{equation}\label{Wchiinterplay}
\begin{tikzpicture}[baseline={([yshift=-.5ex]current bounding box.center)},>=Triangle, thick]
\draw[Wdef, line width=2pt] (20, -1) -- (18,1);
\draw[Wdef, ->, line width=2pt,] (20, -1) -- (19.3, -.3);
\draw[Wdef, ->, line width=2pt,] (20, -1) -- (18.4, .6);
\node[Wdef, above] at (18, 1) {$\rho^\vee_q(\chi)$};
\node[Wdef, below] at (20, -1) {$\chi$};
%
\draw[line width=2pt, Qdef] (18, -1) -- (20,1);
\draw[->, line width=2pt, Qdef] (18, -1) -- (18.7, -0.3);
\draw[->, line width=2pt, Qdef] (18, -1) -- (19.6, 0.6);
\node[Qdef, above] at (20, 1) {$q$};
\node[Qdef, below] at (18, -1) {$q$};
%
%
\draw[line width=2pt, Adef] (8.5, -.5) -- (10,1);
\draw[->, line width=2pt, Adef] (8.5, -.5) -- (9.6, 0.6);
\node[Adef, above] at (10, 1) {$a$};
\fill[Adef] (8.5,-.5) circle (4pt);
%
%
\draw[Wdef, line width=2pt] (10, 0) -- (8,0);
\draw[Wdef, ->, line width=2pt,] (10, -0) -- (9.3, 0);
\draw[Wdef, ->, line width=2pt,] (10, 0) -- (8.3, 0);
\node[Wdef, below] at (8, .7) {$\chi$};
\node at (11.2, 0) {$= \ee^{\ii\chi(a)}$};
\draw[line width=2pt, Adef] (12.7, -.5) -- (14.2,1);
\draw[->, line width=2pt, Adef] (12.7, -.5) -- (13.8, 0.6);
\node[Adef, above] at (14.2, 1) {$a$};
\fill[Adef] (12.7,-.5) circle (4pt);
%
%
\draw[Wdef, line width=2pt] (14.2, -1) -- (12.2,-1);
\draw[Wdef, ->, line width=2pt,] (14.2, -1) -- (12.95, -1);
\node[Wdef, above] at (14, -.85) {$\chi$};
\end{tikzpicture}.
\end{equation}
Note that although all defects are drawn as lines, only the green ones are truly one-dimensional; the blue and red ones denote codimension-1 defects. The left diagram shows that when a Wilson line $W_\chi$ passes through the boundary of an $a$-symmetry defect, it picks up a phase $\ee^{\ii\chi(a)}$. On the other hand, the right diagram illustrates the action of a $q$-symmetry defect on a Wilson line via the group homomorphism $\rho^\vee\colon Q \to \Aut(A^\vee)$, where $\Aut(A^\vee)$ denotes the automorphism group of $A^\vee$. This group homomorphism $\rho^\vee$ is determined by the group homomorphism $\rho$. Consider the following process
\begin{equation}
\begin{tikzpicture}[baseline={([yshift=-.5ex]current bounding box.center)},>=Triangle, thick]
\draw[Adef, line width=2pt] (4, 0) -- (2,0);
\draw[Adef, ->, line width=2pt,] (4, 0) -- (3.3, 0);
\draw[Adef, ->, line width=2pt,] (4, 0) -- (2.4, 0);
\node[Adef, above] at (2, 0.1) {$\rho_q(a)$};
\node[Adef, above] at (4, 0.1) {$a$};
\fill[Adef] (2,0) circle (4pt);
\fill[Adef] (4,0) circle (4pt);
%
\draw[line width=2pt, Qdef] (3, -1) -- (3,1);
\draw[->, line width=2pt, Qdef] (3, -1) -- (3, -0.3);
\draw[->, line width=2pt, Qdef] (3, -1) -- (3, 0.6);
\node[Qdef, above] at (3, 1) {$q$};
\node[Qdef, below] at (3, -1) {$q$};
%
\draw[line width=2pt, Wdef] (1, -1) -- (1,1);
\draw[->, line width=2pt, Wdef] (1, -1) -- (1, -0.3);
\draw[->, line width=2pt, Wdef] (1, -1) -- (1, 0.6);
\node[Wdef, above] at (1, 1) {$\chi$};
\node at (6.5, 0) {$=\ee^{\ii [\chi(\rho_q(a)) - \rho^\vee_{q^{-1}}(\chi)(a)]}$};
%
\draw[line width=2pt, Wdef] (12.2, -1) -- (12.2,1);
\draw[->, line width=2pt, Wdef] (12.2, -1) -- (12.2, -0.3);
\draw[->, line width=2pt, Wdef] (12.2, -1) -- (12.2, 0.6);
\node[Wdef, above] at (12.2, 1) {$\rho^\vee_{q^{-1}}(\chi)$};
\draw[Adef, line width=2pt] (11.2, 0) -- (9.2,0);
\draw[Adef, ->, line width=2pt,] (11.2, 0) -- (10.5, 0);
\draw[Adef, ->, line width=2pt,] (11.2, 0) -- (9.6, 0);
\node[Adef, above] at (9.2, 0.1) {$\rho_q(a)$};
\node[Adef, above] at (11.2, 0.1) {$a$};
\fill[Adef] (9.2,0) circle (4pt);
\fill[Adef] (11.2,0) circle (4pt);
%
\draw[line width=2pt, Qdef] (10.2, -1) -- (10.2,1);
\draw[->, line width=2pt, Qdef] (10.2, -1) -- (10.2, -0.3);
\draw[->, line width=2pt, Qdef] (10.2, -1) -- (10.2, 0.6);
\node[Qdef, above] at (10.2, 1) {$q$};
\node[Qdef, below] at (10.2, -1) {$q$};
\end{tikzpicture}
\end{equation}
where in going from the LHS to the RHS, we hold the four-way junction fixed and move the Wilson line across it. The Wilson line first picks up a phase $\ee^{\ii \chi(\rho_q(a)) }$ when it crosses the left boundary of the $\rho_q(a)$-symmetry defect, and then another phase $\ee^{-\ii \rho^\vee_{q^{-1}}(\chi(a))}$ when crossing the right boundary of the $a$-symmetry defect. Alternatively, one can reach the RHS via a different sequence of moves: starting from the LHS, first move the $q$-symmetry defect past the right boundary of $a$-symmetry defect, then move the Wilson line past the $\rho_q(a)$-symmetry defects, and lastly move the $q$-symmetry defect back to its initial position to arrive at the RHS. This process does not create any phase. The consistency of the two processes therefore implies that
\begin{equation}
\exp[\ii\rho^\vee_q(\chi)(a) ]= \exp[\ii\chi(\rho_{q^{-1}}(a))],
\end{equation}
which fixes the group homomorphism $\rho^\vee$ in terms of the group homomorphism $\rho$.

After gauging $A$, the Wilson lines $W_\chi$ become the symmetry defects of the dual $A^\vee$ $d$-form symmetry. The right diagram of~\eqref{Wchiinterplay} then implies that the $Q$ symmetry defects act on these $A^\vee$ symmetry lines via the group homomorphism $\rho^\vee$. The $Q$ symmetry defects also act on the codimension-2 symmetry defects $V_a$ of the $A$ 1-form symmetry. Recall that these codimension-2 defects are the boundaries of the codimension-1 $A$ symmetry defects that were gauged. Since the $Q$ symmetry defects act on the $A$ symmetry defects by $\rho$ before gauging, they also act on the codimension-2 $A$ symmetry defects by $\rho$ after gauging. All the codimension-1, 2, and ${d+1}$ symmetry defects of the SymTFT form a split ${(d+1)}$-group symmetry ${(G^{(0)}, G^{(1)}, \cdots, G^{(d)}, \al^{(1)}, \al^{(2)}, \cdots, \al^{(d)})}$, where $G^{(k)}$ is the $k$-form symmetry group and $\al^{(k)}$ is the action of $G^{(0)}$ on $G^{(k)}$. When ${d=1}$, ${G^{(1)} = A\times A^\vee}$ and ${\al_q^{(1)}((a,\chi)) = (\rho_q(a), \rho^\vee_q(\chi))}$, while for ${d\neq 1}$,
\begin{equation}
G^{(k)} = \begin{cases}
Q \quad &k=0,\\
A \quad &k=1,\\
A^\vee \quad &k= d,\\
\Z_1 \quad &\text{else},
\end{cases},\hspace{40pt} \al^{(k)} = \begin{cases}
\rho\quad &k=1,\\
\rho^\vee \quad &k=d,\\
\mathrm{id} \quad &\text{else},
\end{cases}
\end{equation}
where $\mathrm{id}$ represents the trivial action.

The extension class $[c]$ of the group extension~\eqref{SESsec2} affects the 't Hooft anomaly of the ${(d+1)}$-group symmetry. Before gauging $A$, we have the following relation:
\begin{equation}
\begin{tikzpicture}[baseline={([yshift=-.5ex]current bounding box.center)},>=Triangle, thick]
\draw[line width=2pt, Adef] (7, 0) -- (6,1);
\draw[->, line width=2pt, Adef] (7, 0) -- (6.1, 0.9);
\node[Adef, above] at (6, 1) {$c(q_1,q_2)$};
\draw[Qdef, line width=2pt] (6, -1) -- (7,0);
\draw[Qdef, ->, line width=2pt,] (6, -1) -- (6.6, -.4);
\node[Qdef, below] at (6, -1) {$q_1$};
%
\draw[Qdef, line width=2pt] (8, -1) -- (7,0);
\draw[Qdef, ->, line width=2pt,] (8, -1) -- (7.4, -.4);
\node[Qdef, below] at (8, -1) {$q_2$};
%
\draw[Qdef, line width=2pt] (7, 0) -- (8,1);
\draw[Qdef, ->, line width=2pt,] (7, 0) -- (7.9, .9);
\node[Qdef, above] at (8, 1) {$q_1 q_2$};
%
\draw[Wdef, line width=2pt] (8, .4) -- (6,.4);
\draw[Wdef, ->, line width=2pt,] (8, .4) -- (6.75, .4);
\node[Wdef, right] at (8, .4) {$\rho^\vee_{[q_1q_2]^{-1}}(\chi)$};
\node[Wdef, left] at (6, .4) {$\chi$};
\fill[Adef] (7, 0) circle (4pt);
\node at (11.6, 0) {$=\ee^{\ii \chi(c(q_1,q_2))}$};
\draw[line width=2pt, Adef] (15, 0) -- (14,1);
\draw[->, line width=2pt, Adef] (15, 0) -- (14.4, 0.6);
\node[Adef, above] at (14, 1) {$c(q_1,q_2)$};
\draw[Qdef, line width=2pt] (14, -1) -- (15,0);
\draw[Qdef, ->, line width=2pt,] (14, -1) -- (14.4, -.6);
\node[Qdef, below] at (14, -1) {$q_1$};
%
\draw[Qdef, line width=2pt] (16, -1) -- (15,0);
\draw[Qdef, ->, line width=2pt,] (16, -1) -- (15.6, -.6);
\node[Qdef, below] at (16, -1) {$q_2$};
%
\draw[Qdef, line width=2pt] (15, 0) -- (16,1);
\draw[Qdef, ->, line width=2pt,] (15, 0) -- (15.6, .6);
\node[Qdef, above] at (16, 1) {$q_1 q_2$};
%
\draw[Wdef, line width=2pt] (16, -.4) -- (14,-.4);
\draw[Wdef, ->, line width=2pt,] (16, -.4) -- (14.75, -.4);
\node[Wdef, right] at (16, -.4) {$\rho^\vee_{[q_1q_2]^{-1}}(\chi)$};
\node[Wdef, left] at (14, -.4) {$\chi$};
\fill[Adef] (15, 0) circle (4pt);
\end{tikzpicture}.
\end{equation}
Since gauging $A$ trivializes its symmetry defects, the trivalent junctions of $Q$ 0-form symmetry defects after gauging are dressed by the $A$ 1-form symmetry defects (which were the boundary of the codimension-1 $A$ symmetry defects). When wrapping the $A^\vee$ $d$-form symmetry defects of the SymTFT around the trivalent junctions, they pick up a phase as illustrated below:
\begin{equation}
\begin{tikzpicture}[baseline={([yshift=-.5ex]current bounding box.center)},>=Triangle, thick]
\draw[Qdef, line width=2pt] (6, -1) -- (7,0);
\draw[Qdef, ->, line width=2pt,] (6, -1) -- (6.4, -.6);
\node[Qdef, below] at (6, -1) {$q_1$};
%
\draw[Qdef, line width=2pt] (8, -1) -- (7,0);
\draw[Qdef, ->, line width=2pt,] (8, -1) -- (7.6, -.6);
\node[Qdef, below] at (8, -1) {$q_2$};
%
\draw[Qdef, line width=2pt] (7, 0) -- (7,1.41);
\draw[Qdef, ->, line width=2pt,] (7, 0) -- (7, 1.25);
\node[Qdef, above] at (7, 1.41) {$q_1 q_2$};
%
\draw[Wdef, ->, line width=2pt,] (6.45, .01) -- (6.45, -.17);
\node[Wdef, left] at (6.55, 0.5) {$\chi$};
\fill[Adef] (7, 0) circle (4pt);
\node at (9.8, 0) {$=\, \ee^{\ii \chi\left(c(q_1,q_2)\right)}$};
%
%
\draw[Qdef, line width=2pt] (11.5, -1) -- (12.5,0);
\draw[Qdef, ->, line width=2pt,] (11.5, -1) -- (12.1, -.4);
\node[Qdef, below] at (11.5, -1) {$q_1$};
%
\draw[Qdef, line width=2pt] (13.5, -1) -- (12.5,0);
\draw[Qdef, ->, line width=2pt,] (13.5, -1) -- (12.9, -.4);
\node[Qdef, below] at (13.5, -1) {$q_2$};
%
\draw[Qdef, line width=2pt] (12.5, 0) -- (12.5,1.41);
\draw[Qdef, ->, line width=2pt,] (12.5, 0) -- (12.5, 1.01);
\node[Qdef, above] at (12.5, 1.41) {$q_1 q_2$};
\fill[Adef] (12.5, 0) circle (4pt);
\draw[Wdef, line width=2pt] (7,0) circle (16pt);
\end{tikzpicture}.
\end{equation}
This means that the $A^\vee$ $d$-form symmetry defects carry fractional $Q$ symmetry charges -- a hallmark of symmetry fractionalization. This symmetry fractionalization is characterized by the 2-cocycle $c(q_1,q_2)$, which specifies which $A$ 1-form symmetry defects decorate the trivalent junctions of $Q$ 0-form symmetry defects~\cite{BBC14104540, DGH220615118, BCD220615401}. Since the $A^\vee$ $d$-form symmetry defects are charged under the $Q$ 0-form symmetry, there is an obstruction to gauging both symmetries simultaneously, giving rise to an 't Hooft anomaly of the full $(d+1)$-group symmetry.

\subsubsection{Mixed anomalies}

Another possible form of interplay between two symmetries is a mixed anomaly. Consider a $0$-form symmetry in ${(d+1)}$-dimensional spacetime described by a finite group $G$. When the symmetry is internal, its anomalies are in one-to-one correspondence with $G$-SPTs in ${((d+1)+1)}$D, which are classified by the cohomology ${H^{d+2}(BG,U(1))}$. Having already discussed the interplays via group extensions, we now focus on the case where $G$ is a product group ${G = A \times Q}$, with $A$ an Abelian finite group. For such a product group, the cohomology ${H^{d+2}(BG,U(1))}$ factorizes by the K{\"u}nneth formula
\begin{equation}\label{KunForm}
\!\! H^{d+2}\left(BG,U(1)\right) = \! \!\bigoplus_{\substack{p+q\,=\, d+2 \\ p,q\,\geq \,0 }} \! \!H^{p}\!\left(BQ,H^{q}\!\left(BA,U(1)\right)\right).
\end{equation}
When interpreted as $G$-SPTs, the cohomology group for a given $q$ corresponds to SPTs built from decorating $q$-dimensional $A$-SPTs onto the $q$-dimensional junctions of $Q$ symmetry defects~\cite{CLV1301, GJ171207950, WNC210413233}. Mixed anomalies between $A$ and $Q$ are captured by those cohomology groups with ${q\neq 0, d+2}$. 

After gauging $A$, the $G$-SPT becomes an $A$ gauge theory enriched by the residual $Q$ symmetry. Importantly, the $Q$ symmetry defects remain decorated, but now by the gauged $A$-SPT defects, which are a class of topological defects in the $A$ gauge theory constructed by gauging lower-dimensional $A$-SPTs~\cite{BSW220805973, BCH220807367}. Through these decoration patterns, the mixed anomaly of $G$ determines how $Q$ enriches the $A$ gauge theory.

Let us now explore this effect when ${d=1}$. In this case, the $A$ gauge theory is ${(2+1)}$-dimensional. It has both electric defect lines $W_\chi(C)$ labeled by $\chi\in A^\vee$ and  magnetic defects lines $V_a(C)$ labeled by ${a\in A}$. There are two possible ways to decorate the $Q$ symmetry defects, which correspond to the following two cohomology groups
\begin{equation}
\begin{aligned}
&[\nu]\in H^{1}\!\left(BQ,H^{2}\!\left(BA,U(1)\right)\right),
\\
&[\eta]\in H^{2}\!\left(BQ,H^{1}\!\left(BA,U(1)\right)\right).
\end{aligned}
\end{equation}
Below, we discuss the effects of these decorations.

A representative $\nu$ of ${[\nu]}$ defines a group homomorphism ${\nu\colon Q\to H^{2}\!\left(BA,U(1)\right)}$. Physically, it dresses a $(1+1)$D gauged $A$-SPT $\nu_q$ defect on a ${q\in Q}$ symmetry surface defect. As a result, acting a $q$-symmetry defect on a magnetic defect line $V_a(C)$ causes
\begin{equation}\label{magDefTraSlant}
V_a(C)\to V_a(C)\, W_{\iota_a\nu_{q}}(C),
\end{equation}
where $\iota_a \nu_q\in H^1(BA,U(1
))$ is an irreducible representation of $A$ defined by the slant product
\begin{equation}
\iota_a\nu_q(\,\t a\,) = \frac{\nu_q(\,a,\t a\,)}{\nu_q(\,\t a,a\,)}.
\end{equation}
This action on $V_a$ follows from the circle compactification in the presence of an $a$ holonomy of the ${1+1}$D $A$-SPT $\nu_q$ producing the ${0+1}$D $A$-SPT $\int_{S^1_a}\nu_q =\iota_a\nu_q$~\cite{T170609769}, which after gauging $A$ becomes $W_{\iota_a\nu_{q}}$ in~\eqref{magDefTraSlant}. Therefore, when $\iota_a\nu$ is non-trivial, the $Q$ symmetry enriches the $A$ gauge theory by acting non-trivially on the $A$ magnetic defect lines, hence implementing an anyon automorphism.\footnote{When $\iota_q\nu$ is trivial, $Q$ symmetry defects can still act non-trivially on the TQFT by implementing \textit{soft} braided tensor autoequivalences of the UMTC~\cite{D13127466, KB250103314}. These are braided tensor autoequivalences that do no transform topological defect lines, but act non-trivially on their trivalent junctions (i.e., the Hom spaces of the UMTC).} 
When $Q$ is Abelian, this enrichment can alternatively be described by the cohomology class ${\om_{2} \in H^{2}\!\left(BA,H^{1}\!\left(BQ,U(1)\right)\right)}$ in the K{\"u}nneth formula, which contributes to the $A$ gauge theory action
\begin{equation}
S\supset\ii \int_{X_3} \cQ\cup \mathcal{A}{}^{*}\om_{2},
\end{equation}
where ${\cQ\in H^1(X_3, Q)}$ is a background $Q$ gauge field and ${\mathcal{A}\in H^1(X_3, A)}$ the dynamical $A$ gauge field. This term enforces $q$-symmetry defects to be dressed by the gauged ${1+1}$D $A$-SPT $q\cdot\om_{2}=\nu_q$.

A representative $\eta$ of $[\eta]$ is a 2-cocycle ${\eta\colon Q\times Q \to H^1(BA,U(1))}$. It dresses the gauged ${0+1}$D $A$ SPT ${\eta(q_1,q_2)}$ on the trivalent line junction formed by fusing ${q_1\in Q}$ and ${q_2 \in Q}$ symmetry surface defects. The gauged ${0+1}$D $A$ SPT ${\eta(q_1,q_2)}$ is the electric defect line ${W_{\eta(q_1,q_2)}}$. Therefore, the magnetic defect lines $V_a$ braid non-trivially with $Q$ symmetry defect junction lines and carry fractional $Q$ symmetry charge.  $[\eta]$ describes the symmetry fractionalization pattern of $Q$ in the $A$ gauge theory. This symmetry enrichment contributes to the $A$ gauge theory's action
\begin{equation}
S\supset\ii \int_{X_3} \mathcal{A} \cup \cQ^{*}\eta_{2}.
\end{equation}
Note that this term is equivalent to the term in~\eqref{SymTFTsec2.1} after replacing $\mathcal{A}$ by $\mathcal{B}$ when ${d=1}$.

\section{SymTFT enriched by lattice translation}\label{translationSec}

In this section, we consider SymTFTs enriched by a discrete spatial translation symmetry in $2+1$D. Such translation symmetries naturally arise from an underlying spatial lattice. The resulting lattice-translation-enriched SymTFTs describe symmetries in $1+1$D that exhibit a non-trivial interplay with the lattice translations.

We will focus on two well-known interplays involving a finite internal symmetry described by the Abelian group $G$. The first is spatially modulated $G$ symmetries in ${1+1}$D~\cite{Gromov:2018nbv, Sala_2020,Gromov_2020,sala2022dynamics,stahl2022multipoleSSB, lake2022dipolarBH, GLS220110589, SYH230708761, PDL240612962}. The second is LSM anomalies between lattice translations and uniform $G$ symmetry~\cite{Lieb61, Affleck1986, Oshikawa1997, Koma2000, Oshikawa2000, Hastings2004, Hastings2005, CB151102263, Ogata2019, Ogata2021, CS221112543, S230805151}. These interplays have been explored throughout the literature, and a fairly general understanding of them has developed. We will, therefore, investigate their SymTFTs at a more general level before exploring examples.\footnote{Examples of SymTFTs for specific modulated symmetries have appeared in~\cite{PLA240918113, SCS241104182, KYH}.} Readers who prefer to see examples first can refer to the example sections before reading the general constructions.

We will show that the SymTFTs for these Abelian finite $G$ symmetries are described by $G$ gauge theory non-trivially enriched by lattice translations. For both types of interplay, lattice translations induce an anyon automorphism of the $G$ gauge theory's topological order. Such translation symmetry enriched topological orders (SETs) have recently gathered much attention~\cite{W0303, K062, EH1306, RS201212280, OKM211002658, PW220407111, DFG220700409, EH220907987, Gorantla221007, OPH230104706, E230203747, EHN231006701, EHN240110677}, exhibiting UV/IR mixing and having anyons with restricted mobility and position-dependent braiding.

Furthermore, we show that as field theories, these SymTFTs are necessarily not topological in spatial directions parallel to the boundary. 
Therefore, they are not topological field theories. Despite not being a topological field theory, since the SymTFT is topological in the interval compactification direction, the interval compactification shown in Fig.~\ref{fig:TopHolo} can still be performed to relate the SymTFT sandwich ${(\mathfrak{B}^{\text{sym}}_\cS, \mathfrak{Z}(\cS), \mathfrak{B}^{\text{phys}}_{\mathfrak{T}^\cS})}$ to the physical theory $\mathfrak{T}^\cS$. The notion of not being topological in one direction is made precise by foliated field theory~\cite{SAW181201613, S200803852, HS210509363, OS221011001, S230413067, CJ231001474, EHN231006701, EHN240110677, EHN240805048, JJ250522261}. This makes the continuum SymTFT a foliated field theory:~it does not depend on a background metric of spacetime but does depend on a background foliation\footnote{Strictly speaking, the continuum SymTFT depends on slightly more data than the foliation structure. A codimension 1 foliation is described by a 1-form foliation field $e_\mu$ that is normal to the leaves of foliation. The foliation field has a scaling redundancy ${e_\mu\sim \exp[-f]\, e_\mu}$ where $f$ is a scalar field. A foliated field theory, however, depends on the norm of the foliation field in a mild way. For flat foliation, the foliation field is closed $\dd e=0$. We can then define a finite spacing between leaves from the foliation field by requiring that $\int e = 1$ between adjacent leaves. In this formulation, the integral $\oint e$ along the foliation field determines the total number of leaves, and the scaling ${e_\mu \to \exp[-f]\, e_\mu}$ amounts to modifying the spacing between leaves.
We refer the reader to Appendix~\ref{FoliationApp} for an introduction to foliation structures.} of spacetime. In this case, if the ${1+1}$D spacetime is the $(x,t)$ plane, then the ${2+1}$D foliation structure describes leaves spanning each $(y,t)$ plane orthogonal to the boundaries.

\subsection{Modulated symmetries}

A finite $G$ symmetry is a spatially modulated symmetry if the total symmetry group is ${G \rtimes_{\varphi} \Z_L}$ where $\Z_L$ denotes lattice translations on a periodic lattice with $L$ number of lattice sites. The group homomorphism ${\varphi \colon \Z_L \to \Aut(G)}$, where $\Aut(G)$ is the automorphism group of $G$, describes how $G$ is spatially modulated through the action of lattice translations on $G$. For example, let us specialize to a class of modulated $\Z_N^{ n}$ symmetries in ${1+1}$D system of $\Z_N$ qudits whose symmetry generators are represented by
\begin{equation}\label{genModSym}
U_q = \prod_j (X_j)^{f_j^{(q)}},
\end{equation}
where $X$ is the $\mathbb{Z}_N$ generalization of the Pauli $X$ operator and ${q = 1,2,\cdots, n}$. The $\Z_N$-valued lattice functions ${F = \{f^{(1)}, f^{(2)},\cdots, f^{(n)}\}}$ are independent,\footnote{More precisely, the functions are independent over the ring ${\Z/N\Z}$ in the sense of Ref.~\cite[Section 3.1]{PDL240612962}.} and are such that each $U_q$ is order $N$. They encode the action of lattice translations on $\Z_N^{ n}$ by
\begin{equation}
T^m U_q T^{-m} = \prod_j (X_j)^{f_{j-m}^{(q)}},
\end{equation}
the right-hand side of which can be decomposed into a product of various $U_q$ operators. Unless specified otherwise, we will assume that the number of lattice sites $L$ in the direction of the spatial modulation is divisible by the period of the spatial modulation (i.e., that ${T^L U_q T^{-L} = U_q}$).
Perhaps the most standard modulated symmetry of this type is the $\Z_N$ dipole symmetry. This is a ${\Z_N\times \Z_N}$ modulated symmetry whose lattice functions are ${f^{(1)}_j = 1}$ and ${f^{(2)}_j = j}$, and the translation operator $T$ acts as ${T U_1 T^{-1} = U_1}$ and ${T U_2 T^{-1} = U_1^{-1} U_2}$.

In what follows, we will construct the SymTFT for the modulated $\Z_N^{ n}$ symmetry generated by~\eqref{genModSym}. For simplicity, we will specialize to lattice functions $F$ such that the algebra of symmetric operators---the so-called bond algebra~\cite{NO08124309}---is generated by two types of local operators.\footnote{In Ref.\ \cite[Section 4.1]{PDL240612962}, 
sufficient conditions on $S$ are derived for the bond algebra to be of this form.} Namely, we will assume that this algebra is
\begin{equation}\label{B{F}modSym}
\mathfrak{B}[F] = \bigg\langle \,
X_j, ~
\prod_{l} Z_{l}^{D_{j, l}}
\,\bigg\rangle,
\end{equation}
where $D_{j,l}$ are $\Z_N$-valued matrix elements of the ${L\times L}$ matrix $D$ that satisfy
\begin{enumerate}
\item ${\sum_{l} D_{j, l}\, f_{l}^{(q)}=0 \bmod N}$,\label{test}
\item ${D_{0,k}= 0}$ if ${k<0}$ or ${k>n}$,
\item $D_{j+k,l}= D_{j, l-k}$,
\item ${\gcd(D_{0,0},N)= \gcd(D_{0,n},N) = 1}$.
\end{enumerate}
The first and second conditions, respectively, ensure that the operator ${\prod_{l} Z_{l}^{D_{j, l}}}$ is symmetric and local. The third condition ensures that the bond algebra $\mathfrak{B}[F]$ is closed under translations and implies that ${D_{j,l}=D_{0,l-j}}$. Lastly, the fourth condition ensures that ${D_{0,0},D_{0,n}\neq 0 \bmod N}$ and is a technical requirement for $\mathfrak{B}[F]$ to be generated by only the two types of operators in~\eqref{B{F}modSym} (see~\cite[Section 4.1]{PDL240612962} for further discussion). We note that from the second and third condition, we can write $D_{i,j}$ in the basis of finite difference matrices ${[\Del]_{i,j} = \del_{i+1,j} - \del_{i,j}}$ as
\begin{equation}\label{DtoDel}
D_{i,j} = \sum_{l=0}^n C_l \, [\Del^l]_{i,j},
\end{equation}
where ${[\Delta^l]_{i,j}=\sum_{a=0}^l (-1)^{l-a}\binom{l}{a} \del_{i+a,j}}$ is defined recursively by ${[\Delta^l]_{i,j}=[\Delta^{l-1}]_{i+1,j}-[\Delta^{l-1}]_{i,j}}$ with the initial condition $[\Delta^0]_{i,j}=\delta_{i,j}$.
The first and fourth condition above put constraints on the allowed coefficients $C_l$ in $D_{i,j}$.

\subsubsection*{Quantum code perspective}

The corresponding SymTFT can be constructed by first extending the modulated symmetry to ${2+1}$D, identifying the associated ${2+1}$D trivial SPT phase,\footnote{In this context, the trivial SPT phase refers to the SPT phase whose fixed-point state is a product state. The corresponding SymTFT is constructed from the trivial SPT because the modulated symmetry operators~\eqref{genModSym} are onsite and, thus, anomaly-free. Indeed, the symmetries they generate can be gauged~\cite{PDL240612962}. However, modulated symmetries can have anomalies, in which case the SPT would be non-trivial.
} and then gauging the modulated symmetry. For modulated symmetries~\eqref{genModSym} in ${1+1}$D whose bond algebra is~\eqref{B{F}modSym}, gauging (without discrete torsion) the entire modulated symmetry induces the gauging map~\cite{PDL240612962}
\begin{equation}
X_j^\dag \to \prod_{l} X_l^{D^\mathsf{T}_{jl}},\hspace{40pt} \prod_l Z_l^{D_{jl}} \to X_{j}.
\end{equation}
This can be implemented using the Gauss operator ${G_j = X_j \prod_{l} (X_{l, l+1})^{D_{jl}^\mathsf{T}}}$, where $X_{l, l+1}$ act on newly introduced $\Z_N$ qudits on the links. This leads to the dual bond algebra
\begin{equation}\label{dualB{F}modSym}
\mathfrak{B}^\vee[F] = \bigg\langle \,
Z_j, ~
\prod_{l} X_{l}^{D^\mathsf{T}_{j, l}}
\,\bigg\rangle,
\end{equation}
and dual modulated symmetry generated by
\begin{equation}\label{dualModSym}
U_q^\vee=\prod_{j} (Z_j)^{f^{(q)}_{-j}}.
\end{equation}
The transpose $D^\mathsf{T}$ has a similar decomposition in terms of the finite difference operators $\Del$ as $D$ does (see~\eqref{DtoDel}). In particular, it is straightforward to show that ${[\Del^l]_{ji} = (-1)^l[\Del^l]_{i-l,j}}$ and, therefore, 
\begin{equation}\label{DTtoDel}
[D^\mathsf{T}]_{ij} = \sum_{l=0}^n (-1)^l\,C_l \,[\Del^l]_{i-l,j}. 
\end{equation}

The ${2+1}$D extension of the modulated symmetry~\eqref{genModSym} to the square lattice is generated by
\begin{equation}\label{genModSym2d}
\prod_{\bm{r}} (X_{\bm{r}})^{f^{(q)}_{r_x}},
\end{equation}
where ${\bm{r} = (r_x, r_y)\in\Z_{L_x}\times \Z_{L_y}}$ is a two-dimensional lattice vector. This symmetry operator is modulated in the $x$-direction, but commutes with translations in the $y$-direction. Therefore, the symmetry generated by~\eqref{genModSym2d} can be gauged using the Gauss operator ${G_{\bm{r}} = X_{\bm{r}} A_{\bm{r}}}$ where
\begin{equation}\label{stabGenModA}
A_{\bm{r}}
=
\begin{tikzpicture}[scale = 0.5, baseline = {([yshift=-.5ex]current bounding box.center)}]  
\draw[line width=0.015in, gray] (-1.5,0) -- (13.5, 0);
\node[fill=white] at (6,0) {$\dots$};
\draw[line width=0.015in, gray] (1.5,-3) -- (1.5, 3);
\draw[line width=0.015in, gray] (4.5,-3) -- (4.5, 3);
\draw[line width=0.015in, gray] (7.5,-3) -- (7.5, 3);
\draw[line width=0.015in, gray] (10.5,-3) -- (10.5, 3);
\fill[black] (10.5,0) circle (6pt);
\node at (11, -.5) {\normalsize $\bm{r}$};
\node at (0, .45) {\normalsize $\textcolor[HTML]{cb181d}{X^{D_{0,n}}}$};
\node at (3, .45) {\normalsize $\textcolor[HTML]{cb181d}{X^{D_{0,n-1}}}$};
\node at (9, .45) {\normalsize $\textcolor[HTML]{cb181d}{X^{D_{0,1}}}$};
\node at (12, .45) {\normalsize $\textcolor[HTML]{cb181d}{X^{D_{0,0}}}$};
\node at (10.5, 1.5) {\normalsize $\textcolor[HTML]{cb181d}{X^{\da}}$};
\node at (10.5, -1.5) {\normalsize $\textcolor[HTML]{cb181d}{X}$};
\end{tikzpicture}
\,\equiv\,
\begin{tikzpicture}[scale = 0.5, baseline = {([yshift=-.5ex]current bounding box.center)}]
\draw[line width=0.015in, gray] (0,0) -- (6, 0);
\draw[line width=0.015in, gray] (3,-3) -- (3, 3);
\node at (4.7, .55) {\normalsize $\textcolor[HTML]{cb181d}{\prod X^{D^\mathsf{T}}}$};
\node at (3, 1.5) {\normalsize $\textcolor[HTML]{cb181d}{X^\da}$};
\node at (3, -1.5) {\normalsize $\textcolor[HTML]{cb181d}{X}$};
\fill[black] (3,0) circle (6pt);
\node at (3.5, -.5) {\normalsize $\bm{r}$};
\end{tikzpicture}~.
\end{equation}
The flux term, made of only $Z$ operators that commutes with $A_{\bm{r}}$, is
\begin{equation}\label{stabGenModB}
B_{\bm{r}}=\begin{tikzpicture}[scale = 0.5, baseline={([yshift=-.5ex]current bounding box.center)}]
\draw[line width=0.015in, gray] (-3, 0) -- (-3, 3) -- (9, 3) -- (9, 0) -- cycle;
\draw[line width=0.015in, gray] (0,0) -- (0, 3);
\draw[line width=0.015in, gray] (3,0) -- (3, 3);
\draw[line width=0.015in, gray] (6,0) -- (6, 3);
\node[fill=white] at (4.5,0) {$\dots$};
\node[fill=white] at (4.5,3) {$\dots$};
\node at (-1.5, 0.4) {\normalsize $\textcolor[HTML]{2171b5}{Z}$};
\node at (-1.5, 3.45) {\normalsize $\textcolor[HTML]{2171b5}{Z^\da}$};
\node at (-2.7, 1.5) {\normalsize $\textcolor[HTML]{2171b5}{Z^{D_{0,0}}}$};
\node at (0.3, 1.5) {\normalsize $\textcolor[HTML]{2171b5}{Z^{D_{0,1}}}$};
\node at (3.3, 1.5) {\normalsize $\textcolor[HTML]{2171b5}{Z^{D_{0,2}}}$};
\node at (6.7, 1.5) {\normalsize $\textcolor[HTML]{2171b5}{Z^{D_{0,n-1}}}$};
\node at (9.3, 1.5) {\normalsize $\textcolor[HTML]{2171b5}{Z^{D_{0,n}}}$};
\fill[black] (-3,0) circle (6pt);
\node at (-3.5, -0.5) {\normalsize $\bm{r}$};
\end{tikzpicture}
\equiv\,
\begin{tikzpicture}[scale = 0.5, baseline={([yshift=-.5ex]current bounding box.center)}]
\draw[line width=0.015in, gray] (-1.5, -1.5) -- (-1.5, 1.5) -- (1.5, 1.5) -- (1.5, -1.5) -- cycle;
\node at (0, -1.1) {\normalsize $\textcolor[HTML]{2171b5}{Z}$};
\node at (0, 2) {\normalsize $\textcolor[HTML]{2171b5}{Z^{\da}}$};
\node at (-1.3, 0.0) {\normalsize $\textcolor[HTML]{2171b5}{\prod Z^{D}}$};
\fill[black] (-1.5,-1.5) circle (6pt);
\node at (-2, -2) {\normalsize $\bm{r}$};
\end{tikzpicture}~.
\end{equation}
The local constraint ${B_{\bm{r}} = 1}$ is the flatness condition involved in gauging finite symmetries.

The SymTFT is found by applying the gauging map implemented by $G_{\bm{r}}$ to the trivial symmetric theory. As a quantum code, the trivial theory has only one type of stabilizer $X_{\bm{r}}$ and the code space is $\mathbb{C}$ (i.e., the single state that satisfies ${X_{\bm{r}} = 1}$ for all ${\bm{r}}$). From the Gauss law ${G_{\bm{r}} = 1}$, the stabilizer $X_{\bm{r}}$ becomes $A_{\bm{r}}$ under the gauging map. The flatness condition ${B_{\bm{r}} = 1}$ is imposed by
further introducing $B_{\bm{r}}$ as a stabilizer. Therefore, the SymTFT can be expressed as the code space of a Calderbank-Shor-Steane (CSS) code of $\Z_N$ qubits on a square lattice whose stabilizers are $A_{\bm{r}}$ and $B_{\bm{r}}$. Notice that for a uniform $\Z_N$ symmetry where ${F = \{1\}}$, then ${D = \Del_x}$ and the stabilizer code reduces to the $\Z_N$ toric code.

It is straightforward to check that $A_{\bm{r}}$ and $B_{\bm{r}}$ commute for all sites $\bm{r}$. Therefore, the code space is spanned by all states ${\ket{\psi}}$ satisfying ${A_{\bm{r}}\ket{\psi} = B_{\bm{r}} \ket{\psi} = \ket{\psi}}$ for all $\bm{r}$. The logical operators act on qudits along a closed loop. They depend only on this loop's homology class, making them topological operators. For instance, with periodic boundary conditions in the $x$ direction, the logical operators winding around the $x$ direction are generated by
\begin{equation}\label{generalem}
W^{(q)} =  \prod_{r_x = 1}^{L} (Z_{\bm{r},x})^{f^{(q)}_{-r_x}},\hspace{40pt} V^{(q)} =  \prod_{r_x = 1}^{L} (X_{\bm{r},y})^{f^{(q)}_{r_x}}.
\end{equation}
The subscript ${\bm{r},\mu}$ denotes the link ${\< \bm{r} , \bm{r}+\hat{\mu}\rangle}$, where $\hat{\mu}$ is the unit vector in the $\mu$-direction. Furthermore, the site $r_y$ is not specified in~\eqref{generalem} since the logical operators are topological and do not depend on $r_y$ in the code space.

As a topological order, the code space has $\Z_N^{ n}$ topological order. However, it is non-trivially enriched by lattice translations in the $x$ direction. Indeed, the logical operators $W^{(q)}$ and $V^{(q)}$ do not commute with translations in the $x$ direction. Instead, such lattice translations act as a non-trivial anyon automorphisms. As discussed in Section~\ref{generalSection}, this arises from the mixing of spatial and internal symmetries before gauging the internal symmetry. 

To verify that this quantum code is the SymTFT for the modulated symmetry, we consider a spatial boundary at ${y = L_y}$. The different choices of boundary degrees of freedom and boundary stabilizers correspond to different symmetries captured by this SymTFT. For instance, the rough boundary has no degrees of freedom on the boundary links ${((r_x,L_y),x)}$ and is specified by the boundary stabilizer
\begin{equation}
B^\text{rough}_{\bm{r}}=\begin{tikzpicture}[scale = 0.5, baseline={([yshift=-.5ex]current bounding box.center)}]
\draw[line width=0.015in, gray] (-3,0) -- (9, 0);
\draw[line width=0.015in, gray] (-3,0) -- (-3, 3);
\draw[line width=0.015in, gray] (0,0) -- (0, 3);
\draw[line width=0.015in, gray] (3,0) -- (3, 3);
\draw[line width=0.015in, gray] (6,0) -- (6, 3);
\draw[line width=0.015in, gray] (9,0) -- (9, 3);
\node[fill=white] at (4.5,0) {$\dots$};
\node at (-1.5, 0.4) {\normalsize $\textcolor[HTML]{2171b5}{Z}$};
\node at (-2.7, 1.5) {\normalsize $\textcolor[HTML]{2171b5}{Z^{D_{0,0}}}$};
\node at (0.3, 1.5) {\normalsize $\textcolor[HTML]{2171b5}{Z^{D_{0,1}}}$};
\node at (3.3, 1.5) {\normalsize $\textcolor[HTML]{2171b5}{Z^{D_{0,2}}}$};
\node at (6.7, 1.5) {\normalsize $\textcolor[HTML]{2171b5}{Z^{D_{0,n-1}}}$};
\node at (9.3, 1.5) {\normalsize $\textcolor[HTML]{2171b5}{Z^{D_{0,n}}}$};
\fill[black] (-3,0) circle (6pt);
\node at (-3.5, -0.5) {\normalsize $\bm{r}$};
\end{tikzpicture}
\equiv\,
\begin{tikzpicture}[scale = 0.5, baseline={([yshift=-.5ex]current bounding box.center)}]
\draw[line width=0.015in, gray] (-1.5,-1.5) -- (1.5, -1.5);
\draw[line width=0.015in, gray] (-1.5,-1.5) -- (-1.5, 1.5);
\draw[line width=0.015in, gray] (1.5,-1.5) -- (1.5, 1.5);
\node at (0, -1.1) {\normalsize $\textcolor[HTML]{2171b5}{Z}$};
\node at (-1.3, 0.0) {\normalsize $\textcolor[HTML]{2171b5}{\prod Z^{D}}$};
\fill[black] (-1.5,-1.5) circle (6pt);
\node at (-2, -2) {\normalsize $\bm{r}$};
\end{tikzpicture}~.
\end{equation}
This stabilizer commutes with all $A_{\bm{r}}$. The logical operator $V^{(q)}$ acting on this boundary commutes with the stabilizer and, therefore, is a symmetry of the ${1+1}$D theory. The boundary logical operator $W^{(q)}$, however, is trivialized. Therefore, this boundary encodes the symmetry generated by $\prod_{r_x = 1}^{L_x} (X_{(r_x,L_y-1),y})^{f_{r_x}^{(q)}}$, which are the modulated symmetry operators~\eqref{genModSym}. On the other hand, the smooth boundary has $\Z_N$ qudits on the boundary links and the boundary stabilizers are
\begin{equation}
A^\text{smooth}_{\bm{r}}
=
\begin{tikzpicture}[scale = 0.5, baseline = {([yshift=-.5ex]current bounding box.center)}]  
\draw[line width=0.015in, gray] (-1.5,0) -- (13.5, 0);
\node[fill=white] at (6,0) {$\dots$};
\draw[line width=0.015in, gray] (1.5,-3) -- (1.5, 0);
\draw[line width=0.015in, gray] (4.5,-3) -- (4.5, 0);
\draw[line width=0.015in, gray] (7.5,-3) -- (7.5, 0);
\draw[line width=0.015in, gray] (10.5,-3) -- (10.5, 0);
\fill[black] (10.5,0) circle (6pt);
\node at (11, -.5) {\normalsize $\bm{r}$};
\node at (0, .45) {\normalsize $\textcolor[HTML]{cb181d}{X^{D_{0,n}}}$};
\node at (3, .45) {\normalsize $\textcolor[HTML]{cb181d}{X^{D_{0,n-1}}}$};
\node at (9, .45) {\normalsize $\textcolor[HTML]{cb181d}{X^{D_{0,1}}}$};
\node at (12, .45) {\normalsize $\textcolor[HTML]{cb181d}{X^{D_{0,0}}}$};
\node at (10.5, -1.5) {\normalsize $\textcolor[HTML]{cb181d}{X}$};
\end{tikzpicture}
\,\equiv\,
\begin{tikzpicture}[scale = 0.5, baseline = {([yshift=-.5ex]current bounding box.center)}]
\draw[line width=0.015in, gray] (0,0) -- (6, 0);
\draw[line width=0.015in, gray] (3,-3) -- (3, 0);
\node at (4.7, .55) {\normalsize $\textcolor[HTML]{cb181d}{\prod X^{D^\mathsf{T}}}$};
\node at (3, -1.5) {\normalsize $\textcolor[HTML]{cb181d}{X}$};
\fill[black] (3,0) circle (6pt);
\node at (3.5, -.5) {\normalsize $\bm{r}$};
\end{tikzpicture}~.
\end{equation}
On this boundary, the logical operators $V^{(q)}$ are trivialized while the logical operators $W^{(q)}$ commute with the stabilizer and act non-trivially. Therefore, the smooth boundary encodes the symmetry generated by $\prod_{r_x = 1}^{L_x} (Z_{(r_x,L_y),x})^{f_{-r_x}^{(q)}}$, which is the dual modulated symmetry~\eqref{dualModSym}. 
In general, there are other boundary conditions where different subsets of logical operators are trivialized. We will consider the complete characterization of gapped boundary conditions later when studying explicit examples.

\subsubsection*{Field theory perspective}

Having constructed the SymTFT for the modulated symmetry~\eqref{genModSym} from a quantum code perspective, we now complement this discussion by considering this SymTFT from a field theory perspective. Consider a cubic Euclidean spacetime lattice whose sites we denote by the integer-valued vector ${\bm{r} \equiv (r_t,r_x,r_y)}$. The $\Z_N$ Pauli operators are related to lattice fields $a_{\bm{r},\mu}$ and $b_{\bm{r},\mu}$ by
\begin{equation}
\begin{aligned}
Z_{\bm{r},x}&= \ee^{\frac{2\pi\ii}N a_{\bm{r},x}}, \hspace{60.6pt} Z_{\bm{r},y} = \ee^{\frac{2\pi\ii}N a_{\bm{r},y}},\\
X_{\bm{r},x}&= \ee^{\frac{2\pi\ii}N b_{\bm{r}+\hat{x}-\hat{y},y}}, \hspace{40pt} X_{\bm{r},y}= \ee^{-\frac{2\pi\ii}N b_{\bm{r},x}}.
\end{aligned}
\end{equation}
The lattice fields $a_{\bm{r},\mu}$ and $b_{\bm{r},\mu}$ are integer-valued. Furthermore, we use the notation ${a_{\bm{r},\mu} = - a_{\bm{r}+\hat{\mu},-\mu}}$ (and similarly for $b_{\bm{r},\mu}$) with ${\hat{\mu}}$ denoting one of the three spacetime basis vector $\hat{t}$, $\hat{x}$, or $\hat{y}$. The lattice field $b_{\bm{r},\mu}$ is most naturally associated with links of the dual lattice. We have shifted the sites of the dual lattice by the vector ${\frac12 (\hat{x} - \hat{y})}$ to the direct spacetime lattice. The change from dual to direct lattice causes the ${\hat{x}-\hat{y}}$ shift for $b_{\bm{r}+\hat{x}-\hat{y},y}$ in $X_{\bm{r},x}$ and the minus sign in front of $b_{\bm{r},x}$ in $X_{\bm{r},y}$.

In terms of these lattice fields, the stabilizer constraints ${A_{\bm{r}} = 1}$ and ${B_{\bm{r}} = 1}$ are 
\begin{align}
(b_{\bm{r}+\hat{y},x} - b_{\bm{r},x}) + \sum_{j=0}^n D_{0,j}\, b_{\bm{r}+\hat{x}-j\hat{x},y} &\equiv \Del_y b_{\bm{r},x} + D_x^\mathsf{T} b_{\bm{r}+\hat{x},y} = 0\bmod N
,\label{constr1}\\
\sum_{j=0}^n D_{0,j} \, a_{\bm{r}+j\hat{x},y} - (a_{\bm{r}+\hat{y},x} - a_{\bm{r},x}) &\equiv D_x a_{\bm{r},y} - \Del_y a_{\bm{r},x} = 0\bmod N.\label{constr2}
\end{align}
The code space can then be described by the Euclidean lattice Lagrangian
\begin{equation}\label{modSymTFTL}
\mathscr{L}_{\bm{r}} = \frac{2\pi \ii}{N}\, \big(
a_{\bm{r},y}\Del_t b_{\bm{r},x} - a_{\bm{r},x}\Del_t b_{\bm{r}+\hat{x}-\hat{y},y} + b_{\bm{r}+\hat{x},t} ( D_x a_{\bm{r},y} - \Del_y a_{\bm{r},x} ) - a_{\bm{r}+\hat{y}-\hat{t},t} ( \Del_y b_{\bm{r},x} + D_x^\mathsf{T} b_{\bm{r}+\hat{x},y} )
\big),
\end{equation}
where, for example, ${\Del_t b_{\bm{r},x} \equiv b_{\bm{r}+\hat{t},x} - b_{\bm{r},x}}$. The first two terms in $\mathscr{L}_{\bm{r}}$ enforce $a_{\bm{r},\mu}$ and $b_{\bm{r},\mu}$ to obey commutation relations consistent with ${ZX = \ee^{2\pi\ii/N}XZ}$. In the last two terms, $a_{\bm{r},t}$ and $b_{\bm{r},t}$ are Lagrange multipliers enforcing the constraints~\eqref{constr1} and~\eqref{constr2}. This lattice Lagrangian has the gauge redundancy
\begin{equation}
\begin{aligned}
a_{\bm{r},t}\sim a_{\bm{r},t} + \Del_t \al_{\bm{r}},\qquad &a_{\bm{r},x}\sim a_{\bm{r},x} + D_x \al_{\bm{r}},\qquad a_{\bm{r},y} \sim a_{\bm{r},y} + \Del_y \al_{\bm{r}} ,\\
b_{\bm{r},t}\sim b_{\bm{r},t} + \Del_t \bt_{\bm{r}},\qquad &b_{\bm{r},x}\sim b_{\bm{r},x} - D_x^\mathsf{T} \bt_{\bm{r}+\hat{x}},\qquad b_{\bm{r},y} \sim b_{\bm{r},y} + \Del_y \bt_{\bm{r}},
\end{aligned}
\end{equation}
and its equations of motion are
\begin{equation}\label{modSymeomL}
\begin{aligned}
D_x a_{\bm{r},y} - \Del_y a_{\bm{r},x} = 0,\qquad &\Del_t a_{\bm{r},y} - \Del_y a_{\bm{r},t} = 0
,\qquad \Del_t a_{\bm{r},x} - D_x a_{\bm{r},t}  = 0 ,\\
\Del_y b_{\bm{r},x} + D_x^\mathsf{T} b_{\bm{r}+\hat{x},y} = 0,\qquad 
&\Del_t b_{\bm{r},y} - \Del_y b_{\bm{r},t} = 0
,\qquad \Del_t b_{\bm{r},x} + D_x^\mathsf{T} b_{\bm{r}+\hat{x},t} = 0.
\end{aligned}
\end{equation} 

As we will see, the Lagrangian~\eqref{modSymTFTL} is the SymTFT for the modulated symmetry~\eqref{genModSym}. It is an anisotropic lattice BF theory. For uniform symmetries, where ${F = \{1\}}$, the $D_x$ matrix satisfies ${D_x a_{\bm{r},y} = \Del_x a_{\bm{r},y}}$ and ${D^\mathsf{T}_x b_{\bm{r}+\hat{x},y} = -\Del_x b_{\bm{r},y}}$, and~\eqref{modSymTFTL} reduces to the level-$N$ lattice BF theory ${\frac{2\pi\ii}N b\cup \dd a}$. 

The theory~\eqref{modSymTFTL} has no local degrees of freedom. Its physical observables are all extended defect lines. Denoting by $C$ an oriented cycle on the spacetime lattice, the gauge-invariant defect lines can be generated from
\begin{equation}\label{TDLsymTFTmod}
W^{(q)}(C) = \exp[\frac{2\pi\ii}N \sum_{(\bm{r},\mu)\subset C} A^{(q)}_{r_x,\mu}\, a_{\bm{r},\mu}],
\hspace{40pt} 
V^{(q)}(C) = \exp[\frac{2\pi\ii}N \sum_{(\bm{r},\mu)\subset C} B^{(q)}_{r_x,\mu}\, b_{\bm{r},\mu}],
\end{equation}
where the integer-valued matrices are
\begin{equation}
A^{(q)}_{r_x,\mu} = \begin{cases}
f^{(q)}_{-r_x}  &\mu = x,\\
\Del_x^{-1}\,f^{(q)}_{-r_x}\, D_x  \qquad\qquad &\mu = t,y,
\end{cases}
\hspace{40pt}
B^{(q)}_{r_x,\mu} = \begin{cases}
f^{(q)}_{r_x}  &\mu = x,\\
-\Del_x^{-1}\,f^{(q)}_{r_x}\, D_x^\mathsf{T} \, T_x  \qquad\qquad &\mu = t,y.
\end{cases}
\end{equation}
The operator $T_x$ transforms ${b_{\bm{r},\mu}\to b_{\bm{r}+\hat{x},\mu}}$ and $\Del_x^{-1}$ is the inverse of $\Del_x$ (i.e., the indefinite sum operator). The gauge invariance of $W^{(q)}(C)$ and $V^{(q)}(C)$ follows from Stokes theorem and property~\ref{test} of $D_x$. However, with periodic boundary conditions in the $x$-direction, there is an additional constraint that allowed topological defects $[W^{(q)}]^n$ and $[V^{(q)}]^n$ must satisfy ${n\,f^{(q)}_{-r_x} = n\, f^{(q)}_{-r_x + L_x}\bmod N}$ and ${n\,f^{(q)}_{r_x} = n\, f^{(q)}_{r_x + L_x}\bmod N}$, respectively, to be gauge invariant.

The defect lines ${W^{(q)}}$ and ${V^{(q)}}$ both satisfy $\Z_N$ fusion rules ${[W^{(q)}]^N = [V^{(q)}]^N = 1}$. Furthermore, using Stokes theorem and the equations of motion~\eqref{modSymeomL}, it is easy to show that they are \textit{topological} defect lines:
\begin{equation}
\< W^{(q)}(C + \pp D) \> = \< W^{(q)}(C) \>,\hspace{40pt}
\< V^{(q)}(C + \pp D) \> = \< V^{(q)}(C) \>.
\end{equation}
While these are topological defects, the translation ${r_x \to r_x + 1}$ transforms them non-trivially whenever ${f^{(q)}_{r_x}\neq 1}$. Therefore, these are modulated 1-form symmetries of the field theory. This makes the SymTFT a $\Z_N^{ n}$ topological order enriched by translations in the $x$-direction. This enrichment is specified by a group homomorphism ${\varphi\colon \Z_{L_x} \to \Aut(\Z_N^{ n})}$. In fact, the same homomorphism appears at the beginning of this section in the definition of the ${1+1}$D modulated symmetry. It describes the action of translation on $V^{(q)}$. The action on $W^{(q)}$ follows from replacing ${t\in \Z_{L_x}}$ with $t^{-1}$ in $\varphi$.

Let us verify that this is the SymTFT for the modulated symmetry~\eqref{genModSym}. Consider a boundary at ${y=L_y}$ with the Dirichlet boundary conditions ${a_{\bm{r},\mu}\mid_{r_y=L_y} = 0}$. On this boundary, the topological defect lines ${W^{(q)} = 1}$ while $V^{(q)}$ remains unchanged. Therefore, this topological boundary encodes a symmetry generated by $V^{(q)}$. For a cycle $C$ running along the $x$-direction, ${V^{(q)}}$ forms the modulated symmetry~\eqref{genModSym}. Therefore, this boundary is the symmetry boundary of the modulated symmetry~\eqref{genModSym}. Similarly, choosing the Neumann boundary conditions ${b_{\bm{r},\mu}\mid_{r_y=L_y} = 0}$ causes ${W^{(q)}}$ to be unchanged on the boundary while ${V^{(q)} = 1}$. Therefore, this symmetry boundary encodes a symmetry generated by ${W^{(q)}}$ which is a modulated symmetry operator with the modulation function $f^{(q)}_{-r_x}$. Therefore, this symmetry boundary corresponds to the dual modulated symmetry~\eqref{dualModSym} obtained by gauging the original modulated symmetry~\eqref{genModSym}.

The above discussion of the SymTFT used lattice field theory for clarity. The continuum limit of~\eqref{modSymTFTL} is naturally formulated in terms of a topological defect network~\cite{Aasen:2020zru}. Often times, the continuum limit can be found by simply replacing lattice derivatives $\Del_\mu$ with $\pp_\mu$ (recall~\eqref{DtoDel} and~\eqref{DTtoDel}) and replacing $\frac{2\pi}{N}a_{\bm{r},\mu}$ and $\frac{2\pi}{N}b_{\bm{r},\mu}$ with $U(1)$ gauge fields $a_\mu$ and $b_\mu$, respectively. The continuum theory, however, is not a topological field theory. Indeed, discrete translations in the $x$-direction are non-trivial since they permute the topological defect lines. Moreover, the number of anyon types depends on $L_x$ when the $x$-direction of space is compact, so the partition function can change as the size of space in the $x$-direction is changed. However, it is still topological in the $y$ and $t$ directions, making it a type of foliated field theory. In particular, there are leaves spanning each $yt$ plane, each of which is dressed by an invertible condensation defect surface~\cite{Roumpedakis:2022aik} that implements the anyon permutation arising from discrete $x$-translations. These leaves and permutations encode in the continuum SymTFT how the ${1+1}$D symmetry is spatially modulated.

\subsubsection{Example: exponential symmetry}
\label{sec:exponential sym}

For the first example, we consider a class of $\Z_N$ exponential symmetries. This is a modulated symmetry which in a system of $\Z_N$ qudits has symmetry operators generated by
\begin{equation}\label{expSymGen}
U = \prod_{j=1}^L (X_j)^{k^j}.
\end{equation}
That is, there is a single modulation function ${F = \{ k^j\}}$ and the parameter ${k\in \Z}$ defines the type of exponential symmetry. We restrict ourselves to $k$ such that ${\gcd(k,N) = 1}$ so $k^{j}$ with ${j<0}$ is well defined in multiplication modulo $N$. For example, when ${N=5}$ and ${k=2}$, then ${k^{-1} = 3}$. This exponential symmetry forms a ${\Z_N\rtimes_\varphi \Z_L}$ symmetry with lattice translations, where the group homomorphism ${\varphi\colon \Z_L\to\Aut(\Z_N)}$ describes the action
\begin{equation}\label{TactinexpSym}
T U T^{-1} = U^{k^{-1}}.
\end{equation}

A simple quantum lattice model commuting with~\eqref{expSymGen} is
\begin{equation}
\sum_{j=1}^L \left( Z_j^{-k}\, Z_{j+1} + h X_j \right) + \text{H.c.}
\end{equation}
Furthermore, the bond algebra of the exponential symmetry is of the type~\eqref{B{F}modSym}~\cite{PDL240612962}. The matrix $D$ satisfying ${D_{ij}k^j = 0 \bmod N}$ is
\begin{equation}\label{Dexp}
D_{i,j} = \del_{i+1,j} - k \, \del_{i,j} \equiv  \Del_{i,j}+ \del_{i,j}\, (1-k).
\end{equation}
Furthermore, the dual symmetry obtained by gauging the entire exponential symmetry is (see~\eqref{dualModSym})
\begin{equation}\label{dualexpSymGen}
U^\vee = \prod_{j=1}^L (X_j)^{k^{-j}}.
\end{equation}
For general $k$, this generates a different exponential symmetry than~\eqref{expSymGen}.

The SymTFT of this exponential symmetry is a $\Z_N$ topological order enriched by lattice translations in the $x$-direction to encode~\eqref{TactinexpSym}. We will discuss it both from the quantum code and field theory perspectives.

\subsubsection*{Quantum code perspective}

The stabilizers defining the quantum code description of the corresponding SymTFT are~\eqref{stabGenModA} and~\eqref{stabGenModB} with $D$ given by~\eqref{Dexp}. That is, for general $k$ coprime to $N$, they are
\begin{align}\label{expSymTFTStabs}
A_{\bm{r}}
=
\begin{tikzpicture}[scale = 0.5, baseline = {([yshift=-.5ex]current bounding box.center)}]
\draw[line width=0.015in, gray] (7.5,0) -- (13.5, 0);
\draw[line width=0.015in, gray] (10.5,-3) -- (10.5, 3);
\node at (9, .45) {\normalsize $\textcolor[HTML]{cb181d}{X}$};
\node at (12, .45) {\normalsize $\textcolor[HTML]{cb181d}{X^{\dag k}}$};
\node at (10.5, 1.5) {\normalsize $\textcolor[HTML]{cb181d}{X^{\da}}$};
\node at (10.5, -1.5) {\normalsize $\textcolor[HTML]{cb181d}{X}$};
\fill[black] (10.5,0) circle (6pt);
\node at (10, -0.5) {\normalsize $\bm{r}$};
\end{tikzpicture}~,
\hspace{40pt}
B_{\bm{r}}=\begin{tikzpicture}[scale = 0.5, baseline={([yshift=-.5ex]current bounding box.center)}]
\draw[line width=0.015in, gray] (-3, 0) -- (-3, 3) -- (0, 3) -- (0, 0) -- cycle;
\node at (-1.5, 0.4) {\normalsize $\textcolor[HTML]{2171b5}{Z}$};
\node at (-1.5, 3.45) {\normalsize $\textcolor[HTML]{2171b5}{Z^\da}$};
\node at (-2.9, 1.5) {\normalsize $\textcolor[HTML]{2171b5}{Z^{\dag k}}$};
\node at (0, 1.5) {\normalsize $\textcolor[HTML]{2171b5}{Z}$};
\fill[black] (-3,0) circle (6pt);
\node at (-3.5, -0.5) {\normalsize $\bm{r}$};
\end{tikzpicture}.
\end{align}
This stabilizer code is of the type studied in \Rf{WCF221100299}. Using~\eqref{TDLsymTFTmod}, the logical operators of the code act on a cycle $C$ of the square lattice, and are generated by
\begin{align}
V[C] &= \prod_{(\bm{r},\mu)\subset C}\left( [X_{\bm{r},y}]^{\si\, k^{r_x}} \del_{\mu, x}
+
[X_{r-\hat{x}+\hat{y},x}]^{-\si\, k^{r_x}} \del_{\mu,y} \right),\\
W[C] &= \prod_{(\bm{r},\mu)\subset C} \left([Z_{\bm{r},x} ]^{\si\, k^{-r_x}} \del_{\mu,x}
+
[ Z_{\bm{r},y} ]^{\si\, k^{-r_x+1} } \del_{\mu,y} \right),
\end{align}
where ${\si}$ captures the orientation of $C$ (i.e., ${\si=\pm 1}$ when $C$ is running in the ${\pm ~ x,y}$ directions.). For example, graphical depictions of these logical operators for a small ${3\times 2}$ rectangular cycle are
\begin{equation*}
\begin{tikzpicture}[scale = 0.5, baseline={([yshift=-.5ex]current bounding box.center)}]
\draw[line width=0.015in, gray] (-6,0) -- (6, 0);
\draw[line width=0.015in, gray] (-6,3) -- (6, 3);
\draw[line width=0.015in, gray] (-3,-3) -- (-3, 6);
\draw[line width=0.015in, gray] (0,-3) -- (-0, 6);
\draw[line width=0.015in, gray] (3,-3) -- (3, 6);
%
\node at (-2.8, -1.5) {\normalsize $\textcolor[HTML]{cb181d}{X^{k^{r_x}}}$};
\node at (0.5, -1.5) {\normalsize $\textcolor[HTML]{cb181d}{X^{k^{r_x+1}}}$};
\node at (3.5, -1.5) {\normalsize $\textcolor[HTML]{cb181d}{X^{k^{r_x+2}}}$};
\node at (4.5, 0.5) {\normalsize $\textcolor[HTML]{cb181d}{X^{\dag k^{r_x+3}}}$};
\node at (4.5, 3.5) {\normalsize $\textcolor[HTML]{cb181d}{X^{\dag k^{r_x+3}}}$};
\node at (3.6, 4.6) {\normalsize $\textcolor[HTML]{cb181d}{X^{\dag k^{r_x+2}}}$};
\node at (0.6, 4.6) {\normalsize $\textcolor[HTML]{cb181d}{X^{\dag k^{r_x+1}}}$};
\node at (-2.6, 4.6) {\normalsize $\textcolor[HTML]{cb181d}{X^{\dag k^{r_x}}}$};
\node at (-4.5, 0.5) {\normalsize $\textcolor[HTML]{cb181d}{X^{k^{r_x}}}$};
\node at (-4.5, 3.5) {\normalsize $\textcolor[HTML]{cb181d}{X^{k^{r_x}}}$};
\fill[black] (-3,0) circle (6pt);
\node at (-3.5, -0.5) {\normalsize $\bm{r}$};
\node at (-7.5, 1.5) {\normalsize $V=$};
\end{tikzpicture}
\hspace{60pt}
\begin{tikzpicture}[scale = 0.5, baseline={([yshift=-.5ex]current bounding box.center)}]
\draw[line width=0.015in, gray] (-3,0) -- (6, 0);
\draw[line width=0.015in, gray] (-3,3) -- (6, 3);
\draw[line width=0.015in, gray] (-3,6) -- (6, 6);
\draw[line width=0.015in, gray] (-3,0) -- (-3, 6);
\draw[line width=0.015in, gray] (0,0) -- (-0, 6);
\draw[line width=0.015in, gray] (3,0) -- (3, 6);
\draw[line width=0.015in, gray] (6,0) -- (6, 6);
%
\node at (-1.5, 0.5) {\normalsize $\textcolor[HTML]{2171b5}{Z^{k^{-r_x}}}$};
\node at (1.5, 0.5) {\normalsize $\textcolor[HTML]{2171b5}{Z^{k^{-r_x-1}}}$};
\node at (4.5, 0.5) {\normalsize $\textcolor[HTML]{2171b5}{Z^{k^{-r_x-2}}}$};
\node at (6.9, 1.5) {\normalsize $\textcolor[HTML]{2171b5}{Z^{k^{-r_x-2}}}$};
\node at (6.9, 4.5) {\normalsize $\textcolor[HTML]{2171b5}{Z^{k^{-r_x-2}}}$};
\node at (4.7, 6.5) {\normalsize $\textcolor[HTML]{2171b5}{Z^{\da k^{-r_x-2}}}$};
\node at (1.7, 6.5) {\normalsize $\textcolor[HTML]{2171b5}{Z^{\da k^{-r_x-1}}}$};
\node at (-1.5, 6.5) {\normalsize $\textcolor[HTML]{2171b5}{Z^{\da k^{-r_x}}}$};
\node at (-2, 1.5) {\normalsize $\textcolor[HTML]{2171b5}{Z^{\da k^{-r_x+1}}}$};
\node at (-2, 4.5) {\normalsize $\textcolor[HTML]{2171b5}{Z^{\da k^{-r_x+1}}}$};
\fill[black] (-3,0) circle (6pt);
\node at (-3.5, -0.5) {\normalsize $\bm{r}$};
\node at (-5, 3) {\normalsize $W=$};
\end{tikzpicture}.
\end{equation*}
For such contractible cycles ${C = \pp D}$, these logical operators can be rewritten as products of $(A_{\bm{r}})^{k^{r_x}}$ and $(B_{\bm{r}})^{k^{-r_x}}$, respectively. Therefore, they are topological operators in the code space. The logical operators $V$ and $W$ with $C$ running in the $+x$ direction at fixed $r_y$ are isomorphic to the exponential symmetry operator~\eqref{expSymGen} and its dual symmetry operator~\eqref{dualexpSymGen}, respectively. For a general cycle $C$, they follow $\Z_N\times \Z_N$ fusion rules. Furthermore, for $V$ running in the $x$ ($y$) direction and $W$ in the $y$ ($x$) direction, the two operators fail to commute by the phase ${\exp[\frac{2\pi\ii}{N}k]}$. Therefore, the quantum code describes a $\Z_N$ topological order.

Under a single lattice translation in the $x$ direction, these logical operators satisfy
\begin{equation}
T_x\,V\,T_x^{-1} = V^{k^{-1}},\hspace{40pt}T_x\, W\,T_x^{-1} = W^{k}.
\end{equation}
These are, respectively, the same transformations as $T_x$ acting on the exponential symmetry and its dual symmetry. This action of translations on the logical operators makes the SymTFT a $\Z_N$ topological order enriched by lattice translations. 

A complementary characterization of this SET is through position-dependent excitations~\cite{PW220407111}. Indeed, let us denote a gapped excitation (i.e., an error of the code) corresponding to ${A_{\bm{r}} = \exp[\frac{2\pi\ii}{N}n]}$ and ${B_{\bm{r}} = \exp[\frac{2\pi\ii}{N}n]}$ by $n\mathfrak{e}_{\bm{r}}$ and $n\mathfrak{m}_{\bm{r}}$, respectively. They satisfy ${N \mathfrak{e}_{\bm{r}} = N \mathfrak{m}_{\bm{r}} = 0}$. Furthermore, these excitations are created by the Pauli operators $X_{\bm{r},\mu}$ and $Z_{\bm{r},\mu}$, which gives rise to the relations
\begin{equation}
\mathfrak{m}_{\bm{r} + \hat{y}} = \mathfrak{m}_{\bm{r}},
\hspace{30pt} 
k^{-1} \mathfrak{m}_{\bm{r} + \hat{x}} =  \mathfrak{m}_{\bm{r}},
\hspace{30pt}
\mathfrak{e}_{\bm{r} + \hat{y}} = \mathfrak{e}_{\bm{r}},
\hspace{30pt} 
\mathfrak{e}_{\bm{r} + \hat{x}} = k^{-1}  \mathfrak{e}_{\bm{r}}.
\end{equation}
These equalities mean that the excitation appearing on left-hand-side of the equations are of the same anyon-type as those appearing on the right-hand-side, i.e.,~they belong to the same superselection sector.
Solving these recurrence relations, we express
the anyon-types at position $\bm{r}$ are given by 
\begin{equation}
\mathfrak{m}_{\bm{r}} = k^{r_x} \mathfrak{m},\hspace{40pt}
\mathfrak{e}_{\bm{r}} = k^{-r_x} \mathfrak{e},
\end{equation}
Since the anyon-type depends on the $x$-component of the excitation's position, the topological order is non-trivially enriched by lattice translations in the $x$-direction. Furthermore, when the lattice has periodic boundary conditions, the anyon-types have to satisfy ${\mathfrak{m}_{\bm{r}} = \mathfrak{m}_{\bm{r}+L_x\hat{x}}}$ and ${\mathfrak{e}_{\bm{r}} = \mathfrak{e}_{\bm{r}+L_x\hat{x}}}$, which gives rise to the additional constraints ${(k^{L_x}-1) \mathfrak{m}= (k^{-L_x}-1) \mathfrak{e}=0}$. Therefore, the number of globally distinguishable excitations on a torus is ${\gcd(k^{L_x}-1,N)\gcd(k^{-L_x}-1,N)}$, which equals the ground state degeneracy on a torus.

The smooth and rough boundary stabilizers of this quantum code are
\begin{equation}
A^\text{smooth}_{\bm{r}}
=
\begin{tikzpicture}[scale = 0.5, baseline = {([yshift=-.5ex]current bounding box.center)}]
\draw[line width=0.015in, gray] (7.5,0) -- (13.5, 0);
\draw[line width=0.015in, gray] (10.5,-3) -- (10.5, 0);
\node at (9, .45) {\normalsize $\textcolor[HTML]{cb181d}{X}$};
\node at (12, .45) {\normalsize $\textcolor[HTML]{cb181d}{X^{\da k}}$};
\node at (10.5, -1.5) {\normalsize $\textcolor[HTML]{cb181d}{X}$};
\fill[black] (10.5,0) circle (6pt);
\node at (10, -.5) {\normalsize $r$};
\end{tikzpicture}~,
\hspace{40pt}
B^\text{rough}_{\bm{r}}=\begin{tikzpicture}[scale = 0.5, baseline={([yshift=-.5ex]current bounding box.center)}]
\draw[line width=0.015in, gray] (-3,0) -- (0, 0);
\draw[line width=0.015in, gray] (-3,0) -- (-3, 3);
\draw[line width=0.015in, gray] (0,0) -- (0, 3);
\node at (-1.5, 0.4) {\normalsize $\textcolor[HTML]{2171b5}{Z}$};
\node at (-2.8, 1.5) {\normalsize $\textcolor[HTML]{2171b5}{Z^{\da k}}$};
\node at (0, 1.5) {\normalsize $\textcolor[HTML]{2171b5}{Z}$};
\fill[black] (-3,0) circle (6pt);
\node at (-3.5, -0.5) {\normalsize $r$};
\end{tikzpicture}~.
\end{equation}
The rough boundary condition causes ${W = 1}$ on the boundary but leaves $V$ on the boundary unchanged. Therefore, this boundary symmetry is generated by $V$, which is the exponential symmetry. The smooth boundary is the opposite, and it is the symmetry boundary for the dual exponential symmetry. Therefore, gauging the exponential symmetry is implemented in the SymTFT by changing the rough boundary to the smooth boundary.

\subsubsection*{Field theory perspective}

The Euclidean lattice Lagrangian defining the field theory description of the exponential symmetry's SymTFT is~\eqref{modSymTFTL} with $D$ given by~\eqref{Dexp}. It can be written as
\begin{align}
\mathscr{L}_{\bm{r}} =& \frac{2\pi\ii}N\, (
\,
a_{\bm{r},y}\Del_t b_{\bm{r},x} - a_{\bm{r},x}\Del_t b_{\bm{r}+\hat{x}-\hat{y},y} + b_{\bm{r}+\hat{x},t} ( \Del_x a_{\bm{r},y} - \Del_y a_{\bm{r},x} ) - a_{\bm{r}+\hat{y}-\hat{t},t} ( \Del_y b_{\bm{r},x} - \Del_x b_{\bm{r},y} )
\,),\nonumber\\
&+ \frac{2\pi\ii}N\,(k-1) (
\,
a_{\bm{r}+\hat{y}-\hat{t},t}  b_{\bm{r}+\hat{x},y} - b_{\bm{r}+\hat{x},t}   a_{\bm{r},y} ),\label{expSymLatLag}
\end{align}
The first line of $\mathscr{L}_{\bm{r}}$ is level-$N$ lattice BF theory and can be compactly written as ${\frac{2\pi\ii}N b\cup \dd a}$. The second line arises from the modulated nature of the exponential symmetry. It is turned off when ${k=1}$ and the $\Z_N$ exponential symmetry becomes a uniform $\Z_N$ symmetry.

The topological defect lines of~\eqref{expSymLatLag} follow from the general expression~\eqref{TDLsymTFTmod}. Using that for exponential symmetry~\eqref{expSymGen}, ${D_x= \Del_x + 1-k}$ and ${f_{r_x} = k^{r_x}}$, the topological defect lines are formed by
\begin{align}
W[C] &= \exp[ \frac{2\pi\ii}N \sum_{(\bm{r},\mu)\subset C} \left( k^{-r_x}\,a_{\bm{r},\mu}\,\del_{|\mu|,x}
\,+\,
k^{-r_x+1} a_{\bm{r},\mu}\,\del_{|\mu|,y}
\,+\,
k^{-r_x+1} a_{\bm{r},\mu}\,\del_{|\mu|,t} \right) ],\\
V[C] &= \exp[\frac{2\pi\ii}N \sum_{(\bm{r},\mu)\subset C} \left( k^{r_x}\,b_{\bm{r},\mu}~\del_{|\mu|, x}
\,+\,
k^{r_x} b_{\bm{r},\mu}\ \del_{|\mu|,y}
\,+\,
k^{r_x} b_{\bm{r},\mu}\, \del_{|\mu|, t} \right) ],
\end{align}
where $C$ is a cycle of the spacetime lattice. The equations of motion of~\eqref{expSymLatLag} make $W$ and $V$ topological defects. They also have non-trivial mutual braiding with a Hopf link configuration equaling the phase ${\exp[\frac{2\pi\ii}{N}k]}$, and are the topological defects of $\Z_N$ topological order. Under the transformation $T_x\colon \bm{r} \to \bm{r}+\hat{x}$, they transform as
\begin{align}\label{expTaction}
W[C] \to W^{k}[C],\hspace{40pt} V[C] \to V^{k^{-1}}[C].
\end{align}
Therefore, $W$ and $V$ are modulated topological defect lines. This makes the SymTFT a $\Z_N$ topological order non-trivially enriched by lattice translations, which induces the anyon automorphism~\eqref{expTaction}.

As a continuum field theory, the lattice translation action on the anyons manifests as a foliation structure on spacetime. Indeed, the continuum limit of~\eqref{expSymLatLag} is the action
\begin{equation}\label{expSymContLag}
S[e] = \frac{\ii N}{2\pi}\int b\wdg \dd a - (k-1)\, a\wdg b \wdg e,
\end{equation}
where the foliation 1-form field ${e = \Lambda\, \dd x}$ with $\La^{-1}$ a lattice spacing---a necessary UV cutoff for~\eqref{expSymContLag}. $S$ is a foliated field theory because it depends on the background foliation field $e$. It is a flat foliation whose leaves are the $(y,t)$ planes of spacetime. The action has the gauge redundancy
\begin{equation}
a \sim a + \dd \al + (1-k) \al \, e,\hspace{40pt} b \sim b + \dd \bt - (1-k) \bt\, e,
\end{equation}
which depends on the foliation structure. The second term in $S[e]$ is an insertion of a condensation defect of level-$N$ BF theory on each leaf of the foliation that acts on topological defect lines as~\eqref{expTaction}. It is turned off when ${k=1}$, in which case the modulated symmetry becomes a uniform $\Z_N$ symmetry and the SymTFT action is ordinary level-$N$ BF theory.

The symmetry boundary at fixed $y$ for the exponential symmetry arises from choosing the Dirichlet boundary condition ${a_t = a_x = 0}$. Indeed, for this choice of boundary condition, the topological defect line ${W = 1}$ while $V$ is unchanged, which is an exponential symmetry defect in the $(x,t)$ plane. On the other hand, the Neumann boundary condition ${b_t = b_x = 0}$ trivializes $V$ while leaving $W$ unchanged. $W$ on the boundary is the symmetry defect for the dual exponential symmetry. Therefore, gauging the exponential symmetry in the SymTFT is implemented by changing the Dirichlet to Neumann boundary condition.

\subsubsection*{Application: Classifying phases}

Gapped and gapless states protected by a symmetry are classified by condensable algebras of the symmetry's SymTFT. Condensable algebras for $\Z_N$ topological order are formed by all possible fusions of $e^{a}$ and $m^b$ such that ${ab = 0\bmod N}$. Here, $e$ and $m$ are bosons with $\ee^{2\pi\ii/N}$ mutual statistics and they corresponds to the logical operators/topological defects $W^{k^{-1}}$ and $V$. The constraint ${ab = 0\bmod N}$ ensures that $e^{a}$ and $m^b$ have trivial mutual statistics. They form the group under fusion
\begin{equation}\label{ZnLagAlg}
\cA_{a,b} = \< e^a, m^b \>,
\end{equation}
generated by $e^a$ and $m^b$, where we assume $a$ and $b$ divide $N$. $\cA_{a,b}$ is a Lagrangian algebra and corresponds to a gapped phase if ${|\cA_{a,b}| = N}$. These Lagrangian algebras are associated with the different symmetry-breaking patterns of $\Z_N$ in ${1+1}$D. Furthermore, the Lagrangian algebra corresponding to the symmetry boundary is $\cA_{1,0}$.

For a uniform $\Z_N$ symmetry, the above condensable algebras would be the end of the story. However, for the exponential symmetry, the SymTFT is enriched by translations, and the condensable algebras may not be invariant under the translation action 
\begin{equation}\label{TxActionem}
T_x\colon (e,m) \to (e^k, m^{k^{-1}}).
\end{equation}
Those not invariant correspond to phases for which translations must be explicitly broken to realize. However, each condensable algebra $\cA_{a,b}$ is, in fact, invariant under~\eqref{TxActionem}. Indeed, as a group, $\cA_{a,b}$ is isomorphic to the product group
\begin{equation}
\cA_{a,b} \cong \{1,e^a,e^{2a},\cdots \} \times \{1,m^b,m^{2b},\cdots \}.
\end{equation}
Then, because $k$ is coprime to $N$, the transformation~\eqref{TxActionem} will generally permute the elements of $\{1,e^a,e^{2a},\cdots \}$ and $\{1,m^b,m^{2b},\cdots \}$, but leaves $\cA_{a,b}$ invariant. Therefore, a $\Z_N$ exponential symmetry has the same number of gapped and gapless states as a $\Z_N$ uniform symmetry.

For a ${\Z_N \times \Z_N}$ exponential symmetry whose symmetry operators with two flavors of qudits per site are ${\prod_j [X^{(1)}_j]^{k^j}}$ and ${\prod_j [X^{(2)}_j]^{q^j}}$, the phases' classification is more interesting. The SymTFT of this exponential symmetry is a ${\Z_N\times\Z_N}$ topological order enriched by lattice translations such that
\begin{equation}\label{TxActione12m12}
T_x\colon (e_1,\, m_1, \, e_2,\, m_2) \to (e^{k}_1,\, m^{k^{-1}}_1\!\!,\, e^{q}_2,\, m^{q^{-1}}_2).
\end{equation}
The Lagrangian algebra defining the symmetry boundary is ${\cA = \< e_1, e_2\>}$. There are $N$ magnetic Lagrangian algebras with respect to this symmetry boundary (those whose overlap with $\cA$ is trivial), and they have the general form ${\cA_n = \< 1, m_1 e_2^n, m_2 e_1^{-n}, \cdots\>}$ with ${n = 0,1,\cdots, N-1}$. For a uniform ${\Z_N\times\Z_N}$ symmetry, they correspond to the $N$ different SPTs. However, not all $\cA_n$ are generally invariant under~\eqref{TxActione12m12}. For example, when ${N=8}$, ${k=3}$, and ${q=5}$, only $\cA_{n}$ with ${n=0, 4}$ is invariant. More generally, $\cA_n$ corresponds to an allowed SPT if ${n( k q - 1) = 0 \bmod N}$. Therefore, depending on $k$ and $q$, there can be fewer translation-invariant ${\Z_N \times \Z_N}$ exponential symmetry SPTs than there are uniform ${\Z_N\times\Z_N}$ symmetry SPTs. This agrees with the observation in \Rf{L231104962}.

\subsubsection*{Application: (anomalous) non-invertible reflection}

As we saw both from the quantum code and field theory perspectives, gauging the exponential symmetry is performed in the SymTFT by changing the rough/Dirichlet boundary to the smooth/Neumann boundary. This is implemented by first performing the unitary transforming
\begin{equation}
    U_{\mathrm{KW}}\colon(
    X_{\bm{r},x},\,
    X_{\bm{r},y},\,
    Z_{\bm{r},x},\,
    Z_{\bm{r},y}
    )
    \to 
    (
    Z_{\bm{r}+\hat{x}-\hat{y},y},\,
    Z^\dag_{\bm{r},x},\,
    X^\dag_{\bm{r}+\hat{x}-\hat{y},y},\,
    X_{\bm{r},x}
    ),
\end{equation}
and then the reflection 
\begin{equation}
    R\colon(
    X_{\bm{r},x},\,
    X_{\bm{r},y},\,
    Z_{\bm{r},x},\,
    Z_{\bm{r},y}
    ) 
    \to 
    (
    X^\dag_{R(\bm{r})-\hat{x},x},\,
    X_{R(\bm{r}),y},\,
    Z^\dag_{R(\bm{r})-\hat{x},x},\,
    Z_{R(\bm{r}),y}
    ),
\end{equation}
where ${R(\bm{r}) = (-r_x,r_y)}$. The operator $RU_{\mathrm{KW}}$ is a symmetry of the SymTFT on a torus, transforming the stabilizers~\eqref{expSymTFTStabs} by
\begin{equation}
    RU_{\mathrm{KW}}\colon (A_{\bm{r}},\, B_{\bm{r}}) \to (B_{R(\bm{r})-\hat{x}-\hat{y}},\, A^\dag_{R(\bm{r})-\hat{x}}).
\end{equation}
Therefore, it also exchanges the rough/Dirichlet and smooth/Neumann boundary conditions and, relatedly, transforms $V$ and $W$ running in the $x$ direction by ${(V, W^\dag)\to (W^{k^{-1}}, V^{k})}$. However, since it implements the gauging map of the exponential symmetry, it is a non-invertible reflection in any ${1+1}$D system described by the SymTFT. An operator for this non-invertible reflection and ${1+1}$D quantum spin models with it as a non-invertible reflection symmetry were constructed in \Rf{PDL240612962}.

The non-invertible reflection as a symmetry is always anomalous for a single $\Z_N$ exponential symmetry (i.e., there are no SPT phases compatible with the non-invertible reflection). Indeed, the only SPT phase of a $\Z_N$ exponential symmetry corresponds to the Lagrangian algebra ${\cA_{0,1} = \<1,m,\cdots\>}$, which is not invariant under ${(m, e)\to (e^{k^{-1}}, m^{-k})}$. For the double $\Z_N$ exponential symmetry generated by ${\prod_j [X^{(1)}_j]^{k^j}}$ and ${\prod_j [X^{(2)}_j]^{q^j}}$, we can deduce necessary, but not sufficient conditions for an anomaly. In particular, when there is only one SPT corresponding to the Lagrangian algebra ${\< 1, m_1, m_2, \cdots\>}$ of the SymTFT, the non-invertible reflection symmetries will all be anomalous. Using the result from the previous section, this occurs whenever the only solution to ${n( k q - 1) = 0 \bmod N}$ is ${n=0}$. For example, this is satisfied when ${k=q=2}$ with ${N=5}$.

\subsubsection{Example: dipole symmetry}\label{ZNdipEx}

We next consider a $\Z_N$ dipole symmetry in ${1+1}$D system of $\Z_N$ qudits, which is generated by the symmetry operators
\begin{equation}
U_1 = \prod_{j=1}^L X_j,\hspace{40pt} U_2 = \prod_{j=1}^L (X_j)^j.
\end{equation}
Along with lattice translations, this forms a ${(\Z_N\times\Z_N)\rtimes_\varphi \Z_L}$ symmetry, where the group homomorphism ${\varphi\colon \Z_L \to \Aut(\Z_N\times\Z_N)}$ describes
\begin{equation}
T U_1 T^{-1} = U_1,\hspace{40pt}   T U_2 T^{-1} = U^{-1}_1 U_2. 
\end{equation}
A simple lattice model with a $\Z_N$ dipole symmetry is the $\Z_N$ dipolar clock model
\begin{equation}
\sum_{j=1}^L \left( Z^\da_{j-1} Z_j^2 Z_{j+1}^\da + h X_j\right) + \text{H.c.}
\end{equation}
The bond algebra of a $\Z_N$ dipole symmetry is of the type~\eqref{B{F}modSym} for all $N$~\cite{PDL240612962}, and the matrix $D$ is
\begin{equation}
\label{eq:D matrix dipole}
D_{ij} = [\Del^2]_{ij} = \del_{i+2,j} -2 \del_{i+1,j} + \del_{i,j}.
\end{equation}
Because ${[D^\mathsf{T}]_{i,j} = D_{i-2,j}}$, the dual symmetry from gauging a $\Z_N$ dipole symmetry is also a $\Z_N$ dipole symmetry.

The quantum code description of the SymTFT of a ${G\times Z(G)}$ dipole-type symmetry, for which ${G = \Z_N}$ is a $\Z_N$ dipole symmetry, was constructed in \Rf{PLA240918113}. Here, we will discuss the SymTFT for $\Z_N$ dipole symmetry and its applications both from a quantum code and field theory perspective.

\subsubsection*{Quantum code perspective}

With the $D$ matrix in Eq.\
\eqref{eq:D matrix dipole}, 
the stabilizers~\eqref{stabGenModA} and~\eqref{stabGenModB} become
\begin{align}
A_{\bm{r}}
=
\begin{tikzpicture}[scale = 0.5, baseline = {([yshift=-.5ex]current bounding box.center)}]
\draw[line width=0.015in, gray] (4.5,0) -- (13.5, 0);
\draw[line width=0.015in, gray] (7.5,-3) -- (7.5, 3);
\draw[line width=0.015in, gray] (10.5,-3) -- (10.5, 3);
\node at (6, .45) {\normalsize $\textcolor[HTML]{cb181d}{X}$};
\node at (9, .45) {\normalsize $\textcolor[HTML]{cb181d}{X^{-2}}$};
\node at (12, .45) {\normalsize $\textcolor[HTML]{cb181d}{X}$};
\node at (10.5, 1.5) {\normalsize $\textcolor[HTML]{cb181d}{X^{\da}}$};
\node at (10.5, -1.5) {\normalsize $\textcolor[HTML]{cb181d}{X}$};
\fill[black] (10.5,0) circle (6pt);
\node at (10, -.5) {\normalsize $r$};
\end{tikzpicture}~,
\hspace{40pt}
B_{\bm{r}}=\begin{tikzpicture}[scale = 0.5, baseline={([yshift=-.5ex]current bounding box.center)}]
\draw[line width=0.015in, gray] (-3, 0) -- (-3, 3) -- (3, 3) -- (3, 0) -- cycle;
\draw[line width=0.015in, gray] (0,0) -- (0, 3);
\node at (-1.5, 0.4) {\normalsize $\textcolor[HTML]{2171b5}{Z}$};
\node at (-1.5, 3.45) {\normalsize $\textcolor[HTML]{2171b5}{Z^\da}$};
\node at (-3, 1.5) {\normalsize $\textcolor[HTML]{2171b5}{Z}$};
\node at (0, 1.5) {\normalsize $\textcolor[HTML]{2171b5}{Z^{\da 2}}$};
\node at (3, 1.5) {\normalsize $\textcolor[HTML]{2171b5}{Z}$};
\fill[black] (-3,0) circle (6pt);
\node at (-3.5, -.5) {\normalsize $r$};
\end{tikzpicture}.
\end{align}
This stabilizer code has been studied in~\cite{EHN240110677, PLA240918113} (albeit in a different unitary frame). Denoting by $C$ an oriented cycle of the square lattice, following Eq.\ \eqref{TDLsymTFTmod} the logical operators are formed by
\begin{align}
V^{(1)}[C] &= \prod_{(\bm{r},\mu)\subset C} \left([X_{\bm{r},y}]^\si \del_{\mu, x}
+
[X^\dag_{\bm{r}-2\hat{x}+\hat{y},x}\,
X_{\bm{r}-\hat{x}+\hat{y},x}]^\si
\del_{\mu,y}\right),\\
W^{(1)}[C] &= \prod_{(\bm{r},\mu)\subset C} \left(
[Z_{\bm{r},x}]^\si \del_{\mu,x}
+
[Z^\da_{\bm{r},y} \, Z_{\bm{r}+\hat{x},y} ]^\si \del_{\mu,y}
\right) 
,\\
V^{(2)}[C] &= \prod_{(\bm{r},\mu)\subset C} \left( [X_{\bm{r},y}  ]^{\si r_x}\del_{\mu, x}
+
[X_{\bm{r}-2\hat{x}+\hat{y},x}^\dag ]^{\si r_x}
[X_{\bm{r}-\hat{x}+\hat{y},x}]^{\si(r_x-1)}
\del_{\mu,y} \right)
,\\
W^{(2)}[C] &= \prod_{(\bm{r},\mu)\subset C} \left( [Z_{\bm{r},x}^\dag]^{\si r_x} \del_{\mu,x}
+
[Z_{\bm{r},y} ]^{\si r_x}
[Z_{\bm{r}+\hat{x},y}^\dag ]^{\si (r_x-1)}
\del_{\mu,y} \right).
\end{align}
Graphical representations of these logical operators for a small contractible cycle are
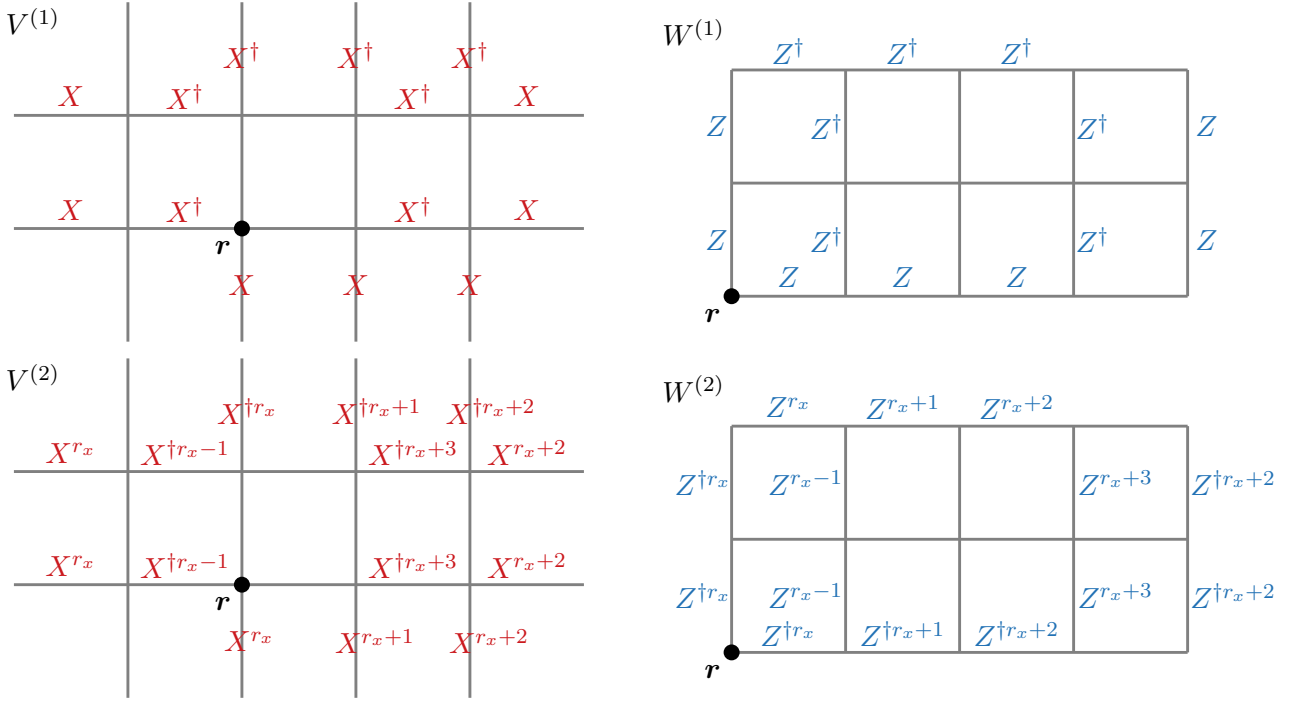
\begin{align*}
&
\begin{tikzpicture}[scale = 0.5, baseline={([yshift=-.5ex]current bounding box.center)}]
\draw[line width=0.015in, gray] (-9,0) -- (6, 0);
\draw[line width=0.015in, gray] (-9,3) -- (6, 3);
\draw[line width=0.015in, gray] (-3,-3) -- (-3, 6);
\draw[line width=0.015in, gray] (-6,-3) -- (-6, 6);
\draw[line width=0.015in, gray] (0,-3) -- (-0, 6);
\draw[line width=0.015in, gray] (3,-3) -- (3, 6);
%
\node at (-3, -1.5) {\normalsize $\textcolor[HTML]{cb181d}{X}$};
\node at (0, -1.5) {\normalsize $\textcolor[HTML]{cb181d}{X}$};
\node at (3, -1.5) {\normalsize $\textcolor[HTML]{cb181d}{X}$};
\node at (4.5, 0.5) {\normalsize $\textcolor[HTML]{cb181d}{X}$};
\node at (1.5, 0.5) {\normalsize $\textcolor[HTML]{cb181d}{X^\dag}$};
\node at (4.5, 3.5) {\normalsize $\textcolor[HTML]{cb181d}{X}$};
\node at (1.5, 3.5) {\normalsize $\textcolor[HTML]{cb181d}{X^\dag}$};
\node at (3, 4.6) {\normalsize $\textcolor[HTML]{cb181d}{X^\dag}$};
\node at (0, 4.6) {\normalsize $\textcolor[HTML]{cb181d}{X^\dag}$};
\node at (-3, 4.6) {\normalsize $\textcolor[HTML]{cb181d}{X^\dag}$};
\node at (-4.5, 0.5) {\normalsize $\textcolor[HTML]{cb181d}{X^\dag}$};
\node at (-4.5, 3.5) {\normalsize $\textcolor[HTML]{cb181d}{X^\dag}$};
\node at (-7.5, 0.5) {\normalsize $\textcolor[HTML]{cb181d}{X}$};
\node at (-7.5, 3.5) {\normalsize $\textcolor[HTML]{cb181d}{X}$};
\fill[black] (-3,0) circle (6pt);
\node at (-3.5, -0.5) {\normalsize $\bm{r}$};
\node at (-8.5, 5.5) {\normalsize $V^{(1)}$};
\end{tikzpicture}
\hspace{25pt}
\begin{tikzpicture}[scale = 0.5, baseline={([yshift=-.5ex]current bounding box.center)}]
\draw[line width=0.015in, gray] (-3,0) -- (9, 0);
\draw[line width=0.015in, gray] (-3,3) -- (9, 3);
\draw[line width=0.015in, gray] (-3,6) -- (9, 6);
\draw[line width=0.015in, gray] (-3,0) -- (-3, 6);
\draw[line width=0.015in, gray] (0,0) -- (-0, 6);
\draw[line width=0.015in, gray] (3,0) -- (3, 6);
\draw[line width=0.015in, gray] (6,0) -- (6, 6);
\draw[line width=0.015in, gray] (9,0) -- (9, 6);
%
\node at (-1.5, 0.5) {\normalsize $\textcolor[HTML]{2171b5}{Z}$};
\node at (1.5, 0.5) {\normalsize $\textcolor[HTML]{2171b5}{Z}$};
\node at (4.5, 0.5) {\normalsize $\textcolor[HTML]{2171b5}{Z}$};
\node at (6.5, 1.5) {\normalsize $\textcolor[HTML]{2171b5}{Z^\da}$};
\node at (6.5, 4.5) {\normalsize $\textcolor[HTML]{2171b5}{Z^\da}$};
\node at (9.5, 1.5) {\normalsize $\textcolor[HTML]{2171b5}{Z}$};
\node at (9.5, 4.5) {\normalsize $\textcolor[HTML]{2171b5}{Z}$};
\node at (-1.5, 6.5) {\normalsize $\textcolor[HTML]{2171b5}{Z^\da}$};
\node at (1.5, 6.5) {\normalsize $\textcolor[HTML]{2171b5}{Z^\da}$};
\node at (4.5, 6.5) {\normalsize $\textcolor[HTML]{2171b5}{Z^\da}$};
\node at (-3.4, 1.5) {\normalsize $\textcolor[HTML]{2171b5}{Z}$};
\node at (-3.4, 4.5) {\normalsize $\textcolor[HTML]{2171b5}{Z}$};
\node at (-.5, 1.5) {\normalsize $\textcolor[HTML]{2171b5}{Z^{\da}}$};
\node at (-.5, 4.5) {\normalsize $\textcolor[HTML]{2171b5}{Z^{\da}}$};
\fill[black] (-3,0) circle (6pt);
\node at (-3.5, -0.5) {\normalsize $\bm{r}$};
\node at (-4, 7) {\normalsize $W^{(1)}$};
\end{tikzpicture}
\\
&\begin{tikzpicture}[scale = 0.5, baseline={([yshift=-.5ex]current bounding box.center)}]
\draw[line width=0.015in, gray] (-9,0) -- (6, 0);
\draw[line width=0.015in, gray] (-9,3) -- (6, 3);
\draw[line width=0.015in, gray] (-3,-3) -- (-3, 6);
\draw[line width=0.015in, gray] (-6,-3) -- (-6, 6);
\draw[line width=0.015in, gray] (0,-3) -- (-0, 6);
\draw[line width=0.015in, gray] (3,-3) -- (3, 6);
%
\node at (-2.8, -1.5) {\normalsize $\textcolor[HTML]{cb181d}{X^{ r_x}}$};
\node at (0.5, -1.5) {\normalsize $\textcolor[HTML]{cb181d}{X^{ r_x+1}}$};
\node at (3.5, -1.5) {\normalsize $\textcolor[HTML]{cb181d}{X^{ r_x+2}}$};
\node at (4.5, 0.5) {\normalsize $\textcolor[HTML]{cb181d}{X^{ r_x+2}}$};
\node at (1.5, 0.5) {\normalsize $\textcolor[HTML]{cb181d}{X^{\dag r_x+3}}$};
\node at (4.5, 3.5) {\normalsize $\textcolor[HTML]{cb181d}{X^{r_x+2}}$};
\node at (1.5, 3.5) {\normalsize $\textcolor[HTML]{cb181d}{X^{\dag r_x+3}}$};
\node at (3.55, 4.6) {\normalsize $\textcolor[HTML]{cb181d}{X^{\dag r_x+2}}$};
\node at (0.55, 4.6) {\normalsize $\textcolor[HTML]{cb181d}{X^{\dag r_x+1}}$};
\node at (-2.85, 4.6) {\normalsize $\textcolor[HTML]{cb181d}{X^{\dag r_x}}$};
\node at (-4.5, 0.5) {\normalsize $\textcolor[HTML]{cb181d}{X^{\dag r_x-1}}$};
\node at (-4.5, 3.5) {\normalsize $\textcolor[HTML]{cb181d}{X^{\dag r_x-1}}$};
\node at (-7.5, 0.5) {\normalsize $\textcolor[HTML]{cb181d}{X^{r_x}}$};
\node at (-7.5, 3.5) {\normalsize $\textcolor[HTML]{cb181d}{X^{r_x}}$};
\fill[black] (-3,0) circle (6pt);
\node at (-3.5, -0.5) {\normalsize $\bm{r}$};
\node at (-8.5, 5.5) {\normalsize $V^{(2)}$};
\end{tikzpicture}
\hspace{25pt}
\begin{tikzpicture}[scale = 0.5, baseline={([yshift=-.5ex]current bounding box.center)}]
\draw[line width=0.015in, gray] (-3,0) -- (9, 0);
\draw[line width=0.015in, gray] (-3,3) -- (9, 3);
\draw[line width=0.015in, gray] (-3,6) -- (9, 6);
\draw[line width=0.015in, gray] (-3,0) -- (-3, 6);
\draw[line width=0.015in, gray] (0,0) -- (-0, 6);
\draw[line width=0.015in, gray] (3,0) -- (3, 6);
\draw[line width=0.015in, gray] (6,0) -- (6, 6);
\draw[line width=0.015in, gray] (9,0) -- (9, 6);
%
\node at (-1.5, 0.5) {\normalsize $\textcolor[HTML]{2171b5}{Z^{\dag r_x}}$};
\node at (1.5, 0.5) {\normalsize $\textcolor[HTML]{2171b5}{Z^{\dag r_x+1}}$};
\node at (4.5, 0.5) {\normalsize $\textcolor[HTML]{2171b5}{Z^{\dag r_x+2}}$};
\node at (7.1, 1.5) {\normalsize $\textcolor[HTML]{2171b5}{Z^{ r_x+3}}$};
\node at (7.1, 4.5) {\normalsize $\textcolor[HTML]{2171b5}{Z^{ r_x+3}}$};
\node at (10.2, 1.5) {\normalsize $\textcolor[HTML]{2171b5}{Z^{\dag r_x +2}}$};
\node at (10.2, 4.5) {\normalsize $\textcolor[HTML]{2171b5}{Z^{\dag r_x+2}}$};
\node at (-1.5, 6.5) {\normalsize $\textcolor[HTML]{2171b5}{Z^{ r_x}}$};
\node at (1.5, 6.5) {\normalsize $\textcolor[HTML]{2171b5}{Z^{ r_x+1}}$};
\node at (4.5, 6.5) {\normalsize $\textcolor[HTML]{2171b5}{Z^{ r_x+2}}$};
\node at (-3.75, 1.5) {\normalsize $\textcolor[HTML]{2171b5}{Z^{\dag r_x}}$};
\node at (-3.75, 4.5) {\normalsize $\textcolor[HTML]{2171b5}{Z^{\dag r_x}}$};
\node at (-1.05, 1.5) {\normalsize $\textcolor[HTML]{2171b5}{Z^{ r_x-1}}$};
\node at (-1.05, 4.5) {\normalsize $\textcolor[HTML]{2171b5}{Z^{ r_x-1}}$};
\fill[black] (-3,0) circle (6pt);
\node at (-3.5, -0.5) {\normalsize $\bm{r}$};
\node at (-4, 7) {\normalsize $W^{(2)}$};
\end{tikzpicture}
\end{align*}
For any contractible cycle, 
the logical operators can be written as products of $A_{\bm{r}}$, $B_{\bm{r}}$, $A_{\bm{r}}^{r_x}$, and $B_{\bm{r}}^{r_x}$, respectively. Therefore, they are all topological operators in the code space, depending only on the homology class of the cycle $C$. 

The logical operators form the group $\Z_N^{4}$ under multiplication. Furthermore, for the operator $V^{(a)}$ running in the $x$ ($y$) direction and the operator $W^{(b)}$ in the $y$ ($x$) direction, one finds the algebra
\begin{equation}\label{ZndipSmat}
\begin{aligned}
V^{(1)} W^{(1)} &= W^{(1)} V^{(1)},
\hspace{72pt}
V^{(1)} W^{(2)} = \ee^{2\pi\ii/N} W^{(2)} V^{(1)}
,\\
V^{(2)} W^{(1)} &= \ee^{2\pi\ii/N} W^{(1)} V^{(2)}
,
\hspace{40pt}
V^{(2)} W^{(2)} = \ee^{2\pi\ii/N} W^{(2)} V^{(2)}.
\end{aligned}
\end{equation}
Therefore, the quantum code describes a ${\Z_N\times\Z_N}$ topological order.\footnote{With the $S$ matrix data~\eqref{ZndipSmat}, the corresponding mutual Chern-Simons theory's $K$ matrix is
\begin{equation*}
K = \begin{pmatrix}
0 & -N & 0 & N \\
-N & 0 & N & 0 \\
0 & N & 0 & 0 \\
N & 0 & 0 & 0
\end{pmatrix}.
\end{equation*}
It is then easy to find an element ${U\in\mathrm{SL}^\pm(4,\Z)}$ that brings ${U K U^\mathsf{T}}$ to the canonical $K$ matrix for ${\Z_N\times\Z_N}$ topological order. For example,
\begin{equation*}
U K U^{\mathsf{T}} = \begin{pmatrix}
0 & N & 0 & 0 \\
N & 0 & 0 & 0 \\
0 & 0 & 0 & N \\
0 & 0 & N & 0
\end{pmatrix}
\quad
\text{with}
\quad
U = \begin{pmatrix}
1 & 0 & 0 & 0 \\
0 & 0 & 0 & 1 \\
0 & 0 & 1 & 0 \\
0 & 1 & 0 & 1
\end{pmatrix} .
\end{equation*}
In this basis, the logical operators are generated by $V^{(1)}$, $[W^{(1)}]^\dag W^{(2)}$, $V^{(2)}$ and $W^{(1)}$.
}
Under translation by one site in the $x$ direction, the logical operators wrapping the $x$ direction transform as
\begin{align}
\begin{split}
T_x V^{(1)} T_x^{-1} &= V^{(1)}, \hspace{78pt} T_x W^{(1)} T_x^{-1} = W^{(1)},\\
T_x V^{(2)} T_x^{-1} &= [V^{(1)}]^{-1} V^{(2)}, \hspace{40pt} T_x W^{(2)} T_x^{-1} = W^{(1)} W^{(2)}.
\end{split}
\end{align}
The lattice translation acts non-trivially on the logical operators, which causes the ${\Z_N\times \Z_N}$ topological order to be non-trivially enriched by lattice translations.

The symmetry enrichment is also characterized in terms of position-dependent excitations~\cite{PW220407111}. We denote by $n\, \mathfrak{e}_{\bm{r}}$ and $n\, \mathfrak{m}_{\bm{r}}$ gapped excitations corresponding to ${A_{\bm{r}} = \exp[\frac{2\pi\ii}{N}n]}$ and ${B_{\bm{r}} = \exp[\frac{2\pi\ii}{N}n]}$, respectively. They satisfy ${N\, \mathfrak{e}_{\bm{r}} = N\, \mathfrak{m}_{\bm{r}} = 0}$. Furthermore, these excitations are created by the Pauli operators $X_{\bm{r},\mu}$ and $Z_{\bm{r},\mu}$, which gives rise to the relations
\begin{equation}
\mathfrak{m}_{\bm{r} + \hat{y}} = \mathfrak{m}_{\bm{r}} ,
\hspace{30pt} 
\mathfrak{m}_{\bm{r} + \hat{x}}  + \mathfrak{m}_{\bm{r}-\hat{x}} = 2 \mathfrak{m}_{\bm{r}},
\hspace{30pt}
\mathfrak{e}_{\bm{r} + \hat{y}} = \mathfrak{e}_{\bm{r}},
\hspace{30pt} 
\mathfrak{e}_{\bm{r} + \hat{x}} + \mathfrak{e}_{\bm{r}-\hat{x}} = 2 \mathfrak{e}_{\bm{r}}.
\end{equation}
Solving these recurrence relations, we find that their anyon-types at position $\bm{r}$ are given by 
\begin{equation}
\mathfrak{m}_{\bm{r}} = \mathfrak{m}_1 - r_x\, \mathfrak{m}_2 ,\hspace{40pt}
\mathfrak{e}_{\bm{r}} = \mathfrak{e}_1 + r_x\, \mathfrak{e}_2.
\end{equation}
Since these excitations depend on the lattice position $\bm{r}$, the ${\Z_N\times\Z_N}$ topological order is non-trivially enriched by translations. Furthermore, under periodic boundary conditions in the $x$-direction, the anyon types further satisfy ${L_x \mathfrak{m}_2 = L_x \mathfrak{e}_2 = 0}$. Therefore, there are $[N \gcd(L_x,N)]^2$ globally distinguishable anyons, and ${[N \gcd(L_x,N)]^2}$ ground states on a spatial torus.

The smooth and rough boundary stabilizers of this quantum code are
\begin{equation}
A^\text{smooth}_{\bm{r}}
=
\begin{tikzpicture}[scale = 0.5, baseline = {([yshift=-.5ex]current bounding box.center)}]
\draw[line width=0.015in, gray] (4.5,0) -- (13.5, 0);
\draw[line width=0.015in, gray] (7.5,-3) -- (7.5, 0);
\draw[line width=0.015in, gray] (10.5,-3) -- (10.5, 0);
\node at (6, .45) {\normalsize $\textcolor[HTML]{cb181d}{X}$};
\node at (9, .45) {\normalsize $\textcolor[HTML]{cb181d}{X^{\da 2}}$};
\node at (12, .45) {\normalsize $\textcolor[HTML]{cb181d}{X}$};
\node at (10.5, -1.5) {\normalsize $\textcolor[HTML]{cb181d}{X}$};
\fill[black] (10.5,0) circle (6pt);
\node at (10, -.5) {\normalsize $r$};
\end{tikzpicture}~,
\hspace{40pt}
B^\text{rough}_{\bm{r}}=\begin{tikzpicture}[scale = 0.5, baseline={([yshift=-.5ex]current bounding box.center)}]
\draw[line width=0.015in, gray] (-3,0) -- (3, 0);
\draw[line width=0.015in, gray] (0,0) -- (0, 3);
\draw[line width=0.015in, gray] (-3,0) -- (-3, 3);
\draw[line width=0.015in, gray] (3,0) -- (3, 3);
\node at (-1.5, 0.4) {\normalsize $\textcolor[HTML]{2171b5}{Z}$};
\node at (-3, 1.5) {\normalsize $\textcolor[HTML]{2171b5}{Z}$};
\node at (0, 1.5) {\normalsize $\textcolor[HTML]{2171b5}{Z^{\da 2}}$};
\node at (3, 1.5) {\normalsize $\textcolor[HTML]{2171b5}{Z}$};
\fill[black] (-3,0) circle (6pt);
\node at (-3.5, -.5) {\normalsize $r$};
\end{tikzpicture}~.
\end{equation}
The rough boundary condition causes ${W^{(1)} = W^{(2)} = 1}$ on the boundary but leaves $V^{(1)}$ and $V^{(2)}$ on the boundary unchanged. Therefore, the symmetry corresponding to this boundary is generated by $V^{(1)}$ and $V^{(2)}$, which is $\Z_N$ dipole symmetry. The smooth boundary is the opposite, and it is the symmetry boundary for the dual dipole symmetry. This is generated by $W^{(1)}$ and $W^{(2)}$, which is the same $\Z_N$ dipole symmetry. Therefore, gauging the $\Z_N$ dipole symmetry is implemented in the SymTFT by changing the rough to smooth boundary.

\subsubsection*{Field theory perspective}

The Euclidean lattice Lagrangian~\eqref{modSymTFTL} describing this code space is
\begin{equation}\label{dipSymLatLag}
\mathscr{L}_{\bm{r}}= \frac{2\pi\ii}N\, (
\,
a_{\bm{r},y}\Del_t b_{\bm{r},x} - a_{\bm{r},x}\Del_t b_{\bm{r}+\hat{x}-\hat{y},y} + b_{\bm{r}+\hat{x},t} ( \Del^2_x a_{\bm{r},y} - \Del_y a_{\bm{r},x} ) - a_{\bm{r}+\hat{y}-\hat{t},t} ( \Del_y b_{\bm{r},x} + \Del^2_x b_{\bm{r}-\hat{x},y} )
\,).
\end{equation}
The topological defect lines of this SymTFT follow from the general expression~\eqref{TDLsymTFTmod}. Using that ${D_x= \Del^2_x}$ and ${f^{(1)}_{r_x} = 1}$, ${f^{(2)}_{r_x} = r_x}$, the topological defect lines are formed by
\begin{align}
V^{(1)}[C] &= \exp[\frac{2\pi\ii}N \sum_{(\bm{r},\mu)\subset C} \left( b_{\bm{r},\mu}~\del_{\mu, x}
\,-\,
\Del_x b_{\bm{r}-\hat{x},\mu}\, \del_{\mu,y}
\,-\,
\Del_x b_{\bm{r}-\hat{x},\mu}\, \del_{\mu, t} \right) ],\\
W^{(1)}[C] &= \exp[ \frac{2\pi\ii}N \sum_{(\bm{r},\mu)\subset C} \left( a_{\bm{r},\mu}\,\del_{\mu,x}
\,+\,
\Del_x a_{\bm{r},\mu}\,\del_{\mu,y}
\,+\,
\Del_x a_{\bm{r},\mu}\,\del_{\mu,t} \right) ],\\
V^{(2)}[C] &= \exp[\frac{2\pi\ii}N \sum_{(\bm{r},\mu)\subset C} \left( r_x\,b_{\bm{r},\mu}~\del_{\mu, x}
\,+\,
(b_{\bm{r}-\hat{x},\mu} -(r_x-1) \Del_x b_{\bm{r}-\hat{x},\mu})\,( \del_{\mu,y}
+ \del_{|\mu|, t} ) \right) ],\\
W^{(2)}[C] &= \exp[ \frac{2\pi\ii}N \sum_{(\bm{r},\mu)\subset C} \left( -r_x\,a_{\bm{r},\mu}\,\del_{\mu,x}
\,+\,
(a_{\bm{r},\mu} -(r_x-1) \Del_x a_{\bm{r},\mu})\,(\del_{\mu,y}
+
\del_{|\mu|,t}) \right) ].
\end{align}
where $C$ is a cycle of the spacetime lattice. The equations of motion of~\eqref{dipSymLatLag} make these topological, and it is straightforward to check that they are the topological defects of a $\Z_N\times\Z_N$ topological order. Under the transformation $T_x\colon \bm{r} \to \bm{r}+\hat{x}$, they transform as
\begin{equation}\label{dipolesymtftfoliationtransformatoin}
V^{(1)}\to V^{(1)}, \qquad
W^{(1)}\to W^{(1)}, \qquad
V^{(2)}\to [V^{(1)}]^{-1} \, V^{(2)}, \qquad
W^{(2)}\to W^{(1)} \, W^{(2)} .
\end{equation}
Therefore, the SymTFT is a ${\Z_N\times\Z_N}$ topological order non-trivially enriched by lattice translations in the $x$-direction.

The symmetry boundary at fixed $y$ for the dipole symmetry arises from choosing the Dirichlet boundary condition ${a_t = a_x = 0}$. Indeed, for this choice of boundary condition, the topological defect lines ${W^{(1)} = W^{(2)} = 1}$ while $V^{(1)}$ and $V^{(2)}$ are unchanged. $V^{(1)}$ and $V^{(2)}$ describe $\Z_N$ dipole symmetry defects in the $(x,t)$ plane. On the other hand, the Neumann boundary condition ${b_t = b_x = 0}$ trivializes $V^{(1)}$ and $V^{(2)}$ while leaving $W^{(1)}$ and $W^{(2)}$ unchanged. Therefore, gauging the exponential symmetry in the SymTFT is implemented by changing the Dirichlet to Neumann boundary condition.

The continuum limit of the lattice Lagrangian~\eqref{dipSymLatLag} is the symmetric tensor gauge theory
\begin{equation}\label{dipSymContLag}
\mathscr{L} = \frac{\ii\, N}{2\pi \La}\, (
\,
a_{y}\pp_t b_{xx} - a_{xx}\pp_t b_{y} + b_{t} ( \pp^2_x a_{y} - \pp_y a_{xx} ) - a_{t} ( \pp_y b_{xx} + \pp^2_x b_{y} )
\,).
\end{equation}
where $\La^{-1}$ is the lattice spacing which serves as a necessary UV cutoff (on dimensionful grounds). The typical $\pp_x$ derivatives of level-$N$ BF theory are instead $\pp^2_x$ derivatives, making the Lagrangian a type of dipolar BF theory~\cite{EHN231006701, EHN240110677, H240308158}. In its current presentation, however, the foliation structure of the continuum SymTFT is not manifest. However, performing a duality transformation (see Appendix~\ref{ZNdipSymTFTFoliApp} for details), the SymTFT can equivalently be formulated as~\cite{EHN240110677}
\begin{equation}\label{FFT}
S[e] = -\frac{\ii\,N}{2\pi} \int \left(\t{a} \wdg \dd b -\t{b}\wdg\dd a - \t{a}\wdg \t{b} \wdg e\right),
\end{equation}
where $a$, $\t{a}$, $b$, and $\t{b}$ are U(1) gauge fields and ${e = \La \dd x}$ is a background foliation field describing a flat foliation whose leaves are $(y,t)$ planes. The gauge redundancy of~\eqref{FFT} is
\begin{equation}
\begin{aligned}
a &\sim a + \dd \al + \t{\al}\, e,
\hspace{40pt} 
b \sim b + \dd \bt + \t{\bt}\, e,\\
\t{a} &\sim \t{a} + \dd \t{\al},
\hspace{72pt} 
\t{b} \sim \t{b} + \dd \t{\bt},
\end{aligned}
\end{equation}
which depends explicitly on the foliation field. The foliation term dresses each leaf with a condensation defect of ${\Z_N\times\Z_N}$ gauge theory that implements the anyon permutation~\eqref{dipolesymtftfoliationtransformatoin}. Turning off this foliation field causes the translation enriched ${\Z_N\times \Z_N}$ gauge theory to become ${\Z_N\times \Z_N}$ gauge theory without any non-trivial translation enrichment.

The presentation~\eqref{FFT} of the $\Z_N$ dipole SymTFT further reveals that this SymTFT can be constructed by gauging the ${\Z_N\times \Z_N}$ symmetry of the invertible theory~\cite{EHN240110677} 
\begin{equation}
\cZ[\, \t a,\t b,e] = \exp[\frac{\ii N}{2\pi} \int \t a \wdg \t b\wdg e].
\end{equation}
This is a field theory description of a ${\Z_N\times \Z_N}$ weak SPT (i.e., an SPT protected by an internal ${\Z_N\times \Z_N}$ symmetry and lattice translations~\cite{FKM0703, CB151102263, YY160100657, C180410122}), with $\t a$ and $\t b$ the background ${\Z_N\times \Z_N}$ gauge fields. In particular, it is ${1+1}$D $\Z_N$ cluster states in each $(y,t)$ plane of the foliation described by $e$. Such invertible foliated field theories are the anomaly-inflow theories for LSM anomalies between internal symmetries and translations~\cite{CB151102263, C180410122}, which makes clear that this SymTFT has a symmetry boundary with an LSM anomaly. We will discuss more on this relation with LSM anomalies in Section~\ref{LSMwithTranslationSec}.

\subsubsection*{Application: Classifying phases}

Having constructed the SymTFT for a $\Z_N$ dipole symmetry, we can now classify ${1+1}$D phases protected by the symmetry (see Fig.~\ref{fig:TopHoloClass}). The condensable algebras of ${\Z_N\times\Z_N}$ topological order are groups and have the general form
\begin{equation}
\{ e_1^{a_i}\, e_2^{b_i}\, m_1^{c_i}\, m_2^{d_i} ~ \mid ~
a_i c_i + b_i d_i =
a_i c_j + b_i d_j + c_i a_j  + d_ib_j  = 0~\mod~N
\},
\end{equation}
The anyons $e_1$, $e_2$, $m_1$, and $m_2$ self bosons. We choose the convention such that they respectively correspond to the logical operators/topological defects $[W^{(1)}]^\dag\, W^{(2)}$, $W^{(1)}$, $V^{(1)}$ and $V^{(2)}$ of the topological order. The Lagrangian algebra corresponding to the symmetry boundary for the $\Z_N$ dipole symmetry is ${\<e_1,e_2\>}$, generated by $e_1$ and $e_2$ via their fusion.

Not all of these condensable algebras are closed under the translation action
\begin{equation}\label{TactionAny}
T\colon e_1,\ e_2,\ m_1,\ m_2 \mapsto e_1 e_2,\ e_2,\ m_1,\ m_1^{-1} m_2.
\end{equation}
Those that are not closed under~\eqref{TactionAny} correspond to phases that can only be realized by explicitly breaking discrete translation symmetry. To investigate how the interplay due to translations affects the classification, we consider only condensable algebras closed under~\eqref{TactionAny}. Including only translation-invariant condensable algebras, the Hasse diagram\footnote{A useful way of organizing condensable algebras when discussing their applications in classifying phases is using a Hasse diagram~\cite{BPS240300905}. A Hasse diagram is a graphical representation of a finite partially ordered set. In this context, it is the set of all condensable algebras ${\{\cA\}}$ of the SymTFT ordered by inclusion. Graphically, it is a graph whose nodes are labeled by condensable algebras with an oriented edge pointing from $\cA_1$ to $\cA_2$ if ${\cA_1}$ is a subalgebra of ${\cA_2}$. The nodes are arranged in rows based on the quantum dimension of the condensable algebra. Following \Rf{BPS240300905}, we arrange the rows in increasing order of the quantum dimensions. This places the smallest condensable algebra on the top row (i.e., $\cA = \{1\}$) and the largest ones on the bottom row (i.e., the Lagrangian algebras). This gives the Hasse diagram a downward orientation, which we denote using arrows. When arrows are not drawn, Hasse diagrams typically have an implied upward orientation.} of the SymTFT for ${N=2}$ is
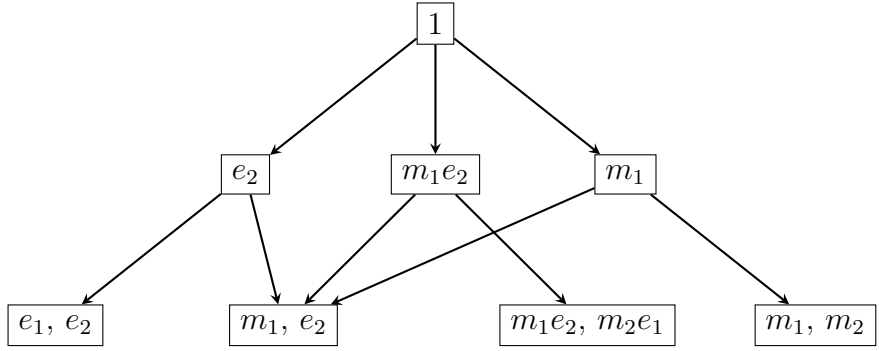
\begin{equation*}
\begin{tikzpicture}[vertex/.style={draw}]
\node[vertex] (1)  at (0,0) {$1$};
\node[vertex] (2)  at (-2.5,-2) {$e_2 $};
\node[vertex] (3) at (0, -2) {${m_1 e_2}$} ;
\node[vertex] (4) at (2.5, -2) {$ m_1$} ;
\node[vertex] (5) at (-5, -4) {$e_1,\,e_2$} ;
\node[vertex] (6) at (-2, -4) {$m_1,\,e_2$} ;
\node[vertex] (7) at (2, -4) {$m_1e_2,\,m_2e_1$} ;
\node[vertex] (8) at (5, -4) {$m_1,\,m_2$} ;
\draw[-stealth] (1) edge [thick] node[label=left:] {} (2);
\draw[-stealth] (1) edge [thick] node[label=left:] {} (3);
\draw[-stealth] (1) edge [thick] node[label=left:] {} (4);
\draw[-stealth] (2) edge [thick] node[label=left:] {} (5);
\draw[-stealth] (2) edge [thick] node[label=left:] {} (6);
\draw[-stealth] (3) edge [thick] node[label=left:] {} (6);
\draw[-stealth] (3) edge [thick] node[label=left:] {} (7);
\draw[-stealth] (4) edge [thick] node[label=left:] {} (6);
\draw[-stealth] (4) edge [thick] node[label=left:] {} (8);
\end{tikzpicture}
\end{equation*}
where each box (node) is labeled by the generator(s) of that condensable algebra (which is a group in this case). The condensable algebras not closed under translations that are not included in the Hasse diagram are
\begin{equation}\label{nonTInvAZN}
\begin{aligned}
&\<e_1\>,
\hspace{30pt}
\<e_1e_2\>,
\hspace{30pt}
\<m_2\>,
\hspace{30pt}
\<m_1 m_2\>,
\hspace{30pt}
\<e_1 m_2\>,
\hspace{30pt}
\<m_1 m_2 e_1 e_2\>,\\
&\<e_1 , \,m_2\>,
\hspace{30pt}
\<e_1 e_2,\, m_1 m_2\>,
\end{aligned}
\end{equation}
with the last two being Lagrangian algebras.

Physically, the nodes of the Hasse diagram label states of matter, and the arrows between them describe symmetry-allowed deformations between phases. Therefore, the bottom row characterizes gapped states, and all other rows characterize gapless states. In the above Hasse diagram, there are four gapped states and four gapless states, corresponding to gapped SPT and SSB states or gapless SPT and SSB---gSPT and gSSB---states. For the $\Z_2$ dipole symmetry boundary ${\cA = \<e_1,\, e_2\>}$, the Hasse diagram is 
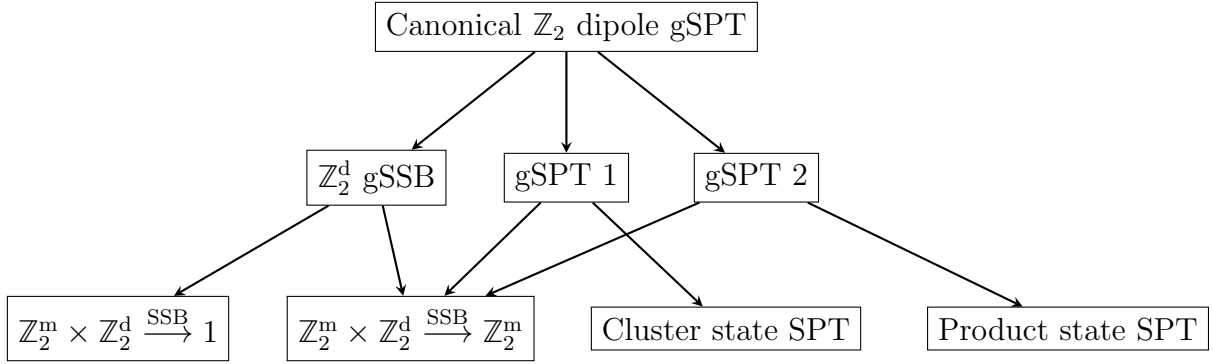
\begin{equation*}
\begin{tikzpicture}[vertex/.style={draw}]
\node[vertex] (1)  at (0,0) {Canonical $\Z_2$ dipole gSPT};
\node[vertex] (2)  at (-2.5,-2) {$\Z_2^\mathrm{d}$ gSSB};
\node[vertex] (3) at (0, -2) {gSPT 1} ;
\node[vertex] (4) at (2.5, -2) {gSPT 2} ;
\node[vertex] (5) at (-5.9, -4) {${\Z_2^{\mathrm{m}} \times \Z_2^{\mathrm{d}}\ssb 1}$} ;
\node[vertex] (6) at (-2.05, -4) {${\Z_2^{\mathrm{m}} \times \Z_2^{\mathrm{d}}\ssb \Z_2^{\mathrm{m}}}$} ;
\node[vertex] (7) at (2.1, -4) {Cluster state SPT} ;
\node[vertex] (8) at (6.6, -4) {Product state SPT} ;
\draw[-stealth] (1) edge [thick] node[label=left:] {} (2);
\draw[-stealth] (1) edge [thick] node[label=left:] {} (3);
\draw[-stealth] (1) edge [thick] node[label=left:] {} (4);
\draw[-stealth] (2) edge [thick] node[label=left:] {} (5);
\draw[-stealth] (2) edge [thick] node[label=left:] {} (6);
\draw[-stealth] (3) edge [thick] node[label=left:] {} (6);
\draw[-stealth] (3) edge [thick] node[label=left:] {} (7);
\draw[-stealth] (4) edge [thick] node[label=left:] {} (6);
\draw[-stealth] (4) edge [thick] node[label=left:] {} (8);
\end{tikzpicture}
\end{equation*}
We denote by ${\Z_2^{\mathrm{m}}}$ and ${\Z_2^{\mathrm{d}}}$ the uniform and modulated $\Z_2$ subgroups, respectively, of the $\Z_2$ dipole symmetry (i.e., $\mathrm{m}$ for monopole and $\mathrm{d}$ for dipole). Therefore, in translation-invariant theories, there are two SSB and two SPT gapped states characterized by a $\Z_2$ dipole symmetry. Furthermore, there is one gapless SSB state with two superselection sectors (i.e., two universes from an emergent $\Z_2$ 1-form symmetry) resulting from the spontaneously broken $\Z_2^\mathrm{d}$ symmetry, and two gapless SPT states.
These two gSPT states differ in their decorated domain wall patterns by a relative $0+1$d $\Z_2$ SPT. Interestingly, at the level of the symmetry, there is no obstruction to reach any of these three gapless states starting from the ${\Z_2^{\mathrm{m}} \times \Z_2^{\mathrm{d}}\ssb \Z_2^{\mathrm{m}}}$ state using a single deformation.

A $\Z_N$ dipole symmetry has the same number of SPTs as a uniform ${\Z_N\times \Z_N}$ symmetry~\cite{L231104962}. Indeed, in the SymTFT, they correspond to the Lagrangian condensable algebras ${\< m_1 e_2^n, m_2 e_1^{-n}\>}$ where ${n\in\{0,1,\cdots,N-1\}}$, and are all invariant under the translation action~\eqref{TactionAny}. A ${\Z_N\times \Z_N}$ dipole symmetry, however, has fewer SPTs than a uniform ${\Z_N^{4}}$ symmetry~\cite{L231104962}. Let us see how this arises through the ${\Z_N\times \Z_N}$ dipole SymTFT, which is simply the ${\Z_N}$ dipole SymTFT stacked with itself.

We denote the anyons of the ${\Z_N\times \Z_N}$ dipole SymTFT by $\{1,e_1,m_1,e_2,m_2,\t{e}_1,\t{m}_1,\t{e}_2,\t{m}_2,\cdots\}$ where the subscript on each anyon follow the same convention as for the ${\Z_N}$ dipole SymTFT. The symmetry boundary corresponding to a ${\Z_N\times \Z_N}$ dipole symmetry corresponds to the Lagrangian subgroup ${\< e_1,e_2,\t{e}_1,\t{e}_2\>}$. The magnetic Lagrangian algebras with respect to this symmetry boundary are\footnote{The magnetic Lagrangian algebras for the ${\Z_N\times \Z_N}$ dipole symmetry can be derived by starting with the group ${\<m_1 e_1^{\,C_{11}}  e_2^{\,C_{12}} \t{e}_1^{\,\,C_{13}} \t{e}_2^{\,\,C_{14}}
,~
m_2 e_1^{\,C_{21}}  e_2^{\,C_{22}} \t{e}_1^{\,\,C_{23}} \t{e}_2^{\,\,C_{24}}
,~
\t{m}_1 e_1^{\,C_{31}}  e_2^{\,C_{32}} \t{e}_1^{\,\,C_{33}}  \t{e}_2^{\,\,C_{34}}
,~
\t{m}_2 
e_1^{\,C_{41}}  e_2^{\,C_{42}} \t{e}_1^{\,\,C_{43}} \t{e}_2^{\,\,C_{44}} \>}$ and then applying the constraints ${C_{ii} = 0~\mod~N}$ (without an implied summation) and ${C_{ij} + C_{ji} = 0~\mod~N}$ to ensure each anyon has bosonic mutual and self statistics.}
\begin{equation}\label{ZN4noT}
\<\, m_1 e_2^{C_{12}}\t{e}_1^{\,\,C_{13}} \t{e}_2^{\,\,C_{14}}
,\,
m_2e_1^{-C_{12}} \t{e}_1^{\,\,C_{23}} \t{e}_2^{\,\,C_{24}}
,\,\t{m}_1e_1^{-C_{13}} e_2^{-C_{23}} \t{e}_2^{\,\,C_{34}}
,\,
\t{m}_2
e_1^{-C_{14}} e_2^{-C_{24}}\t{e}_1^{\,\,-C_{34}}
\, \>,
\end{equation}
with ${C_{ij}\in\Z_N}$. There are six $\Z_N$-valued parameters labeling these (i.e., ${C_{12}, C_{13}, C_{14}, C_{23}, C_{24}, C_{34}}$), which correspond to the ${N^6 = |H^2(B\Z_N^{4},U(1))|}$ different uniform $\Z_N^4$ SPTs. However, not all of these are closed under translations. Those that are satisfy ${C_{13} = 0~\mod~N}$ and ${C_{14} + C_{23}=0~\mod~N}$. These translation-invariant magnetic Lagrangian subgroups are
\begin{equation}
\<\, m_1\,e_2^{\,C_{12}}\, \t{e}_2^{\,\,-C_{23}}
,\,
m_2\,e_1^{\,-C_{12}}\,\t{e}_1^{\,\,C_{23}}\, \t{e}_2^{\,\,C_{24}}
,\,
\t{m}_1\, e_2^{-\,C_{23}}\, \t{e}_2^{\,\,C_{34}}
,\,
\t{m}_2\,
\,e_1^{\,C_{23}}\, e_2^{\,-C_{24}}\,\t{e}_1^{\,\,-C_{34}}
\,\>,
\end{equation}
and they correspond to ${\Z_N\times\Z_N}$ dipolar SPTs. Because there are four $\Z_N$-valued parameters (i.e., ${C_{12}, C_{23}, C_{24}, C_{34}}$), there are ${N^4 = |H^2(B\Z_N^{4},U(1))/(H^2(B\Z_N^{ 2},U(1)))^2|}$ SPTs protected by a ${\Z_N\times\Z_N}$ dipole symmetry. This matches the classification from \Rf{L231104962}.

\subsubsection{Classifying modulated SPTs}\label{classModSPTsSec}

From the exponential and dipole symmetry examples presented in Section~\ref{sec:exponential sym} and Section~\ref{ZNdipEx}, respectively, we saw how the translation-enriched SymTFT is useful for classifying modulated SPTs~\cite{HLL230910036, L231104962, LHY240313880, KLH250315834}. Here, we will discuss aspects of this classification for more general finite modulated symmetries in ${1+1}$D.

Consider a ${2+1}$D SymTFT $\mathfrak{Z}(\cS)$ for the symmetry $\cS$, and suppose the SymTFT is enriched by lattice translations $T_x$ by one site in the $x$-direction. If $\cS$ is a modulated symmetry, then its corresponding symmetry boundary of $\mathfrak{Z}(\cS)$ has defects that are non-trivially acted on by $T_x$. That is, given the Lagrangian algebra $\cL_\cS$ of this symmetry boundary, there is an anyon ${a\not\in\cL_\cS}$ for which $T_x(a)\in\mathcal{L}_\cS$ but ${T_x(a)\not\cong a}$ on the symmetry boundary. It is important that ${T_x(a)\not\cong a}$ holds not only in the SymTFT bulk but also on the symmetry boundary. It is possible that ${T_x(a)\not\cong a}$ in the bulk, but ${T_x(a)\cong a}$ when $a$ resides on the symmetry boundary. This possibility occurred, for example, in Section~\ref{ZNdipEx} when translations acted on the anyon $e_1$ by ${T_x(e_1) = e_1e_2}$ in the bulk, but acted trivially on $e_1$ on the ${1\oplus m_1\oplus e_2\oplus m_1e_2}$ condensed boundary since $e_2$ is condensed on the boundary. We will denote by $\rho_{T}$ the automorphism of $\cS$ induced by restricting the translation anyon automorphism $T_x$ to the symmetry boundary.

The classification of modulated $\cS$-SPTs without lattice translation symmetry is the same as the classification of uniform $\cS$-SPTs. Indeed, let us first forget about the enriching translation symmetry and consider the SymTFT $\mathfrak{Z}(\cS)$ by itself. In this case, the magnetic Lagrangian algebras for $\cL_\cS$ are Lagrangian algebras of $\mathfrak{\cS}$ that overlap trivially with $\cL_\cS$. They are classified by the Fiber functors of the fusion category $\cS$~\cite{TW191202817} and are in one-to-one correspondence to uniform $\cS$-SPT.

When modulated SPTs are discussed in the literature, it is often implicitly assumed that translation symmetry is preserved. A translation-invariant modulated $\cS$-SPT, which we will refer to as just a modulated $\cS$-SPT from here on, corresponds to a $T_x$-stable\footnote{Recall from Section~\ref{gappedBdySETsec} that a $T_x$-stable condensable algebra $\cA$ is one for which $T_x(\cA)\cong \cA$.} magnetic Lagrangian algebra for $\cL_\cS$. Those that are not $T_x$-stable correspond to $\cS$-SPTs that explicitly break lattice translations. The $T_x$ stable magnetic Lagrangian algebras form a subset in the set of all magnetic Lagrangian algebra of $\cL_\cS$. Consequently, the number of modulated $\cS$-SPTs is always less than or equal to the number of uniform $\cS$-SPTs \cite{L231104962}.

Let us contextualize this classification to the invertible finite symmetry case, where ${\cS}$ is a finite group $G$ and the SymTFT is ${2+1}$D $G$ gauge theory. The electric anyons are labeled by irreps $\Ga\in\mathrm{Rep}(G)$ of $G$, and the magnetic anyons by conjugacy classes ${[g]}\in \mathrm{Cl}(G)$ of $G$. The ${G}$ symmetry boundary has the electric Lagrangian algebra ${\cL_G = \bigoplus_{\Ga\in\mathrm{Rep}(G)} d_\Ga\, \Ga}$ condensed, where $d_\Ga$ the dimension of the irrep $\Ga$. The $T_x$ action on the magnetic anyons induces a group automorphism ${\rho_T\in \Aut(G)}$ on the symmetry boundary that describes the semi-direct product structure between the modulated $G$ symmetry with lattice translations.

The different magnetic Lagrangian algebras for $\cL_{G}$ are classified by the projective representations ${H^2(BG,U(1))}$ of $G$ and correspond to different $G$-SPTs. There is always the magnetic Lagrangian algebra $\mathcal{L}={\bigoplus_{[g]\in\mathrm{Cl}(G)}[g]}$, and any other magnetic Lagrangian algebras will differ by their condensed anyons and algebra structure. The action of $T_x$ on the magnetic Lagrangian algebras induces an action of $T_x$ on ${H^2(BG,U(1))}$, which is naturally given by the pullback of ${\rho_T\in\Aut(G)}$. For a representative 2-cocycle ${\om(g_1,g_2)\in Z^2(BG,U(1))}$, this action is ${\rho_T^*\, \om(g_1,g_2) = \om(\rho_T(g_1), \rho_T(g_2))}$. Therefore, the $T_x$-stable magnetic Lagrangian algebras, which correspond to modulated $G$-SPTs, are classified by the $\rho_T$-invariants of ${H^2(BG,U(1))}$:\footnote{This agrees with the classification found using matrix product state~\cite{LN} and real-space/defect network constructions for modulated SPTs~\cite{JCQ190708596, B250806604}. We thank Shang-Qiang Ning and Daniel Bulmash for related discussions.}
\begin{equation}\label{TxInvariants}
\{[\om] \in H^2(BG,U(1)) \mid \rho_T^*[\om] = [\om]\}.
\end{equation}

\subsection{LSM anomaly with translations}\label{LSMwithTranslationSec}

Another way translation and internal symmetries can interplay is through LSM anomalies. An internal symmetry has an LSM anomaly involving translations if it cannot realize a translation-invariant SPT state. However, if the internal symmetry is anomaly-free, there is an SPT state that is symmetric with respect to the internal symmetry but explicitly breaks the translations. Such LSM anomalies have many similarities with 't Hooft anomalies and can be interpreted as mixed 't Hooft anomalies between internal and translation symmetries~\cite{Cho2017, MT170707686, CS221112543, AMF230800743, S230805151}. 

Like 't Hooft anomalies, LSM anomalies between invertible internal symmetries and lattice translations have an anomaly inflow mechanism. The anomaly inflow theory in one-higher dimension is a crystalline SPT protected by the internal and lattice translation symmetry~\cite{TE161200846, CB151102263, Jian2018, C180410122, ET190708204}. For example, a well known LSM-anomaly in a ${1+1}$D system of qubits is realized by the ${\Z_2\times\Z_2}$ symmetry operators ${U_X = \prod_{j=1}^L X_j}$ and ${U_Z = \prod_{j=1}^L Z_j}$~\cite{CGX10083745,Ogata2019,YO201009244}. A manifestation of this anomaly is the projective algebra ${U_XU_Z = (-1)^L U_Z U_X}$ with a lattice-size-dependent phase factor arising from the local projective representation ${X_j Z_j = - Z_j X_j}$. This is interpreted as ${\Z_2\times\Z_2}$ being projectively represented in the presence of translation defects~\cite{CS221112543, S230805151}. The inflow theory is a ${2+1}$D weak ${\Z_2\times\Z_2}$ SPT.\footnote{A weak $G$ SPT is an SPT protected by an internal $G$ symmetry and lattice translations. The adjective ``weak'' is used to emphasize that weak SPTs are fragile to spatial disorder that explicitly breaks the discrete translation symmetry~\cite{FKM0703, CB151102263, YY160100657, C180410122}. In line with this terminology, sometimes SPTs protected by only an internal symmetry are called strong SPTs.} It is constructed by layering ${1+1}$D $\Z_2\times \Z_2$ SPTs on each $(y,t)$ plane orthogonal to the $(x,t)$ boundary, which by themselves correspond to the ${1+1}$D cluster state (i.e., the ${1+1}$D invertible theory ${\cZ[A,B] = (-1)^{\int A\cup B}}$).

Knowing the inflow theory of these LSM anomalies, we can find the SymTFT by gauging the internal symmetry of the weak SPT. In what follows, we consider the LSM anomaly involving one-dimensional lattice translations and an internal symmetry described by the finite Abelian group $G$. These LSM anomalies are classified by ${H^2(BG,U(1))}$,\footnote{From the crystalline equivalence principle~\cite{TE161200846, ZYQ201215657, D210202941, MCB221002452}, these LSM anomalies are classified by ${H^2(\,BG,\,H^1(B\Z,\,U(1))\,)}$ arising in the K{\"u}nneth decomposition of ${H^3(\,B[G\times \Z],U(1)\,)}$. See~\Rf{S230805151} for a derivation of this classification using topological defects.} which corresponds to the different local $G$ projective representation of the onsite $G$ symmetry operators. The corresponding inflow theory is a ${2+1}$D weak $G$ SPT, with ${1+1D}$ $G$ SPTs corresponding to the cohomology class of ${H^2(BG,U(1))}$ dressing the $(y,t)$ planes of ${2+1}$D spacetime. Since $G$ is finite Abelian, it has the canonical isomorphism as the product group ${G \cong \prod_{I=1}^n \Z_{N_I}}$ and ${H^2(BG,U(1))\cong \prod_{I<J} \Z_{N_{IJ}}}$ with $N_{IJ}\equiv \gcd(N_I,N_J)$. The ${1+1D}$ $G$ SPTs can be characterized by the anti-symmetric matrix $K$ whose elements ${K_{IJ}\in \{0,1,\cdots, N_{IJ}-1\}}$ for ${I<J}$ describe the $\Z_{N_{I}}$ symmetry charge carried by $\Z_{N_{J}}$ symmetry defects in the SPT state. The matrix $K$ appears in the ${1+1}$D LSM anomaly through the local projective representation ${U_I^{(j)}U_J^{(j)} = \ee^{2\pi\ii \frac{K_{IJ}}{N_{IJ}}} U_J^{(j)}U_I^{(j)}}$ of the onsite $\Z_{N_I}$ and $\Z_{N_J}$ symmetry operators ${U_I = \prod_{j=1}^L U_I^{(j)}}$ and ${U_J = \prod_{j=1}^L U_J^{(j)}}$, respectively.

In what follows, we will construct the SymTFT from both a quantum code and Euclidean field theory perspective.

\subsubsection*{Quantum code perspective}

From a quantum code perspective, the ${2+1}$D weak ${G \cong \prod_{I=1}^n \Z_{N_I}}$ SPT is described by a quantum code whose code space is one-dimensional on all lattices. To construct this code, we consider a square lattice on a torus and place $\Z_{N_I}$ qudits ($I = 1,2,\cdots, n$) on each site ${\bm{r}\sim \bm{r} + L_x\,\hat{x}\sim \bm{r} + L_y\,\hat{y}}$. They are acted on by the respective clock and shift operators $Z_{\bm{r}}^{(I)}$ and $X_{\bm{r}}^{(I)}$, and the total Hilbert space is ${\otimes_{\bm{r}}\C^{N_1 N_2\cdots N_n}}$. The code space is specified by the mutually commuting stabilizers\footnote{Using the stabilizers ${A^{(I)}_{\bm{r}}}$, one can write down a commuting projector Hamiltonian model whose unique gapped ground state is the SPT state. In particular, the Hamiltonian is ${H = -\sum_{\bm{r},I} P_{\bm{r}}^{(I)}}$ with ${P_{\bm{r}}^{(I)} = N_I^{-1}\sum_{j=1}^{N_I}\, [A^{(I)}_{\bm{r}}]^j}$ the projector onto the ${A^{(I)}_{\bm{r}} = 1}$ states.}
\begin{equation}\label{weakSPTStab}
A^{(I)}_{\bm{r}} = X_{\bm{r}}^{(I)} \prod_{J=1}^n \,[\cO_{\bm{r}}^{(I,J)}]^{\frac{N_J}{N_{IJ}}K_{IJ}}
\hspace{20pt}
\text{where}
\hspace{20pt}
\cO^{(I,J)}_{\bm{r}} = \begin{cases}
Z_{\bm{r}-\hat{y}}^{(J)\,\dag}\,Z_{\bm{r}}^{(J)} &\quad  J<I,
\\
Z_{\bm{r}}^{(J)\,\dag}\, Z_{\bm{r}+\hat{y}}^{(J)} &\quad  J>I.
\end{cases}
\end{equation}
The factor of ${N_J/N_{IJ}}$ is included in front of ${K_{IJ}}$ to ensure each $A^{(I)}_{\bm{r}}$ is an order $N_I$ operator. There is a single state $\ket{\psi}$ satisfying ${A^{(I)}_{\bm{r}}\ket{\psi} = \ket{\psi}}$ for all $\bm{r}$ and $I$. This is an SPT state, satisfying ${U_I \ket{\psi} = \ket{\psi}}$ for each $\Z_{N_I}$ symmetry operator ${U_I = \prod_{\bm{r}} X_{\bm{r}}^{(I)}}$. The corresponding SPT entangler is
\begin{equation}
U_{K} = \prod_{\bm{r}} U_{K}^{\<\bm{r},\bm{r}+\hat{y}\>}
\hspace{15pt}
\text{where}
\hspace{15pt}
U_{K}^{\<\bm{r},\bm{r}+\hat{y}\>} = \sum_{\{\bm{g}\}} 
\ee^{2\pi\ii \sum_{I<J} \frac{K_{IJ}}{ N_{IJ}} \, g^{(I)}_{\bm{r}}\left( g^{(J)}_{\bm{r}+\hat{y}} - g^{(J)}_{\bm{r}}\right) } \ketbra{\bm{g}}{\bm{g}},
\end{equation}
which satisfies ${U_K X_{\bm{r}}^{(I)} U_K^{-1} = A_{\bm{r}}^{(I)}}$.

To gain some intuition for this general quantum code, let us consider ${n=2}$, ${N_1 = N_2 = N}$, and ${K_{12} = 1}$. In this case, the stabilizers become ${A^{(1)} = [Z^{(2)}_{\bm{r}}]^\dag X^{(1)}_{\bm{r}} Z^{(2)}_{\bm{r}+\hat{y}}}$ and ${A^{(2)} = Z^{(1)}_{\bm{r}-\hat{y}} X^{(2)}_{\bm{r}} [Z^{(1)}_{\bm{r}}]^\dag}$, which for a fixed $r_x$ are the stabilizers for the $\Z_N$ cluster state. Therefore, the state $\ket{\psi}$ in this example is a weak $\Z_N$ cluster state, protected by ${\Z_N\times\Z_N}$ and lattice translations in the $x$ direction.

As previously mentioned, this weak $G$ SPT state is characterized by the anti-symmetric matrix $K_{IJ}$, which specifies the decorated domain wall pattern. Indeed, consider inserting a $U_J$ symmetry defect (i.e., non-dynamical domain wall) along the non-contractible cycle of the dual lattice passing through the links ${\<\,(r_x,L_y),\,(r_x,1)\,\>}$. This gives rise to the twisted boundary condition ${Z^{(J)}_{\bm{r}+L_y\hat{y}} = \ee^{-\frac{2\pi\ii}{N_J}}Z^{(J)}_{\bm{r}}}$ and modifies the stabilizers
\begin{equation}
A^{(I)}_{\bm{r}} \xrightarrow{~\text{Insert $U_J$ defect}~} \begin{cases}
[\ee^{-2\pi\ii  \frac{K_{IJ}}{N_{IJ}}}]^{\del_{r_y, L_y
}} A^{(I)}_{\bm{r}} &\quad  I<J,\\
A^{(I)}_{\bm{r}} &\quad  I=J,\\
[\ee^{-2\pi\ii \frac{K_{IJ}}{N_{IJ}} }]^{\del_{r_y, 1
} } A^{(I)}_{\bm{r}} &\quad  I>J.
\end{cases}
\end{equation}
The modified stabilizers cause the code space to change. It is still one-dimensional, but the state $\ket{\psi_J}$ in the code space now satisfies
\begin{equation}
A^{(I)}_{\bm{r}} \ket{\psi_J} = 
\begin{cases}
[\ee^{2\pi\ii  \frac{K_{IJ}}{N_{IJ}}}]^{\del_{r_y, L_y
}} \ket{\psi_J} &\quad  I<J,\\
\ket{\psi_J} &\quad  I=J,\\
[\ee^{2\pi\ii  \frac{K_{IJ}}{N_{IJ}}}]^{\del_{r_y, 1
}} \ket{\psi_J} &\quad  I>J.
\end{cases}
\end{equation}
Therefore, $\ket{\psi_J}$ satisfies ${U_I \ket{\psi_J} = \ee^{2\pi\ii \frac{K_{IJ}}{N_{IJ}}\, L_x} \ket{\psi_J}}$. Consequently, for the SPT state $\ket{\psi}$, each segment of the ${U_J}$ domain wall running in the ${\pm \hat{x}}$ direction carries $\pm\frac{N_I}{N_{IJ}} K_{IJ}$ units of $\Z_{N_I}$ symmetry charge.

To construct the SymTFT, we now gauge the ${G}$ symmetry of the stabilizer code. This is implemented by the gauging map 
\begin{equation}
X_{\bm{r}}^{(I)}
\to~
\begin{tikzpicture}[scale = 0.45, baseline = {([yshift=-.5ex]current bounding box.center)}]
\draw[line width=0.015in, gray] (7.5,0) -- (13.5, 0);
\draw[line width=0.015in, gray] (10.5,-3) -- (10.5, 3);
\node at (9, .5) {\normalsize $\textcolor[HTML]{cb181d}{X^{(I)\dag}}$};
\node at (12, .5) {\normalsize $\textcolor[HTML]{cb181d}{X^{(I)}}$};
\node at (10.6, 2) {\normalsize $\textcolor[HTML]{cb181d}{X^{(I)}}$};
\node at (10.7, -2) {\normalsize $\textcolor[HTML]{cb181d}{X^{(I)\dag}}$};
\fill[black] (10.5,0) circle (6pt);
\node at (10, -0.5) {\normalsize $\bm{r}$};
\end{tikzpicture}~,
\hspace{40pt}
Z^{(I)}_{\bm{r}}Z^{(I)\dag}_{\bm{r}+\hat{x}}
\to
Z^{(I)}_{\bm{r},x},
\hspace{40pt}
Z^{(I)}_{\bm{r}}Z^{(I)\dag}_{\bm{r}+\hat{y}}
\to
Z^{(I)}_{\bm{r},y},
\end{equation}
for each flavor of qudit. This maps the quantum code to a new quantum code with $\Z_{N_I}$ qudits on the edges of the square lattice and whose stabilizers are
\begin{equation}\label{gaugedWeakSPTStab}
A_{\bm{r}}^{(I)} = \begin{tikzpicture}[scale = 0.5, baseline = {([yshift=-.5ex]current bounding box.center)}]
\draw[line width=0.015in, gray] (7.5,0) -- (13.5, 0);
\draw[line width=0.015in, gray] (10.5,-3) -- (10.5, 3);
\node at (9, .5) {\normalsize $\textcolor[HTML]{cb181d}{X^{(I)\dag}}$};
\node at (12, .5) {\normalsize $\textcolor[HTML]{cb181d}{X^{(I)}}$};
\node at (12.75, 2.2) {\normalsize $\textcolor[HTML]{cb181d}{X^{(I)}}\,\textcolor[HTML]{2171b5}{\prod_{J=I}^{n} Z^{(J)\frac{N_J}{N_{IJ}}K_{JI}}}$};
\node at (12.55, -2.2) {\normalsize $\textcolor[HTML]{cb181d}{X^{(I)\dag}}\,\textcolor[HTML]{2171b5}{\prod_{J=1}^{I-1} Z^{(J)\frac{N_J}{N_{IJ}}K_{JI}}}$};
\fill[black] (10.5,0) circle (6pt);
\node at (10, -0.5) {\normalsize $\bm{r}$};
\end{tikzpicture}
\hspace{40pt}
B^{(I)}_{\bm{r}}=\begin{tikzpicture}[scale = 0.5, baseline={([yshift=-.5ex]current bounding box.center)}]
\draw[line width=0.015in, gray] (-3, 0) -- (-3, 3) -- (0, 3) -- (0, 0) -- cycle;
\node at (-1.5, 0.5) {\normalsize $\textcolor[HTML]{2171b5}{Z^{(I)}}$};
\node at (-1.5, 3.5) {\normalsize $\textcolor[HTML]{2171b5}{Z^{(I)\dag}}$};
\node at (-2.7, 1.5) {\normalsize $\textcolor[HTML]{2171b5}{Z^{(I)\dag}}$};
\node at (0.15, 1.5) {\normalsize $\textcolor[HTML]{2171b5}{Z^{(I)}}$};
\fill[black] (-3,0) circle (6pt);
\node at (-3.5, -0.5) {\normalsize $\bm{r}$};
\end{tikzpicture}.
\end{equation}
These are mutually commuting operators, and each $A_{\bm{r}}^{(I)}$ and $B_{\bm{r}}^{(I)}$ are order $N_I$. $A_{\bm{r}}^{(I)}$ arises as the image of~\eqref{weakSPTStab} under the gauging map. $B_{\bm{r}}^{(I)}$ arises from the cokernel of the gauging map and generates all local operators in the cokernel. The stabilizer condition ${B^{(I)}_{\bm{r}} = 1}$ is the flatness condition enforced when gauging a finite symmetry. 

The logical operators of this code are all topological loop operators. The logical operators built out of only $Z^{(I)}$ operators take the simple form
\begin{equation}
W^{(I)}(C) = \prod_{(\bm{r},\mu)\subset C} [Z_{(\bm{r},\mu)}^{(I)}]^{\si(C)},
\end{equation}
where ${\si(C)=\pm 1}$ captures the orientation for the cycle $C$ of the square lattice. They are topological because of the flatness condition ${B^{(I)}_{\bm{r}} = 1}$. The logical operators involving $X$ operators are more complicated. For a cycle $C_x$ running in the $x$-direction at fixed $y$, there are the logical operators
\begin{align}
V^{(I)}(C_x) &= \prod_{(\bm{r},\mu)\subset C_x}
\left(
X_{\bm{r},y}^{(I)} \,
\prod_{J=1}^{I-1} [Z_{\bm{r},x}^{(J)}]^{r_x \frac{N_J}{N_{IJ}} K_{IJ}}\, \prod_{M=I}^{n} [Z_{\bm{r}+\hat{y},x}^{(M)}]^{r_x \frac{N_M}{N_{IM}} K_{IM}}
\right).
\end{align}
For a contractible cycle ${C = \pp D}$, the logical operator $V^{(I)}$ can be written as a product of stabilizers ${[A^{(I)}_{\bm{r}}]^{\dag} \prod_{J=1}^{I-1} [B_{\bm{r}-\hat{y}}^{(J)\frac{N_J}{N_{IJ}}K_{IJ}r_x}]
\prod_{M=I}^{n} [B_{\bm{r}}^{(M)\frac{N_M}{N_{IM}}K_{IM}r_x}]
}$ for sites $\bm{r}$ in $D$. Therefore, since this equals $1$ in the code space, $V^{(I)}$ for a general cycle is a topological operator. These logical operators imply that the quantum code has ${\prod_{J=1}^n\Z_{N_{J}}}$ topological order. Indeed, each $W^{(I)}$ and $V^{(I)}$ are order $N_I$ operators, and for $V^{(I)}$ running in the $x$ ($y$) direction and $W^{(J)}$ in the $y$ ($x$) direction, the operators fail to commute by the phase ${\ee^{\frac{2\pi\ii}{N_I} \del_{IJ}}}$.

The quantum code's topological order has a non-trivial interplay with lattice translations. In particular, a lattice translation $T_x$ by one site in the $x$-direction transforms the logical operators in the code space by
\begin{equation}
T_x\colon V^{(I)} \to V^{(I)}\, \prod_{J=1}^n[W^{(J)}]^{-\frac{N_J}{N_{IJ}} K_{IJ}}.
\end{equation}
Therefore, the quantum code has non-trivial symmetry-enriched topological order, with the $T_x$ lattice symmetry inducing a non-trivial anyon automorphism. This interplay between crystalline symmetry and topological order originates from the quantum code being constructed by gauging a \textit{weak} SPT.

The symmetry-enriched topological order can alternatively be characterized in terms of position-dependent anyons~\cite{PW220407111}. A similar analysis was carried out in \Rf{EHN240110677} for the special case of ${n =2}$, ${N_1 = N_2}$, and ${K_{12} = N_1-1}$. We denote a gapped excitation (i.e., an error) corresponding to ${A^{(I)}_{\bm{r}} = \ee^{\frac{2\pi\ii}{N_I}n}}$ and ${B^{(I)}_{\bm{r}} = \ee^{\frac{2\pi\ii}{N_I}n}}$ by $n\mathfrak{e}_I(\bm{r})$ and $n\mathfrak{m}_I(\bm{r})$, respectively. They satisfy the $\Z_{N_I}$ relations ${N_I \mathfrak{e}_I(\bm{r}) = N_I \mathfrak{m}_I(\bm{r}) = 0}$. These excitations are created by the Pauli operators $X^{(I)}_{\bm{r},\mu}$ and $Z^{(I)}_{\bm{r},\mu}$, which gives rise to the additional relations
\begin{equation}
\begin{aligned}
&\mathfrak{e}_I(\bm{r}) - \mathfrak{e}_I(\bm{r}+\hat{y}) = 0,
\qquad
\mathfrak{e}_I(\bm{r}) - \mathfrak{e}_I(\bm{r}+\hat{x}) = 0,
\qquad
\mathfrak{m}_I(\bm{r}) - \mathfrak{m}_I(\bm{r}+\hat{y}) = 0,
\\
&\mathfrak{m}_I(\bm{r}) - \mathfrak{m}_I(\bm{r}-\hat{x}) - \sum_{J=I}^n \frac{N_J}{N_{IJ}} K_{IJ}\mathfrak{e}_J(\bm{r}) - \sum_{J=1}^{I-1} \frac{N_J}{N_{IJ}} K_{IJ}\mathfrak{e}_J(\bm{r}+\hat{y})  = 0.
\end{aligned}
\end{equation}
Solving these recurrence relations, we find
\begin{equation}
\mathfrak{e}_I(\bm{r}) = \mathfrak{e}_I,\hspace{40pt}
\mathfrak{m}_I(\bm{r}) = \mathfrak{m}_I + r_x \sum_{J=1}^n \frac{N_J}{N_{IJ}} K_{IJ} \mathfrak{e}_J.
\end{equation}
Therefore, the excitations depend on the lattice position $\bm{r}$, implying that the topological order is non-trivially enriched by translations. 
Furthermore, with periodic boundary conditions in the $x$-direction, ${\mathfrak{m}_I(\bm{r}) = \mathfrak{m}_I(\bm{r} + L_x \hat{x})}$ gives rise to the constraint that ${\sum_{J=1}^n\frac{N_J}{N_{IJ}} K_{IJ} L_x \mathfrak{e}_J = 0}$ for all ${I}$. Therefore, the number of globally distinguishable anyons and ground degeneracy on a spatial torus depends on $L_x$, indicating that the SymTFT is not topological in the $x$ direction.

To verify that this stabilizer code and its translation symmetry-enriched topological order is the SymTFT for the translation LSM anomaly, we introduce a spatial boundary in the $y$-direction. In particular, we consider the quantum code's rough boundary, which is characterized by the boundary stabilizers 
\begin{equation} 
B^{(I); \text{rough}}_{\bm{r}}=\begin{tikzpicture}[scale = 0.5, baseline={([yshift=-.5ex]current bounding box.center)}]
\draw[line width=0.015in, gray] (-3,0) -- (0, 0);
\draw[line width=0.015in, gray] (-3,0) -- (-3, 3);
\draw[line width=0.015in, gray] (0,0) -- (0, 3);
\node at (-1.5, 0.5) {\normalsize $\textcolor[HTML]{2171b5}{Z^{(I)}}$};
\node at (-2.7, 1.5) {\normalsize $\textcolor[HTML]{2171b5}{Z^{(I)\dag}}$};
\node at (0.15, 1.5) {\normalsize $\textcolor[HTML]{2171b5}{Z^{(I)}}$};
\fill[black] (-3,0) circle (6pt);
\node at (-3.5, -0.5) {\normalsize $\bm{r}$};
\end{tikzpicture}.
\end{equation}
On this boundary, the $W^{(I)}$ operators become $1$ while the $V^{(I)}$ operators become
\begin{equation}
V^{(I)}_\text{rough} = \prod_{r_x = 1}^{L_x} \left(X^{(I)}_{(r_x,L_y-1),y} \prod_{J=1}^{I-1} Z_{(r_x,L_y-1),y}^{(J)\, \frac{N_J}{N_{IJ}} K_{IJ}}\right).
\end{equation}
The operators $V^{(I)}_\text{rough}$ are order $N_I$ and form a uniform $\prod_{J=1}^n \Z_{N_I}$ symmetry. Despite $V^{(I)}$ corresponding to modulated operators in the bulk, $T_x$ acts trivially on $V^{(I)}_\text{rough}$ since ${W^{(I)}_\text{rough} = 1}$. Furthermore, $V^{(I)}$ on the boundary are onsite and furnish a local projective representation that causes
\begin{equation}
V^{(I)}_\text{rough}\, V^{(J)}_\text{rough} = \exp[2\pi\ii \frac{K_{IJ}}{N_{IJ}} L_x]\, V^{(J)}_\text{rough} V^{(I)}_\text{rough}.
\end{equation}
Therefore, the rough boundary encodes the $\prod_{J=1}^n \Z_{N_I}$ symmetry with an LSM anomaly involving lattice translations.

\subsubsection*{Field theory perspective}

From a Euclidean field theory perspective, the SymTFT is constructed by gauging the ${G \cong \prod_{I=1}^n \Z_{N_I}}$ symmetry of an invertible field theory. This invertible field theory describes a weak ${2+1}$D $G$ SPT, which is the inflow theory for the LSM anomaly. In what follows, we consider three-dimensional Euclidean spacetime to be a cubic lattice. The inflow theory for the LSM anomaly can then be described by the lattice partition function
\begin{equation}
\cZ[A^{(I)}] = \exp[2\pi\ii \sum_{I<J} \frac{K_{IJ}}{N_{IJ}}
\sum_{\bm{r}}(
A^{(I)}_{\bm{r},t}   A^{(J)}_{\bm{r}+\hat{t},y}
-
A^{(I)}_{\bm{r},y}  A^{(J)}_{\bm{r}+\hat{y},t} 
)],
\end{equation}
where ${A^{(I)}_{\bm{r},\mu}}$ is a background $\Z_{N_I}$ lattice gauge field associated to the edge ${\<\bm{r} , \bm{r} +\hat{\mu}\>}$. It satisfies the flatness condition ${\dd A^{(I)}_{\bm{r},\mu\nu} \equiv \Del_{\mu} A^{(I)}_{\bm{r},\nu} -  \Del_{\nu} A^{(I)}_{\bm{r},\mu} = 0}$. 

The matrix $K$ specifies the cohomology class of ${H^2(BG,U(1))}$ characterizing the weak SPT (i.e., $K$ determines ${[\om]\in H^2(BG,\R/\Z)}$ such that $\cZ = \ee^{2\pi\ii\int_{M_2} A^* \om}$ for ${[A]\in H^1(M_2,G)}$). Indeed, consider the backgrounds ${A^{(I)}_{\bm{r},\mu} = \del_{r_t,0}\del_{\mu,t}}$ and ${A^{(J)}_{\bm{r},\mu} = \del_{r_y,0}\del_{\mu,y}}$ for fixed $I$ and $J$, with all other ${A^{(\bullet)}_{\bm{r},\mu} = 0}$. This choice of backgrounds correspond to $\Z_{N_I}$ and $\Z_{N_J}$ symmetry defects inserted along $(x,y)$ and $(t,x)$ planes of spacetime.\footnote{Recall that finite Abelian $G$ gauge fields are in one-to-one correspondence to $G$ symmetry defects by Poincar{\'e} duality ${H^{1}(M_D,G)\cong H_{D-1}(M_D,G)}$.} For these backgrounds, the partition function ${\cZ[A^{(I)}]  = \ee^{2\pi\ii \frac{K_{IJ}}{N_{IJ}} L_x}}$. Each intersection point of these $\Z_{N_I}$ and $\Z_{N_J}$ symmetry defects contributes the phase ${\ee^{2\pi\ii \frac{K_{IJ}}{N_{IJ}}}}$ to $\cZ[A^{(I)}]$. Therefore, for this invertible theory, a $\Z_{N_J}$ domain wall in a $(t,x)$ plane carries ${\frac{N_I}{N_{IJ}} K_{IJ} L_x}$ units of $\Z_{N_I}$ symmetry charge.

To construct the SymTFT, we now make the background gauge fields dynamical (i.e., gauge the $G$ symmetry). Denoting the dynamical gauge fields by lowercase $a^{(I)}$, the lattice Lagrangian for the SymTFT is 
\begin{align}\label{LSMSymTFTLag}
\mathscr{L}_{\bm{r}} =& 2\pi\ii\left(\sum_{I=1}^n\frac{1}{N_I}
[b^{(I)}\cup \dd a^{(I)}]_{\bm{r}} +  \sum_{I<J} \frac{K_{IJ}}{N_{IJ}}(
a^{(I)}_{\bm{r},t}   a^{(J)}_{\bm{r}+\hat{t},y}
-
a^{(I)}_{\bm{r},y}  a^{(J)}_{\bm{r}+\hat{y},t} 
)\right).
\end{align}
We refer the reader to the first line of~\eqref{expSymLatLag} for an explicit expression of ${b^{(I)}\cup \dd a^{(I)}}$. The newly introduced lattice fields $b^{(I)}$ are Lagrange multipliers enforcing the flatness condition on $a^{(I)}$. 

This description of the SymTFT has the gauge redundancy
\begin{equation}
a^{(I)}_{\bm{r},\mu} \sim a^{(I)}_{\bm{r},\mu} + \Del_\mu \al^{(I)}_{\bm{r}},
\hspace{30pt}
b^{(I)}_{\bm{r},\mu} \sim b^{(I)}_{\bm{r},\mu} + \Del_\mu \bt^{(I)}_{\bm{r}}  + \del_{\mu,x}\,N_I \bigg(\sum_{J<I}\frac{K_{IJ}}{N_{IJ}}\al^{(J)}_{\bm{r}-\hat{t}} + \sum_{J>I}\frac{K_{IJ}}{N_{IJ}}\al^{(J)}_{\bm{r}+\hat{y}}\bigg),
\end{equation}
and its equations of motion are
\begin{equation}\label{modSymeomLlsm}
\dd a^{(I)}_{\bm{r},\mu\nu} = 0,
\hspace{30pt}
\dd b^{(I)}_{\bm{r},\mu \nu} =  \bigg(\sum_{J<I}\frac{N_I }{N_{IJ}} K_{IJ} \, a^{(J)}_{\bm{r}-\hat{t},\mu} + \sum_{J>I}\frac{N_I }{N_{IJ}} K_{IJ} \, a^{(J)}_{\bm{r}+\hat{y},\mu}\bigg)\del_{\nu, x} - \{\mu\leftrightarrow \nu\}.
\end{equation}
When ${K_{IJ} = 0}$, the SymTFT is just  a ${\prod_{J=1}^n \Z_{N_J}}$ gauge theory. Otherwise, for non-zero $K$, the holonomies of $b$ are modified due to the twist term in~\eqref{LSMSymTFTLag}. Indeed, denoting by $C$ a cycle of the Euclidean spacetime lattice, the gauge-invariant defect lines are formed by the electric defect lines
\begin{equation}
W^{(I)}(C) = \exp[\ii \sum_{(\bm{r},\mu)\subset C} a^{(I)}_{\bm{r},\mu}  ],
\end{equation}
and magnetic defect lines
\begin{equation}
V^{(I)}(C) = \exp[\ii \sum_{(\bm{r},\mu)\subset C} b^{(I)}_{\bm{r},\mu} + f^{(I)}_{\bm{r},\mu} \left(\sum_{J<I}\frac{N_I }{N_{IJ}}K_{IJ}\,a^{(J)}_{\bm{r}-\hat{t},\mu} + \sum_{J>I}\frac{N_I }{N_{IJ}}K_{IJ}\,a^{(J)}_{\bm{r}+\hat{y},\mu}\right) ]
,
\end{equation}
where ${f^{(I)}_{\bm{r},x} = r_x}$ and ${f^{(I)}_{\bm{r},t} = f^{(I)}_{\bm{r},y} =  r_x -1}$. 

The defect lines ${W^{(I)}}$ and ${V^{(I)}}$ both satisfy $\Z_{N_I}$ fusion rules, and from the equations of motion~\eqref{modSymeomLlsm}, they are topological defect lines. Furthermore, $W^{(I)}(C)$ and $V^{(J)}(C)$ are bosonic lines and have ${\exp[\frac{2\pi\ii}{N_I}\del_{IJ}]}$ mutual braiding. Therefore, the SymTFT has ${\prod_{J=1}^n \Z_{N_J}}$ topological order. While these are topological defects, the lattice translation ${T_x}$ in the $x$ direction transforms them non-trivially whenever there is a ${K_{IJ}\neq 0}$. In particular, they transform as 
\begin{equation}\label{LSMcontTact}
T_x\colon V^{(I)} \to V^{(I)}\, \prod_{J=1}^n[W^{(J)}]^{-\frac{N_J}{N_{IJ}} K_{IJ}}
\end{equation}
Therefore, these are modulated 1-form symmetries of the SymTFT. This makes the SymTFT a ${\prod_{J=1}^n \Z_{N_J}}$ topological order non-trivially enriched by translations in the $x$-direction.

To verify that this is the SymTFT for the finite Abelian $G$ symmetry with LSM anomaly, we consider the Dirichlet boundary condition ${a_{\bm{s},x} = a_{\bm{s},t} = 0}$ where ${\bm{s} = (r_x,L_y,r_t)}$. This boundary trivializes the $W^{(I)}$ topological defects. The $V^{(I)}$ topological defects, however, are non-trivial. Indeed, consider the cycle $C_{\text{bdy}}$ in the $(t,x)$ plane formed by lattice vectors ${\bm{s}-\hat{y}}$. Using the Dirichlet boundary conditions and the equations of motion ${a^{(I)}_{\bm{s}-\hat{y},t} =   -  \Del_t a^{(I)}_{\bm{s}-\hat{y},y}}$ and ${a^{(I)}_{\bm{s}-\hat{y},x}  = -\Del_x a^{(I)}_{\bm{s}-\hat{y},y}}$, the boundary magnetic defects simplify to
\begin{equation}
V^{(I)}(C_{\text{bdy}}) = \exp[\ii \sum_{(\bm{r},\mu)\subset C_{\text{bdy}}} b^{(I)}_{\bm{r},\mu} - f^{(I)}_{\bm{r},\mu} \left(\sum_{J<I}\frac{N_I }{N_{IJ}}K_{IJ}\,\Del_{\mu} a^{(I)}_{\bm{r}-\hat{t},y}\right) ].
\end{equation}
When $C_{\text{bdy}}$ runs along a fixed $r_t$, this simplifies to
${V^{(I)} = \exp[\ii \sum_{r_x} b^{(I)}_{\bm{r},x} + \sum_{J<I}\frac{N_I }{N_{IJ}}K_{IJ}\,a^{(I)}_{\bm{r}-\hat{t},y} ]}$. Despite $V^{(I)}$ corresponding to modulated topological defect lines in the bulk, $T_x$ acts on them trivially on the boundary since ${W^{(I)} = 1}$ on the boundary. In a Hamiltonian formalism, where time is continuous, the local on-site operators of $V^{(I)}$ form a projective representation described by $K_{IJ}$ since $b^{(I)}_{x}$ and $a^{(I)}_y$ fail to commute. Therefore, $V^{(I)}(C_{\text{bdy}})$ are the symmetry defects for a finite Abelian symmetry with LSM anomaly involving translations.

The above discussion of the SymTFT used lattice field theory for clarity. Its continuum limit is 
\begin{equation}\label{LSMSymTFTcont}
S = \frac{\ii}{2\pi}\int \sum_{I}\, N_I \, a^{(I)}\wdg \dd b^{(I)}-\frac{\ii}{2\pi}\int\,\sum_{I<J} \text{lcm}(N_I,N_J)\, K_{IJ} \,a^{\,(I)} \wdg a^{\,(J)} \wdg e,
\end{equation}
where the 1-form ${e = \La \dd x}$ and $\La$ is a UV cutoff related to the lattice spacing. The background field $e_\mu$ is sometimes viewed as the background gauge field for lattice translation. Here, we find that this ``background gauge field for lattice translation'' is more precisely a differential form related to leaves of a foliation of spacetime. Indeed,~\eqref{LSMSymTFTcont} is not a TQFT. Because translations in the $x$-direction act non-trivially on the topological defect lines, the partition function can change as the size of space in the $x$-direction is changed. However, it is still topological in the $y$ and $t$ directions, making it a type of foliated field theory. Each leaf of the foliation is dressed by an invertible condensation surface defect of $G$ gauge theory that implements the anyon automorphism~\eqref{LSMcontTact}. These dressed leaves encode in the continuum SymTFT how the ${1+1}$D symmetry has an LSM anomaly with translations.

\subsubsection{LSM theorem from the SymTFT}

The SymTFT can be used to prove that the LSM anomaly of ${G \cong \prod_{I=1}^n \Z_{N_I}}$ is an obstruction to translation-invariant SPT states (i.e., the LSM theorem). We denote by $e_I$ and $m_I$ the anyons corresponding to the logical operators/topological defects $W^{(I)}$ and $V^{(J)}$, respectively. Because $W^{(I) = 1}$ on the LSM anomaly symmetry boundary, the Lagrangian condensable algebra corresponding to the symmetry boundary is ${\cA_{\txt{LSM}} = \<e_1,e_2,\cdots, e_n\>}$, generated by the $e_I$ anyons via their fusion. In this notation, the symmetry defects are $m_I$, which form all of the anyons not in $\cA_{\txt{LSM}}$. They are uniform symmetries since the translation action on $m_I$ is trivial when the $e_J$ anyons are condensed.

$G$-SPT phases correspond to Lagrangian algebras that have a trivial overlap with $\cA_{\txt{LSM}}$. These condensable algebras $\cA_{C}$ are characterized by an anti-symmetric matrix $C$ and are generated by all fusion combination of $m_I \bigotimes_J e_J^{\, N_J\, C_{IJ}/N_{IJ}}$ anyons:
\begin{equation}\label{condAlgSPTGen}
\cA_C = \< m_I \bigotimes_J e_J^{\, N_J\, C_{IJ}/N_{IJ}} \mid I = 1,2,\cdots, n\>.
\end{equation}
However, for these to correspond to translation-invariant SPTs, they must be invariant under the anyon automorphism
\begin{equation}\label{anyonAutoLSMTAct}
m_I \to m_I\, e_1^{-\frac{N_1}{N_{I1}}K_{I1}} e_2^{-\frac{N_2}{N_{I2}}K_{I2}} \cdots\, e_n^{-\frac{N_n}{N_{In}}K_{In}} \qquad e_I \to e_I,
\end{equation}
induced in the SymTFT by a single lattice translation in the $x$-direction. Under this transformation, the condensable algebra ${\cA_{C}\to \cA_{C-K}}$. Therefore, lattice translations change the decorated domain wall pattern and all $G$-SPTs can only be realized by explicitly breaking lattice translation symmetry. This is precisely the LSM anomaly: no translation-invariant $G$ SPTs exist.

Another signature of the LSM anomaly is that a translation invariant SPT state can be realized if $G$ is explicitly broken to a particular sub-symmetry $G_{\text{sub}}$. This can also be seen using the SymTFT. Indeed, explicitly breaking $G$ to $G_{\text{sub}}$ modifies the SymTFT by forgetting a subset of the $G$ topological order anyons. If the translation action becomes trivial on this subset of anyons, then there exists translation-invariant $G_{\text{sub}}$ SPT states. For example, when ${n=3}$ and ${K_{13} = 0}$, the translation action on anyons can be trivialized by forgetting the $e_2$ and $m_2$ anyons. Therefore, by explicitly breaking $\Z_{N_1}\times \Z_{N_2}\times \Z_{N_3}$ with this LSM anomaly to $\Z_{N_1}\times \Z_{N_3}$, the LSM anomaly is trivialized and there exist translation-invariant $\Z_{N_1}\times \Z_{N_3}$-SPTs.

\subsubsection{Dual symmetries from gauging}\label{LSMtoModsymbyGauging}

The SymTFT can be used to explore gauging internal symmetries with LSM anomalies. For example, gauging the ${G \cong \prod_{I=1}^n \Z_{N_I}}$ symmetry without discrete torsion is implemented by changing the symmetry boundary's condensable algebra ${\cA_{\txt{LSM}} = \<1,e_1,e_2,\cdots, e_n\>}$ to ${\cA_{m} = \<1,m_1,m_2,\cdots, m_n\>}$. The Lagrangian condensable algebra $\cA_{m}$ describes a symmetry boundary for a uniform $G$ symmetry whose symmetry defects correspond to the anyons $e_I$. However, $\cA_{m}$ is not invariant under the anyon automorphism~\eqref{anyonAutoLSMTAct}. Therefore, due to the LSM anomaly, gauging $G$ causes the ordinary lattice translations to explicitly break. However, there is a non-invertible lattice translation operator $\mathsf{D}_T$ that arises from the gauging. It is implemented a first acting $\mathsf{D}_{m\to\text{LSM}}$ that performs the Kramers-Wannier transformation switching ${\cA_{m}\to \cA_{\txt{LSM}}}$, then acting by lattice translations $T$, and then acting $\mathsf{D}_{\text{LSM}\to m}$ to apply another Kramers-Wannier like transformation to switch back ${\cA_{\txt{LSM}}\to\cA_{m}}$. In the SymTFT without boundary, $\mathsf{D}_{m\to\text{LSM}}$ and ${\cA_{\txt{LSM}}\to\cA_{m}}$ are invertible anyon automorphisms, but in the $1+1$D theory they are non-invertible. The non-invertible operator ${\mathsf{D}_T = \mathsf{D}_{\text{LSM}\to m} \, T \, \mathsf{D}_{m \to \text{LSM}}}$ generalizes the non-invertible translations discussed in~\cite{S230805151, PLA240918113}.

One can also use the SymTFT to gauge sub-symmetries of $G$. For example, we gauge a $\Z_{N_I}$ sub-symmetry of $G$ by replacing each $e_I$ in $\cA_{I}$ by $m_I$. The resulting Lagrangian condensable algebra ${\cA_{I} = \<e_1,\cdots, e_{I-1}, m_I, e_{I+1},\cdots  e_n\>}$ is invariant under the translation action~\eqref{anyonAutoLSMTAct}. Because $\cA_{I}$ does not include $e_I$, it describes a modulated symmetry. Indeed, its symmetry defects are formed by $m_{J}$ for ${J\neq I}$ and $e_I$. Using that the anyons in $\cA_I$ are condensed, these symmetry defects transform as ${m_{J}\to m_J e_I^{N_I K_{IJ}/N_{IJ}}}$, $e_I\rightarrow e_I$ under translations. Therefore, gauging $\Z_{N_I}$ causes the uniform $G$ symmetry with LSM anomaly to become a modulated $G$ symmetry. This line of thinking can be used to show that most Lagrangian condensable algebras of the SymTFT correspond to symmetry boundaries for modulated symmetries. This generalizes the relation between LSM anomalies and dipole symmetries from~\cite{AMF230800743, S230805151, PLA240918113}.

The modulated symmetry described by the  $\cA_{I}$ boundary is sometimes anomaly-free, admitting SPT phases. Possible SPT phases correspond to Lagrangian condensable algebras that have a trivial overlap with $\cA_{I}$. Such condensable algebras $\cA_C$ are labeled by an antisymmetric ${n\times n}$ matrix $C$ and are of the form~\eqref{condAlgSPTGen} but with ${e_{I} \leftrightarrow m_I}$. This condensable algebra is invariant under the lattice translation's action when
\begin{equation}\label{SPTcond}
\frac{N_{I}}{N_{Ia}N_{Ib}}(C_{Ia}K_{Ib} - K_{Ia}C_{Ib}) - \frac1{N_{ab}}K_{ab}\in \Z\quad \forall \quad a,b\neq I.
\end{equation}
For example, this is always satisfied when ${n=2}$. Therefore, every ${\Z_{N_1}\times \Z_{N_2}}$ symmetry with a translation LSM anomaly is dual to an anomaly-free modulated symmetry by gauging $\Z_{N_1}$ or $\Z_{N_2}$. However, such dual symmetries are not always anomaly-free. For example, when ${n=3}$, if there is an $a$ and $b$ with ${K_{ab} \neq 0}$ such that ${N_{Ia} = N_{Ib} = 1}$ while ${N_{ab}\neq 1}$, then~\eqref{SPTcond} cannot be satisfied. Therefore, this example has no translation-invariant $\cA_C$, and its symmetries have no translation-invariant SPT states.

\section{SymTFT enriched by reflection}
\label{sec reflection SymTFT}

We now turn our attention to reflection symmetry, focusing on systems where it interacts non-trivially with internal symmetries. This interaction may appear as an LSM-like anomaly or a non-trivial group extension of reflections by internal symmetries. Interestingly, as we show, these two cases are connected through discrete gauging and, thus, share a common SymTFT enriched by reflection symmetry. This differs from the translation-enriched SymTFTs in Section~\ref{translationSec} as the reflection symmetry will display non-trivial symmetry fractionalization~\cite{ZLV150101395, QF150506201, SHF160408151, BBJ161207792, QJW171009391}, which will play an essential role in the following discussion.

\subsection{Example: LSM anomalies and extensions with reflections}\label{LSMreflection1dSec}

Before constructing the reflection-enriched SymTFT, it is instructive to consider a ${1+1}$D models that realize the symmetry interplays we will consider. 

\subsubsection*{Lattice model perspective}

We will start with a lattice model realizing an LSM anomaly between reflection and internal symmetries. Consider a $1+1$D lattice of $L$ sites with a single qubit on each site $j$ acted on by the Pauli operators $Z_j$ and $X_j$. Suppose the system enjoys a $\Z_2$ internal symmetry generated by
\ie\label{eq:internal_Z2}
U_X=\prod_{j=1}^L X_j~.
\fe
Further, we assume there is a $\Z_2^R$ reflection symmetry generated by
\ie\label{eq:reflection_symmetry_operator}
U_R=R\,\prod_{j=1}^L Z_j\,,
\fe
where $R$ is the site-centered reflection operator satisfying ${RX_j R^{\dag} = X_{-j}}$ and ${RZ_j R^{\dag} = Z_{-j}}$. An example of a Hamiltonian with this ${\Z_2\times \Z_2^R}$ symmetry is
\ie\label{eq:reflection_LSM_Hamiltonian}
H=\sum_{j=1}^L(Z_j Y_{j+1}-Y_j Z_{j+1}),
\fe
where ${Y_j = \ii X_j Z_j}$. While $U_R$ commutes with this Hamiltonian, notice that $R$ does not. This Hamiltonian has many more symmetries other than the internal $\Z_2$ symmetry and reflection symmetry, which we will not consider here.

The ${\Z_2\times\Z_2^R}$ symmetry generated by~\eqref{eq:internal_Z2} and~\eqref{eq:reflection_symmetry_operator} has an LSM anomaly, i.e., there is no nondegenerate and gapped ground state with this ${\Z_2\times\Z_2^R}$ symmetry. There are many ways to see this. One way to argue for the LSM is to use that the $U_R$ operator acts like an internal symmetry at the reflection centers, e.g., at ${j=L}$. Then, at these reflection-center sites, the symmetry operators $U_X$ and $U_R$ realize a non-trivial, local projective representation of ${\Z_2\times\Z^R_2}$. Since this projective representation cannot be trivialized by any regrouping or redefinition of the lattice degrees of freedom, by the lattice homotopy conjecture introduced in \Rf{Po2017}, there is an LSM anomaly for ${\Z_2\times\Z^R_2}$. Another way to argue for the LSM anomaly is through its relation to a well-known LSM anomaly between the ${\Z_2\times\Z_2}$ internal symmetry generated by $\prod_{j=1}^L X_j$ and $\prod_{j=1}^L Z_j$ and the reflection symmetry generated $R$~\cite{Po2017, AMF230800743}.\footnote{This a type-III-like LSM anomaly. It manifests through, for example, the ${\prod_{j=1}^L X_j}$ and $R$ symmetry operators furnishing a projective representation of ${\Z_2\times\Z_2}$ in the presence of a ${\prod_{j=1}^L Z_j}$ symmetry defect. Indeed, inserting the ${\prod_{j=1}^L Z_j}$ symmetry defect at the link ${\<n,n+1\>}$ causes the reflection operator $R$ to become the twisted operator ${R_\mathrm{tw} = \left(\prod_{j=-n}^n Z_j\right)\, R}$. This satisfies the projective algebra ${R_\mathrm{tw}\left(\prod_{j=1}^L X_j\right) = -\left(\prod_{j=1}^L X_j\right)\, R_\mathrm{tw}}$, which defines the projective ${\Z_2\times\Z_2}$ representation. This projective representation similarly arises for the ${\prod_{j=1}^L Z_j}$ and $R$ symmetry operators after inserting a ${\prod_{j=1}^L X_j}$ symmetry defect.} Since the $U_R$ operator is nothing but the product of $\prod_{j=1}^L Z_j$ and $R$, there is a type-II-like LSM anomaly for ${\Z_2\times \Z^R_2}$ symmetry induced from the type-III-like LSM anomaly of the ${\Z_2\times \Z_2\times \Z_2}$ symmetry generated by ${\prod_{j=1}^L X_j}$, ${\prod_{j=1}^L Z_j}$, and $R$.

As we now demonstrate, a consequence of the LSM anomaly between $\Z_2$ and $\Z_2^R$ is that the dual symmetry arising from gauging $\Z_2$ is part of a $\Z_4$ reflection symmetry.\footnote{Similar group extensions arising from gauging internal symmetries with LSM anomalies exist for lattice translations, as discussed in Section~\ref{translationSec} and Refs.~\citeonline{AMF230800743, S230805151, PDL240612962, PLA240918113}.} To gauge the $\Z_2$ symmetry, we introduce a qubit onto each link ${\langle j,j+1\rangle}$ of the lattice and denoted by $X_{j,x}$ and $Z_{j,x}$ its Pauli operators. The site and link qubit operators satisfy the Gauss law
\ie\label{eq:gauss_law_reflection}
G_j= X_{j-1,x} X_j X_{j,x}=1~.
\fe
Because ${U_X = \prod_{j=1}^L G_j}$, the Gauss law trivializes the original $\Z_2$ symmetry. Gauging also leads to a new, dual $\Z_2$ symmetry generated by
\ie\label{Z4Rop1d}
U^\vee_X=\prod_{j=1}^L Z_{j,x}~.
\fe
Indeed, the Hamiltonian~\eqref{eq:reflection_LSM_Hamiltonian} after minimal coupling becomes ${\sum_j(Z_j Z_{j,x} Y_{j+1}-Y_j Z_{j,x} Z_{j+1})}$, which commutes with $U_X^\vee$. The original $\Z_2^R$ reflection symmetry operator~\eqref{eq:reflection_symmetry_operator} is not gauge invariant but can be made gauge invariant upon minimal coupling. Assuming that the number of sites $L$ is even, the gauge-invariant version of $U_R$ is
\ie\label{eq:reflection_symmetry_operator_after}
U_R^\vee
 = R\,\prod_{j=1}^L Z_j \, \prod_{n=1}^{L/2}  Z_{2n,x}
 =R\,\prod_{j=1}^L Z_j ({Z}_{j,x})^{\,j+1}~.
\fe
The $R$ operator acts on the site qubit operators just as before and now acts on the link qubit operators as ${R\, {X}_{j,x}R^{\dag} = {X}_{-j-1,x}}$ and ${R\, {Z}_{j,x}R^{\dag} = {Z}_{-j-1,x}}$. Although the reflection symmetry generated by $U_R$ is a $\Z_2^R$ symmetry, the reflection symmetry operator $U_R^\vee$ after gauging is a $\Z_4$ operator, satisfying
\ie
(U_R^\vee)^2=U_X^\vee~.
\fe
This extension of the $\Z_2^R$ symmetry by the dual $\Z_2$ symmetry is a consequence of the LSM anomaly.

\subsubsection*{Field theory perspective}

The LSM anomaly between the $\Z_2$ internal and $\Z_2^R$ reflection symmetries is characterized by an anomaly inflow theory, which is an invertible TFT in one higher dimension. We denote by $X_3$ the ${2+1}$D spacetime manifold for the anomaly inflow theory. The action of the inflow theory is constructed from the background gauge fields ${A\in H^1(X_3,\Z_2)}$ of internal $\Z_2$ symmetry and the first Stiefel-Whitney class $w_1\in H^1(X_3,\Z_2)$ of $X_3$ (i.e., the connection of the orientation bundle on $X_3$). The $\Z_2^R$ reflection symmetry couples to the first Stiefel-Whitney class $w_1$. The ${\Z_2\times\Z_2^R}$ LSM anomaly is then characterized by the SPT action
\ie
\ii\pi \int_{X_3} A\cup \mathrm{Bock}(w_1)\, ,
\fe
where $\cup$ is the cup product and ${\mathrm{Bock}\colon H^1(X_3,\Z_2)\to H^2(X_3,\Z_2)}$ is the Bockstein homomorphism. Introducing the $\Z$-lift $\hat{w}_1$ of $w_1$, the action can also be written as
\ie\label{eq:SPT_reflection}
\frac{\ii\pi}{2} \int_{X_3} A\cup \del \hat{w}_1.
\fe

Following the reasoning from \Rf{T171209542}, the symmetry extension of $\Z_2^R$ to $\Z_4^R$ after gauging $\Z_2$ can be understood using~\eqref{eq:SPT_reflection}. Gauging the internal $\Z_2$ symmetry promotes its background gauge field $A$ to a dynamical gauge field, which we denote by $a$. The SPT action~\eqref{eq:SPT_reflection} then depends on a dynamical field $a$ as ${\frac{\ii\pi}{2} \int_{X_3} a\cup \del \hat{w}_1}$. The dependence of $a$ in the bulk can be removed using the background gauge field $A^\vee$ of the dual $\Z_2$ symmetry. On the boundary of the SPT, $A^\vee$ couples to $a$ through ${\ii\pi \int_{\pp X_3} a\cup A^\vee}$, which is equal to ${\ii\pi\int_{X_3}  \del(a\cup  A^\vee)= \ii\pi\int_{X_3}  a\cup \del A^\vee}$ by Stokes' theorem. The dependence on $a$ in the bulk can then be canceled by requiring $A^\vee$ satisfies
\ie
\del A^\vee + \frac{1}{2}\del \hat{w}_1=0 \bmod 2~.
\fe
This constraint implies that ${\hat{w}_1+2A^\vee}$ is a $\Z_4$ cocycle and, therefore, the reflection symmetry is extended to a $\Z_4^R$ symmetry by the $\Z_2$ dual symmetry.

\subsection{Reflection enriched SymTFT}\label{reflectionEnrichedSymTFT}

We now construct the reflection-enriched SymTFT for the ${\Z_2\times\Z_2^R}$ symmetry with LSM anomaly and the $\Z_4^R$ reflection symmetry and discuss its applications. Since these two symmetries are related by discrete gauging, they share the same SymTFT. We will focus on the SymTFT as a quantum lattice model (i.e., stabilizer code). The field theory for this symmetry-enriched SymTFT is the non-orientable TFT discussed above constructed by promoting $A$ into a dynamical gauge field in~\eqref{eq:SPT_reflection}.

\subsubsection{Stabilizer code}\label{reflectionEnrichedSymTFTStabCode}

We construct the lattice model for the SymTFT by gauging the SPT for the $\Z_4^R$ reflection symmetry \eqref{eq:reflection_symmetry_operator_after}. Since the $\Z_4^R$ reflection symmetry is anomaly-free, this SPT is a trivial paramagnet. It is defined on a square lattice with a qubit at each site ${\bm{r} \equiv (r_x,r_y)}$ and horizontal link ${\<\bm{r},\bm{r}+\hat{x}\>}$ of the lattice. Their respective Pauli operators are denoted by ${(Z_{\bm{r}}, X_{\bm{r}})}$ and ${(Z_{\bm{r},x}, X_{\bm{r},x})}$ (we follow the same notation for sites and edges of the square lattice used in Section~\ref{translationSec}). Its Hamiltonian is
\begin{equation}
    H_{\text{para}} = -\sum_{\bm{r}} (Z_{\bm{r}} + Z_{\bm{r},x}),
\end{equation}
where $\sum_{\bm{r}}$ sums over all lattice sites. The $\Z_4^R$ reflection symmetry operator~\eqref{Z4Rop1d} when acting in two spatial dimensions is\footnote{Here we abuse notation by using the same notation for the symmetry operators in ${2+1}$D as we did in ${1+1}$D.} 
\ie\label{2dZ4ReflectionOp}
U_R^\vee=R\, \prod_{\bm{r}} Z_{\bm{r}} (Z_{\bm{r},x})^{r_x+1}\,,
\fe
where $R$ is the site-centered reflection that satisfies
\begin{equation}
\begin{aligned}
    R\, X_{(r_x,r_y)}R^\dag &= X_{(-r_x,r_y)},
\qquad 
\hspace{28pt}
R\, Z_{(r_x,r_y)} R^\dag = Z_{(-r_x,r_y)},
\\ 
R\, X_{(r_x,r_y),\,x}R^\dag &= X_{(-r_x-1,r_y),\,x},
\qquad 
R\, Z_{(r_x,r_y),\,x}R^\dag = Z_{(-r_x-1,r_y),\,x}~.
\end{aligned}
\end{equation}
The reflection operator $U_R^\vee$ commutes with the paramagnet Hamiltonian $H_\text{para}$.

We now gauge the internal $\Z_2$ subgroup of the $\Z_4^R$ reflection symmetry, which is generated by
\ie
U_X^\vee=(U_R^\vee)^2=\prod_{\bm{r}} Z_{\bm{r},x}~.
\fe
Gauging maps the paramagnet with qubits on sites and horizontal links to a topologically ordered model with two qubits on each site and one qubit on each plaquette. We use the notation ${(X, Z)}$ for the plaquette-qubit operators and ${(\t X,\t Z)}$ for the new site-qubit operators. This gauging map does not transform the original site qubit operators (i.e., it transforms ${X_{\bm{r}} \to X_{\bm{r}}}$ and ${Z_{\bm{r}} \to Z_{\bm{r}}}$) and its effect on the horizontal link qubits is
\begin{equation}
    \begin{tikzpicture}[scale = 0.5, baseline = {([yshift=-.5ex]current bounding box.center)}]
\draw[line width=0.015in, gray] (-3, 0) -- (-3, 3) -- (0, 3) -- (0, 0) -- cycle;
\node at (-1.5, 0) {\normalsize $\textcolor[HTML]{cb181d}{X}$};
\node at (-1.5, 3) {\normalsize $\textcolor[HTML]{cb181d}{X}$};
\node at (1, 1.5) {\normalsize $\to$};
\draw[line width=0.015in, gray] (2, 0) -- (2, 3) -- (5, 3) -- (5, 0) -- cycle;
\node at (3.5, 1.5) {\normalsize $\textcolor[HTML]{cb181d}{Z}$};
\draw[line width=0.015in, gray] (-3-3, 0-2.5) -- (3-3, 0-2.5);
\draw[line width=0.015in, gray] (0-3, -1-2.5) -- (0-3, 1-2.5);
\node at (-1.5-3, 0-2.5) {\normalsize $\textcolor[HTML]{cb181d}{X}$};
\node at (1.5-3, 0-2.5) {\normalsize $\textcolor[HTML]{cb181d}{X}$};
\node at (4-3, 0-2.5) {\normalsize $\to$};
\draw[line width=0.015in, gray] (5-3, 0-2.5) -- (11-3, 0-2.5);
\draw[line width=0.015in, gray] (8-3, -1-2.5) -- (8-3, 1-2.5);
\node at (5.4, -2) {\normalsize $\textcolor[HTML]{cb181d}{\t Z}$};
\end{tikzpicture}
\qquad\quad
    Z_{\bm{r},x} \to A_{\bm{r},x} = \begin{tikzpicture}[scale = 0.5, baseline = {([yshift=-.5ex]current bounding box.center)}]
\draw[line width=0.015in, gray] (-3, 0) -- (-3, 3) -- (0, 3) -- (0, 0) -- cycle;
\draw[line width=0.015in, gray] (-3, 0) -- (-3, -3) -- (0, -3) -- (-0, 0) -- cycle;
\fill[black] (-3,0) circle (6pt);
\node at (-3.4, 0.5) {\normalsize $\textcolor[HTML]{cb181d}{\t X}$};
\node at (.4, .5) {\normalsize $\textcolor[HTML]{cb181d}{\t X}$};
\node at (-1.5, 1.5) {\normalsize $\textcolor[HTML]{cb181d}{X}$};
\node at (-1.5, -1.5) {\normalsize $\textcolor[HTML]{cb181d}{X}$};

\node at (-3.5, -0.5) {\normalsize $\bm{r}$};
\end{tikzpicture}.
\end{equation}
After gauging, the paramagnet Hamiltonian becomes
\begin{equation}\label{eq:hamitlonian_reflection_SymTFT}
    H = -\sum_{\bm{r}} (Z_{\bm{r}} + A_{\bm{r},x} + B_{\bm{r},y}),
\end{equation}
where the stabilizer $B_{\bm{r},y}$ energetically enforces the gauging's flatness condition and given by
\ie
B_{\bm{r},y}=\begin{tikzpicture}[scale = 0.5, baseline={([yshift=-.5ex]current bounding box.center)}]
\draw[line width=0.015in, gray] (-3, 0) -- (-3, 3) -- (0, 3) -- (0, 0) -- cycle;
\draw[line width=0.015in, gray] (-6, 0) -- (-6, 3) -- (-3, 3) -- (-3, 0) -- cycle;
\node at (-3, -0.75) {\normalsize $\textcolor[HTML]{2171b5}{\t Z}$};
\node at (-3, 3.65) {\normalsize $\textcolor[HTML]{2171b5}{\t Z}$};
\node at (-1.5, 1.65) {\normalsize $\textcolor[HTML]{2171b5}{Z}$};
\node at (-4.5, 1.65) {\normalsize $\textcolor[HTML]{2171b5}{Z}$};
\fill[black] (-3,0) circle (6pt);
\node at (-3.5, 0.5) {\normalsize $\bm{r}$};
\end{tikzpicture}~.
\fe
The last two terms in the Hamiltonian~\eqref{eq:hamitlonian_reflection_SymTFT} is the Hamiltonian of a toric code, which captures the topological order of the SymTFT. On the other hand, the first term causes one species of site qubits to form a product state at low-energies, decoupling them from the IR. The reflection symmetry operator~\eqref{2dZ4ReflectionOp} after gauging the internal $\Z_2$ symmetry becomes
\ie\label{eq:reflection_symTFT}
U_R^\vee=R\, \prod_{\bm{r}} Z_{\bm{r}}\, \t X_{\bm{r}},
\fe
where $R$ is the site-centered reflection acting in the expected way on all three species of qubits. This operator satisfies ${(U_R^\vee)^2 = 1}$ on an infinite square lattice. 

The Hamiltonian~\eqref{eq:hamitlonian_reflection_SymTFT} has $\Z_2$ topological order enriched by the reflection symmetry~\eqref{eq:reflection_symTFT}. It has four types of anyons: the trivial anyon $1$, the $e$ anyon, the $m$ anyon, and the ${f=e\times m}$ anyon. The $e$ (resp. $m$) anyon resides on horizontal (resp. vertical) edges of the lattice where ${A_{\bm{r},x}\neq 1}$ (resp. ${B_{\bm{r},y}\neq 1}$). The $e$ anyons carry fractional $U_R^\vee$ symmetry charge. This fractionalization pattern can be diagnosed by the eigenvalue of the reflection symmetry for a reflection-symmetric configuration of $e$ anyons~\cite{ZLV150101395, BBJ161207792, QJW171009391}. Indeed, a reflection-symmetric pair of $e$ anyons is created by the operator
\ie
W_e(r_y,l)=\prod_{r_x = -l}^{l}\t Z_{\bm{r}},
\fe
with ${l\in\Z}$. This string operator is charged under the reflection symmetry as ${U_R^\vee\, W_e\,(U_R^\vee)^\dag=-W_e}$. Since a pair of $e$ anyons carries integer reflection symmetry charge, each $e$ anyon individually carries fractional charge. On the other hand, the $m$ anyons do not carry fractional reflection symmetry charge. Indeed, a reflection-symmetric pair of $m$ anyons is created by the string operator ${\prod_{r_x=-l}^{l+1}X_{r_x-\frac{1}{2},r_y+\frac{1}{2}}}$, which commutes with the reflection symmetry: ${U_R^\vee \, W_m= W_m U_R^\vee}$.

In summary, the SymTFT has a $\Z_2$ topological order enriched by a reflection symmetry that fractionalizes on the $e$ anyons but not on the $m$ anyons.

\subsubsection{Gapped boundary conditions}\label{reflectionboundaryExSec}

Let us now discuss the gapped boundary conditions of this SymTFT. Since the reflection-enriched SymTFT has an underlying $\Z_2$ topological order, there are two classes of gapped boundaries: the $e$ condensing and the $m$ condensing boundary. 

Let us first consider the $m$ condensing boundary (i.e., the smooth boundary of the toric code). To do so, we consider a quiche with a top boundary along the row ${r_y=0}$ of the square lattice. The lattice sites of this square lattice quiche are ${(r_x,r_y)\in \Z\times \Z_{\leq 0}}$. The boundary condition is specified by the boundary stabilizers
\begin{equation}
    A^{(\text{smooth})}_{r_x,x} = \hspace{-10pt}\begin{tikzpicture}[scale = 0.5, baseline = {([yshift=-2ex]current bounding box.center)}]
\draw[line width=0.015in, gray] (-3, 0) -- (-3, -3) -- (0, -3) -- (-0, 0) -- cycle;
\fill[black] (-3,0) circle (6pt);
\node at (-3.4, 0.5) {\normalsize $\textcolor[HTML]{cb181d}{\t X}$};
\node at (.4, .5) {\normalsize $\textcolor[HTML]{cb181d}{\t X}$};
\node at (-1.5, -1.5) {\normalsize $\textcolor[HTML]{cb181d}{X}$};

\node at (-1.9, 0.6) {\normalsize $(r_x,0)$};
\end{tikzpicture}
\end{equation}
The $m$ anyon logical operator operator on the top boundary can be written as
\ie
W^{(\text{smooth})}_m=\prod_{r_x} X_{r_x+\frac12,-\frac{1}{2}}=\prod_{r_x} A^{(\text{smooth})}_{r_x,x}.
\fe
Therefore, ${W^{(\text{smooth})}_m = 1}$ in the ground state subspace, indicating that this boundary is, in fact, the $m$ condensing boundary. On the other hand, the $e$ anyon logical operator on the boundary $W^{(\text{smooth})}_e$ and reflection operator $U_R^{(\text{smooth})}$ in the presence of this boundary are, respectively,
\ie
W^{(\text{smooth})}_e=\prod_{r_x}\t Z_{(r_x,0)}\,
,
\qquad
U_R^{(\text{smooth})} = R \prod_{r_y\leq 0, r_x}Z_{\bm{r}}\, \t X_{\bm{r}}\,.
\fe
These operators commute with all stabilizers and generate ${\Z_2\times \Z_2^R}$ symmetries. Furthermore, by the same reasoning in Section~\ref{LSMreflection1dSec}, the internal $\Z_2$ symmetry generated by ${W^{(\text{smooth})}_e}$ and $\Z_2^R$ reflection symmetry generated by ${U_R^{(\text{smooth})}}$ have an LSM anomaly. In particular, $W^{(\text{smooth})}_e$ and $U_R^{(\text{smooth})}$ form a local projective representation at the reflection-center row ${r_x = 0}$. Therefore, the $m$ condensing boundary of the reflection-enriched SymTFT encodes the ${\Z_2\times \Z_2^R}$ symmetry with LSM anomaly.

On the other hand, the $e$ condensing boundary (i.e., the rough boundary of the toric code) truncates the lattice along the row of vertical edges connecting sites at ${r_y=0}$ and ${r_y=1}$. The lattice sites are still ${(r_x,r_y)\in \Z\times \Z_{\leq 0}}$, but now there are qubits on the boundary plaquettes centered at ${(r_x+\frac12,\frac12)}$. The boundary stabilizers are
\ie
B^{(\text{rough})}_{r_x,y}=\begin{tikzpicture}[scale = 0.5, baseline={([yshift=1.5ex]current bounding box.center)}]
\draw[line width=0.015in, gray] (-6, 0) -- (-6, 2);
\draw[line width=0.015in, gray] (-3, 0) -- (-3, 2);
\draw[line width=0.015in, gray] (0, 0) -- (0, 2);
\draw[line width=0.015in, gray] (-6, 0) -- (0, 0);
\node at (-2.5, 0.6) {\normalsize $\textcolor[HTML]{2171b5}{\t Z}$};
\node at (-1.5, 1.65) {\normalsize $\textcolor[HTML]{2171b5}{Z}$};
\node at (-4.5, 1.65) {\normalsize $\textcolor[HTML]{2171b5}{Z}$};
\fill[black] (-3,0) circle (6pt);
\node at (-4, -0.6) {\normalsize $(r_x,0)$};
\end{tikzpicture}~.
\fe
The $e$ anyon logical operator on this boundary can be written as
\begin{equation}
    W^{(\text{rough})}_e = \prod_{r_x} \t{Z}_{(r_x,0)} = \prod_{r_x}B^{(\text{rough})}_{r_x,y}.
\end{equation}
Therefore, ${W^{(\text{rough})}_e=1}$ in the ground state subspace, which confirms that this is the $e$ condensing boundary. The $m$ anyon logical operator on the boundary is unaffected by the boundary stabilizer. It is given by
\ie
W^{(\text{rough})}_m=\prod_{r_x} X_{r_x+\frac12,\frac{1}{2}}~,
\fe 
and generates a $\Z_2$ symmetry on the boundary. The reflection symmetry operator in the presence of the $e$ condensing boundary is different from that of the $m$ condensing boundary. Indeed, the reflection operator $U_R^{(\text{smooth})}$ is not a symmetry. However, it can be modified on the boundary to become a symmetry, resulting in the conserved reflection operator 
\begin{equation}
    U_R^{(\text{rough})} = R~ \bigg(\prod_{r_y\leq 0, r_x}Z_{\bm{r}}\, \t X_{\bm{r}}\bigg) \bigg(\prod_{r_x} (X_{r_x+\frac12,\frac12})^{r_x} \bigg) \equiv U_R^{(\text{smooth})}\, \prod_{r_x} (X_{r_x+\frac12,\frac12})^{r_x}.
\end{equation}
This is a $\Z_4^R$ reflection operator, extended by the $\Z_2$ symmetry $W_m$ on the boundary:
\ie
(U_R^{(\text{rough})})^2= W^{(\text{rough})}_m~.
\fe

Gauging the $\Z_2$ symmetry generated by the boundary logical operator is implemented by changing boundary conditions. For instance, switching from the $m$ condensing to $e$ condensing boundary corresponds to gauging the $\Z_2$ symmetry generated by $W_e^{(\mathrm{smooth})}$. Therefore, we see from the reflection-enriched SymTFT that a ${\Z_2\times \Z_2^R}$ symmetry with LSM anomaly is related to a $\Z_4^R$ reflection symmetry by gauging the internal $\Z_2$ sub-symmetry. This agrees with what we found in subsection~\ref{LSMreflection1dSec} when working explicitly with qubit models in ${1+1}$D.

\subsubsection{LSM theorem from SymTFT}
\label{sec:reflection LSM from SymTFT}

The LSM anomaly of the ${\Z_2\times \Z_2^R}$ symmetry can also be diagnosed using the symmetric sandwich construction discussed generally in Section~\ref{gappedSymPhasesSec}. The $m$ and $e$ condensing boundaries constructed above correspond to the Lagrangian algebras ${\cL_m = 1\oplus m}$ and ${\cL_e = 1\oplus e}$, respectively, of the $\Z_2$ topological order. Because $e$ anyons carry fractional reflection charge, $\cL_e$ is not a reflection-symmetric Lagrangian algebra. However, $\cL_m$ is reflection-symmetric, and the only symmetric sandwich is one whose top and bottom boundaries are the $m$ condensing boundary. Therefore, the only $\Z_2^R$-enriched gapped phase with a ${\Z_2\times \Z_2^R}$ symmetry has a two-fold ground state degeneracy and spontaneously breaks the internal $\Z_2$ symmetry. The fact that there are no ${\Z_2\times \Z_2^R}$ SPTs is exactly the LSM anomaly between internal $\Z_2$ and reflection symmetries.

\section{SymTFT enriched by time-reversal}
\label{sec TRS SymTFT}

In this section, we consider ${2+1}$D SymTFTs enriched by time-reversal symmetry (TRS). Similar to Section~\ref{sec reflection SymTFT}, we will focus on an example that shows how time-reversal enriched SymTFTs can capture the interplays between $\Z_2$ symmetries and $\Z_2^T$ TRS. In particular, we consider a $\Z_2\times\Z^T_2$ symmetry with mixed anomaly and a $\Z^T_4$ symmetry and show they have the same time-reversal enriched-SymTFT. These interplays are captured in the SymTFT by the fractionalization of the TRS, which gives rise to a local Kramers degeneracy~\cite{LS12051244, BBC14104540, BBJ161207792}.

\subsection{Example: anomalies and extensions with time-reversal}

Before constructing the SymTFT, we first explore an explicit example of these interplays between $\Z_2$ and $\Z_2^T$ symmetries in ${1+1}$D. 

\subsubsection*{Lattice model perspective}

Consider a ${1+1}$D lattice model of qubits with a single qubit on each site $j$. We assume the number of lattice sites $L$ is even and the Pauli operators satisfy periodic boundary conditions ${X_{j+L} \equiv X_j}$
and ${Z_{j+L} \equiv Z_j}$. For this example, we consider the ${\Z_2\times \Z_2^T}$ symmetry generated by the operators
\begin{equation}\label{eq:anomalous Z2xZT2 symmetry}
U = \prod_{j=1}^L Z_j,
\qquad
U_T = \mathsf{K}\,\prod_{n=1}^{L/2} \left( \ee^{\ii\frac{\pi}{4} X_{2n-1}Y_{2n} }\, \ee^{-\ii\frac{\pi}{4} Y_{2n}X_{2n+1} }\right)
\,,
\end{equation}
where $\mathsf{K}$ is the complex conjugation operator. These operators are symmetries of, for example, the XX model
\begin{align}
\label{eq:Ham with anomalous TRS}
H_\mathrm{XX} = \sum_{j=1}^{L} (X_j\,X_{j+1} + Y_j\,Y_{j+1})\,.
\end{align}
The XX model has many other interesting symmetries that we will not consider.\footnote{The XX model has a plethora of interesting symmetries, including non-invertible symmetries and Onsager symmetries, which are infinite-dimensional Lie group symmetries~\cite{PCS241218606}.} It is obvious that the unitary operator $U$ is a symmetry operator and commutes with the XX model Hamiltonian, i.e., ${[H_\mathrm{XX},U] = 0}$. The anti-unitary operator $U_T$ satisfies
\begin{equation}
   U_T X_j U^\dag_T= \begin{cases}X_j & j \text { odd,} \\ X_{j-1} X_j X_{j+1} & j \text { even,}\end{cases}
   \qquad
   U_T Y_j U^\dag_T = \begin{cases}-Y_{j-1} Y_j Y_{j+1} & j \text { odd,} \\ -Y_j & j \text { even,}\end{cases}
\end{equation}
and using these relations, it is straightforward to verify that $U_T$ is also a symmetry operator of the XX model.

The ${\Z_2\times \Z_2^T}$ symmetry generated by~\eqref{eq:anomalous Z2xZT2 symmetry} has a mixed anomaly between $\Z_2$ and $\Z_2^T$. This follows from the fact that the unitary ${\Z_2\times\Z_2}$ symmetry generated by $U$ and the unitary part of $U_T$ (i.e., $U_T$ without $\mathsf{K}$) has a mixed anomaly~\cite{CPS240912220, PCS241218606}. A consequence of the mixed anomaly is that the $\Z_2^T$ TRS operator becomes a $\Z_4^T$ TRS operator after gauging the unitary $\Z_2$ symmetry. Indeed, the gauging map for the symmetry generated by $U$ can be implemented using the Kramers-Wannier transformation
\begin{equation}\label{KWtra}
    Z_j \to Z_j Z_{j+1},\qquad X_j X_{j+1} \to X_{j+1}.
\end{equation}
After gauging, there is a dual $\Z_2$ symmetry generated by ${U^\vee = \prod_{j=1}^L X_j}$ and the TRS operator $U_T$ under~\eqref{KWtra} becomes
\begin{equation}
    U_T^\vee = \mathsf{K}\,\prod_{n=1}^{L/2} \left( \ee^{\ii\frac{\pi}{4} Y_{2n}Z_{2n+1} }\, \ee^{-\ii\frac{\pi}{4} Z_{2n}Y_{2n+1} }\right)
\end{equation}
This TRS operator satisfies
\begin{align}
(U^\vee_T)^2
=\prod_{n=1}^{L/2} \left( \ee^{\ii\frac{\pi}{2} Y_{2n}Z_{2n+1} }\, \ee^{-\ii\frac{\pi}{2} Z_{2n}Y_{2n+1} }\right)=\prod_{n=1}^{L/2} \left(\ii Y_{2n}Z_{2n+1}\right) \left(-\ii Z_{2n}Y_{2n+1} \right)=
U^\vee,
\end{align}
and, therefore, is a $\Z^T_4$ symmetry operator. As a consequence of the mixed anomaly, the $\Z_2^T$ symmetry before gauging has been extended by the dual $\Z_2$ symmetry after gauging to become a $\Z^T_4$ symmetry.

\subsubsection*{Field theory perspective}

The relationship between the anomalous $\Z_2\times \Z_2^T$ symmetry and $\Z_4^T$ symmetry can also be understood through an anomaly inflow theory in one higher dimension. This is the same invertible TFT that characterized the LSM anomaly between internal $\Z_2$ symmetry and reflection in Sec.~\ref{LSMreflection1dSec}. In particular, the invertible TFT is given by the SPT action
\ie\label{eq:SPT_TRS}
\frac{\ii\pi}{2} \int_{X_3} A\cup \del \hat{w}_1.
\fe
where $X_3$ is a ${2+1}$D spacetime manifold, ${A\in H^1(X_3,\Z_2)}$ is the background gauge field for the internal $\Z_2$ symmetry, and $\hat{w}_1$ is the $\Z$-lift of the first Stiefel-Whitney class $w_1\in H^1(X_3,\Z_2)$ of $X_3$. Using the same argument for the reflection-symmetry with LSM anomaly, we find that the SPT action captures the extension of time-reversal symmetry to $\Z^T_4$ when the internal $\Z_2$ symmetry is gauged (i.e., when $A$ is made dynamical).

\subsection{Time-reversal enriched SymTFT}

We now consider the SymTFT for the anomalous ${\Z_2\times \Z_2^T}$ symmetry and $\Z_4^T$ symmetry. Like for the reflection-enriched SymTFT in Section~\ref{reflectionEnrichedSymTFT}, this SymTFT has a $\Z_2$ topological order enriched by $\Z_2^T$ TRS that fractionalizes on the $e$ anyons. In the field theory language, this SymTFT is obtained by promoting the background gauge field $A$ in the SPT action \eqref{eq:SPT_TRS} to a dynamical one. Here, we will consider the corresponding stabilizer code Hamiltonian, which is the same as that constructed in Appendix~\ref{Z2 enriched TC App}. 

\subsubsection{Stabilizer code}

On each site $\bm{r}$ of the square lattice resides a $\Z_4$ qudit while on each link resides a qubit. The $\Z_4$ qudit operators are denoted by $\cX_{\bm{r}}$ and $\cZ_{\bm{r}}$, while the Pauli operators of the qubits are $\si^x_{\bm{r},\mu}$ and $\si^z_{\bm{r},\mu}$. The physical Hilbert space is subject to the local constraint
\begin{equation}\label{refEnrGausslaw}
\cX_{\bm{r}}^2 \, A_{\bm{r}} = 1, \hspace{20pt}\text{where} \hspace{20pt}A_{\bm{r}} = \begin{tikzpicture}[scale = 0.5, baseline = {([yshift=-.5ex]current bounding box.center)}]
\draw[line width=0.015in, gray] (7.5,0) -- (13.5, 0);
\draw[line width=0.015in, gray] (10.5,-3) -- (10.5, 3);
\node[above] at (9, 0) {\normalsize $\textcolor[HTML]{cb181d}{\si^x}$};
\node[above] at (12,0) {\normalsize $\textcolor[HTML]{cb181d}{\si^x}$};
\node at (10.5, 1.5) {\normalsize $\textcolor[HTML]{cb181d}{\si^x}$};
\node at (10.5, -1.5) {\normalsize $\textcolor[HTML]{cb181d}{\si^x}$};
\fill[black] (10.5,0) circle (6pt);
\node at (10, -0.5) {\normalsize $\bm{r}$};
\end{tikzpicture}~.
\end{equation}

The stabilizer code Hamiltonian on the constrained physical Hilbert space is given by
\begin{align}
\label{eq:TRS enriched TO}
H
=
-\sum_{\bm{r}} (\cX_{\bm{r}} + \cX_{\bm{r}}^\dag + B_{\bm{r}} ),
\hspace{20pt}\text{where} \hspace{20pt}
B_{\bm{r}}=
\begin{tikzpicture}[scale = 0.5, baseline={([yshift=-.5ex]current bounding box.center)}]
\draw[line width=0.015in, gray] (-3, 0) -- (-3, 3) -- (0, 3) -- (0, 0) -- cycle;
\node[below] at (-1.5, 0) {\normalsize $\textcolor[HTML]{2171b5}{\si^z}$};
\node[above] at (-1.5, 3) {\normalsize $\textcolor[HTML]{2171b5}{\si^z}$};
\node at (-3.5, 1.5) {\normalsize $\textcolor[HTML]{2171b5}{\si^z}$};
\node at (0.6, 1.5) {\normalsize $\textcolor[HTML]{2171b5}{\si^z}$};
\fill[black] (-3,0) circle (6pt);
\node at (-3.5, -0.5) {\normalsize $\bm{r}$};
\end{tikzpicture}\,.
\end{align}
It has $\Z_2$ topological order. One way to see this is to note that it has an anomalous ${\Z_2\times \Z_2}$ (non-topological) 1-form symmetry generated by ${A_{\bm{r}}}$ and $B_{\bm{r}}$, which is spontaneously broken. A pair of $e$ anyons is created by the string operator ${\cZ_{\bm{r}} \sigma^z_{\bm{r},\mu} \cZ_{\bm{r}+\hat{\mu}}^\dag}$, and a pair of $m$ anyons by $\si^x_{\bm{r},\mu}$. The Hamiltonian~\eqref{eq:TRS enriched TO} also commutes with the TRS operator
\begin{align}
\label{eq:TRS for TO}
U_T = \mathsf{K}\, \prod_{\bm{r}} \cX_{\bm{r}},
\end{align}
where $\mathsf{K}$ is the complex conjugation. We work in the basis where $\cX$, $\sigma^z$, and $\sigma^x$ are real and hence commute with $\mathsf{K}$ while $\mathsf{K}\,\cZ\,\mathsf{K}=\cZ^\dag$. On an infinite lattice, the TRS operator $U_T$ generates a $\Z_2^T$ symmetry since
\begin{align}
(U_T)^2 = \prod_{\bm{r}} \cX^2_{\bm{r}} = \prod_{\bm{r}}A_{\bm{r}} = 1.
\end{align}

The $\Z_2^T$ symmetry is fractionalized on the $e$ anyons. We can probe this fractionalization by inserting a single, static $e$ anyon at the site $\bm{r}_0$. The presence of this non-dynamical anyon modifies each observable $\cO$ of the model to ${W_{e,\bm{r}_0} \, \cO\, W_{e,\bm{r}_0} ^\dag}$, where
${W_{e,\bm{r}_0} = \cZ_{\bm{r}_0}\prod_{\ell\subset\cP}\sigma^z_{\ell}}$ is an infinite $e$ anyon string operator with $\cP$ a path from $\bm{r}_0$ to infinity. For example, the Hamiltonian~\eqref{eq:TRS enriched TO} becomes
\begin{align}
\label{eq:TRS enriched Hamiltonian w e anyon}
H_e
=
W_{e,\bm{r}_0} \, H \, W_{e,\bm{r}_0}^\dag
=
H + (1-\ii)\cX_{\bm{r}_0} 
+
(1+\ii) \cX^\dag_{\bm{r}_0}
,
\end{align}
which describes the same TRS-enriched $\Z_2$ topological order, but now with a single $e$ anyon trapped at site $\bm{r}_0$. On the other hand, the TRS operator~\eqref{eq:TRS for TO} becomes
\begin{align}
U_{T;e}
=
W_{e,\bm{r}_0} \, U_T \, W_{e,\bm{r}_0}^\dag
=
\mathrm{i}
\cZ^2_{\bm{r}_0}
U_T.
\end{align}
While $U_{T;e}$ commutes with the modified Hamiltonian $H_e$, the original TRS operator $U_T$ does not. This modified TRS operator satisfies
\begin{align}
U^2_{T;e}
=
-1,
\end{align}
and furnishes the non-trivial projective representation of the $\Z^T_2$ symmetry. Therefore, there is a Kramers degeneracy arising from inserting a single $e$ anyon, which is sometimes called a ``local Kramers degeneracy'' and is a key signature of TRS fractionalization~\cite{LS12051244, BBC14104540, BBJ161207792}. We note another manifestation of the $\Z_2^T$ symmetry fractionalization is that the string operator $W_{e,\bm{r}_0}$ for a single $e$ anyon satisfies $U_T W_{e,\bm{r}_0} U_T^\dag = \ii W_{e,\bm{r}_0}$. Hence, a single $e$ anyon can be regarded as carrying a fractional charge $\ii$ of the $\Z_2^T$ symmetry.

\subsubsection{Gapped boundary conditions}

There are two classes of gapped boundaries for this TRS-enriched $\Z_2$ toric code: the $e$ condensing and $m$ condensing boundaries. Just as for the reflection-enriched SET example discussed in Section~\ref{reflectionboundaryExSec}, as symmetry boundaries, they lead to two different symmetries.

The $m$ condensing is the smooth boundary of the $\Z_2$ topological order. It is obtained by truncating the lattice along the ${r_y=0}$ row of sites, yielding a lattice formed by sites ${(r_x,r_y)\in \Z\times \Z_{\leq 0}}$. The $B_{\bm{r}}$ stabilizers are unaffected by this truncation, but the $A_{\bm{r}}$ operator in the local constraint~\eqref{refEnrGausslaw} is modified on this boundary. The local constraint on the ${r_y = 0}$ boundary sites is
\begin{equation}
\cX_{(r_x,0)}^2 \, A^{(\mathrm{smooth})}_{r_x} = 1, \hspace{20pt}\text{where} \hspace{20pt}A^{(\mathrm{smooth})}_{r_x} = \begin{tikzpicture}[scale = 0.5, baseline = {([yshift=-.5ex]current bounding box.center)}]
\draw[line width=0.015in, gray] (7.5,0) -- (13.5, 0);
\draw[line width=0.015in, gray] (10.5,-3) -- (10.5, 0);
\node[below] at (9, 0) {\normalsize $\textcolor[HTML]{cb181d}{\si^x}$};
\node[below] at (12,0) {\normalsize $\textcolor[HTML]{cb181d}{\si^x}$};
\node at (10.5, -1.5) {\normalsize $\textcolor[HTML]{cb181d}{\si^x}$};
\fill[black] (10.5,0) circle (6pt);
\node[above right] at (10, 0) {\normalsize $(r_x,0)$};
\end{tikzpicture}~.
\end{equation}
The $m$ anyon logical operator on this boundary can be written as
\begin{equation}
    W^{(\text{smooth})}_m = \prod_{r_x} \sigma^x_{(r_x,-1),y} = \prod_{r_x}\cX^2_{(r_x-1,0)}.
\end{equation}
Therefore, ${W^{(\text{smooth})}_m=1}$ in the ground state subspace, which confirms that this is the $m$ condensing boundary. The $e$ logical operator on the boundary is unaffected by the boundary stabilizer. It is given by
\ie
W^{(\text{smooth})}_e=\prod_{r_x} \si^z_{(r_x,0),x}~,
\fe 
and generates a $\Z_2$ symmetry on the boundary. With this boundary condition, the TRS operator is unchanged. However, it now squares to 
\begin{align}
U^2_T \equiv \prod_{r_x} \left( A^{(\mathrm{smooth})}_{r_x} \prod_{r_y<0} A_{(r_x,r_y)} \right) = 1.
\end{align}
Therefore, this symmetry boundary encodes a ${\Z_2\times \Z_2^T}$ symmetry. In fact, as we will show momentarily, this ${\Z_2\times \Z_2^T}$ symmetry has a mixed anomaly.

The $e$-condensing boundary is obtained by choosing the rough boundary conditions. For this stabilizer code model, this is one where the lattice is truncated along the ${r_y=1}$ row of sites, yielding a lattice formed by sites ${(r_x,r_y)\in \Z\times \Z_{\leq 1}}$ with qubits on the boundary ${r_y = 1}$ sites and \textit{no} degrees of freedom on the boundary horizontal links. The local constraint~\eqref{refEnrGausslaw} is unchanged and implemented on all bulk sites ${r_y\leq 0}$. The boundary plaquette stabilizers are modified to be
\begin{align}
B^{(\mathrm{rough})}_{r_x}=
\begin{tikzpicture}[scale = 0.5, baseline={([yshift=-.5ex]current bounding box.center)}]
\draw[line width=0.015in, gray] (-3, 0) -- (-3, 3);
\draw[line width=0.015in, gray] (-3, 0) -- (0, 0);
\draw[line width=0.015in, gray] (0, 3) -- (0, 0);
\node[above] at (-1.5, 0) {\normalsize $\textcolor[HTML]{2171b5}{\si^z}$};
\node[left] at (-3, 1.5) {\normalsize $\textcolor[HTML]{2171b5}{\si^z}$};
\node[right] at (0, 1.5) {\normalsize $\textcolor[HTML]{2171b5}{\si^z}$};
\fill[black] (-3,0) circle (6pt);
\node[below] at (-3, 0) {\normalsize $(r_x,0)$};
\node[above] at (-3, 3) {\normalsize $\cZ$};
\node[above] at (0, 3) {\normalsize $\cZ^\dag$};
\end{tikzpicture}.
\end{align}

The $e$ anyon logical operator on this boundary can be written as
\begin{equation}
    W^{(\text{rough})}_e = \prod_{r_x} \sigma^z_{(r_x,0),x} = \prod_{r_x} B^{(\mathrm{rough})}_{r_x}.
\end{equation}
Therefore, ${W^{(\text{rough})}_e=1}$ in the ground state subspace, which confirms that this is the $e$ condensing boundary. The $m$ logical operator on the boundary is unaffected by the boundary stabilizer and given by
\ie
W^{(\text{rough})}_e=\prod_{r_x} \si^x_{(r_x,0),y}~,
\fe 
and generates a $\Z_2$ symmetry on the boundary. With this boundary condition, the TRS operator is unchanged and squares to 
\begin{equation}
    U^2_T 
=
\prod_{\bm{r}}
\cX^2_{\bm{r}}
=
\prod_{r_x} (\cX_{(r_x,1)}^2\, \sigma^x_{(r_x,0),y}).
\end{equation}
Therefore, in the ground state subspace, ${U_T^2 = W^{(\text{rough})}_e}$ and the $\Z^T_2$ bulk symmetry is extended to $\Z^T_4$ symmetry at the boundary.

Gauging the $\Z_2$ symmetry corresponding to the boundary logical operator is implemented by changing boundary conditions. Thus, we see from the SymTFT perspective how anomalous ${\Z_2\times \Z_2^T}$ symmetry is related to a $\Z_4^T$ TRS by gauging the internal $\Z_2$ sub-symmetry.

\subsubsection{Mixed anomaly from SymTFT}
\label{sec:TRS mixed anomaly from SymTFT}

The mixed anomaly of the ${\Z_2\times \Z_2^T}$ symmetry is diagnosed using the symmetric sandwich construction discussed generally in Section~\ref{gappedSymPhasesSec}. The $m$ and $e$ condensing boundaries constructed above correspond to the Lagrangian algebras ${\cL_m = 1\oplus m}$ and ${\cL_e = 1\oplus e}$, respectively, of the $\Z_2$ topological order, and only $\cL_m$ is time-reversal-symmetric. Therefore, the only symmetric sandwich is one whose top and bottom boundaries are the $m$ condensing boundary. Thus, the only $\Z_2^T$-enriched gapped phase with a ${\Z_2\times \Z_2^T}$ symmetry has a two-fold ground state degeneracy and spontaneously breaks the internal $\Z_2$ symmetry. The fact there are no ${\Z_2\times \Z_2^T}$ SPTs is the mixed anomaly between internal $\Z_2$ and $\Z_2^T$ symmetries.

\section{Outlook}

In this paper, we have extended the SymTFT framework to go beyond internal symmetries and incorporate spacetime symmetries. To do so, we considered the SymTFT of the internal symmetry, i.e., a topological order, and enriched it with spacetime symmetries to construct an SET. To illustrate this framework and its applications to gauging and diagnosing anomalies, we focused on invertible internal symmetries and crystalline/time-reversal symmetries in ${1+1}$D (see Section~\ref{SummarySection} for a detailed summary). Many interesting follow-up directions arise from our work that further develop this symmetry-enriched SymTFT framework. Here, we discuss three particular tantalizing extensions.

While we considered spatial translations in this paper (see Section~\ref{translationSec}), we did not consider temporal translations. Discrete temporal translations play a crucial role in, for example, periodically driven systems~\cite{Bukov04032015} and time crystals~\cite{Zaletel:2023aej}, and it would be interesting to develop a SymTFT perspective of such phenomena. Enriching the SymTFT by ordinary temporal translations could make it become a Floquet enriched topological order~\cite{PM161003485}, which would naturally capture temporally modulated symmetries. More generally, the SymTFT could be a type of dynamical code, e.g., Floquet codes~\cite{HH210702194}. Extending the SymTFT to out-of-equilibrium phenomena has received limited attention~\cite{MMB231217176}, and considering Floquet enriched topological orders and dynamical codes as SymTFTs offers an interesting starting point for future work.

Another exciting direction to consider is spacetime symmetries in greater than ${1+1}$D. Crystalline symmetries in higher than one spatial dimension are much richer, now including discrete rotations and various reflection symmetries. With richer crystalline symmetries come richer interplays with internal symmetries. Even the interplays involving spatial translations can become richer. For example, $d$-dimensional translations can fractionalize in the ${((d+1)+1)}$D SymTFT only when ${d>1}$. Furthermore, $d$-dimensional translations can form magnetic translations, leading to an extension with a $U(1)$ (higher-form) symmetry, 
only when ${d>1}$.

Lastly, in this paper, we showed how anomalies can be detected using the SymTFT. For instance, we explored LSM anomalies involving translations in Section~\ref{translationSec} and spatial reflections in Section~\ref{sec reflection SymTFT}. However, there are weaker LSM-like constraints called SPT-LSM theorems that provide obstructions to trivial SPTs~\cite{L170504691, YJV170505421, LRO170509298, ET190708204, JCQ190708596, PLA240918113}. One way or another, these SPT-LSM theorems arise from lattice translations, and it would be interesting to see their manifestations in the symmetry-enriched SymTFT framework developed here.

\section*{Acknowledgments}

We thank 
Maissam Barkeshli,
Daniel Bulmash,
I{\~n}aki Garc{\'i}a Etxebarria,
Wenjie Ji,
Ryohei Kobayashi,
Shang-Qiang Ning,
David Penneys,
Abhinav Prem, 
Nathan Seiberg, 
Xiao-Gang Wen,
and 
Carolyn Zhang
for helpful discussion.
We further thank 
Da-Chuan Lu
and
Zhengdi Sun 
for comments on the manuscript.
This research was supported in part by the Heising-Simons Foundation, the Simons Foundation, and grant no.\ NSF PHY-2309135 to the Kavli Institute for Theoretical Physics (KITP). 
Furthermore, this research was conducted while visiting the Okinawa Institute of Science and Technology (OIST) through the Theoretical Sciences Visiting Program (TSVP).
SDP is supported by the National Science Foundation (NSF) Graduate Research Fellowship (GRF) under grant no.\ 2141064.
{\"OMA} is supported by NSF under grant no.\ DMR-2022428
and Swiss National Science Foundation (SNSF)
under grant no.\ P500PT-214429.
H.T.L. is supported by the U.S. Department of Energy, Office of Science, Office of
High Energy Physics of U.S. Department of Energy under grant Contract Number DE-SC0012567
(High Energy Theory research) and by the Packard Foundation award for Quantum Black Holes from
Quantum Computation and Holography.

\appendix

\section{Introduction to SymTFT}\label{SymTFTReview}

In this Appendix, we review basic aspects of the symmetry topological field theory (SymTFT) framework. The SymTFT ${\mathfrak{Z}(\cS)}$ of a symmetry $\cS$ in ${(d+1)}$D is a ${(d+2)}$D topological theory used to separate the kinematic aspects of $\cS$ from the dynamics of a ${(d+1)}$D theory $\mathfrak{T}^\cS$ with symmetry $\cS$. The formal development and applications of the SymTFT perspective of symmetries is being intensely pursued~\cite{GW14125148, KZ150201690, KZ170501087, FT180600008, KZ190504924, Pulmann:2019vrw, TW191202817, JW191213492, LB200304328, KZ200514178, GK200805960, AFM200808598, ABE211202092, CW220303596, CW220506244, MT220710712, FT220907471, Kaidi:2022cpf, KNZ230107112, ZC230401262, BS230517159, BS240106128, AB240110165, BZM240212347, ABD240214813, ABB240406601, WYP240419004, H240509611, BIT240509754, JC240702488,C240801490, Cordova:2024iti,  GarciaEtxebarria:2024jfv, Choi:2024tri, Bhardwaj:2024igy, 
VRS240906647, PLA240918113}, which we briefly review here. The reader is refereed to \Rfs{CW221214432, S230518296} for a more thorough introduction.

We denote the braided fusion $d$-category describing the topological defects of the SymTFT $\mathfrak{Z}(\cS)$ by $\mathcal{Z}(\cS)$. When ${d=1}$ and $\cS$ is a fusion category, $\mathcal{Z}(\cS)$ is the Drinfeld center of $\cS$ and $\mathfrak{Z}(\cS)$ is the Turaev-Viro TFT $\mathsf{TV}(\cS)$~\cite{Turaev:1992hq}. A defining feature of general $\mathfrak{Z}(\cS)$ is the existence of a gapped boundary condition $\mathfrak{B}^{\text{sym}}_\cS$ on which the topological defects in $\mathcal{Z}(\cS)$ become the symmetry defects of $\cS$. For ${2+1}$D SymTFTs that are non-trivial topological orders, these gapped boundaries are classified by the Lagrangian algebras of $\cZ(\cS)$~\cite{DMN10092117, K13078244, LWW1414, CCW170704564}. The topological defects in $\mathfrak{Z}(\cS)$ that can end on this boundary correspond to the symmetry charges of $\cS$. This gapped boundary encodes the kinematic aspects of the symmetry $\cS$. It, along with the SymTFT ${\mathfrak{Z}(\cS)}$, depend only on $\cS$ and are independent of the details of $\mathfrak{T}^\cS$. On the other hand, the dynamics of $\mathfrak{T}^\cS$ are encoded in a separate boundary $\mathfrak{B}^{\text{phys}}_{\mathfrak{T}^\cS}$ of the SymTFT. These two boundaries and the SymTFT bulk make up the data ${(\mathfrak{B}^{\text{sym}}_\cS, \mathfrak{Z}(\cS), \mathfrak{B}^{\text{phys}}_{\mathfrak{T}^\cS})}$ and form the sandwich picture shown in Fig.~\ref{fig:TopHolo}.

The SymTFT ${\mathfrak{Z}(\cS)}$ is required to be topological in the additional dimension absent from $\mathfrak{T}^\cS$. Since it is topological in this new dimension, the sandwich ${(\mathfrak{B}^{\text{sym}}_\cS, \mathfrak{Z}(\cS), \mathfrak{B}^{\text{phys}}_{\mathfrak{T}^\cS})}$ can be bijectively mapped to the theory $\mathfrak{T}^\cS$ by performing an interval compactification (see Fig.~\ref{fig:TopHolo}). It is typically the case that ${\mathfrak{Z}(\cS)}$ is topological in all directions, and is a TFT. However, strictly speaking, it only needs to be topological in the compactification direction (e.g., see \Rfs{CJ231001474, JJ250522261}). For some of the SymTFTs 
constructed in the bulk of the paper, ${\mathfrak{Z}(\cS)}$ is topological in the compactification direction but not in any other direction. 

There are two equivalent formulations of the SymTFT ${\mathfrak{Z}(\cS)}$ used in the literature. The first formulates the SymTFT using quantum codes, and the second uses Euclidean field theory. Let us review both of these perspectives.

In the quantum code perspective, the SymTFT is described by the code space of a quantum code, and its topological defects are related to the code's topological logical operators. The topological aspect of ${\mathfrak{Z}(\cS)}$ manifests in the logical operators being topological in at least the direction in which the interval compactification is performed. However, in most cases they are topological in all directions. This quantum code can be represented as the ground state subspace of a commuting projector Hamiltonian model. Fig.~\ref{fig:LatticeTopHolo} shows how the SymTFT sandwich on a spatial lattice is organized in this quantum code description.

\begin{figure}
\centering
\begin{tikzpicture}[scale=.75, baseline={([yshift=-.5ex]current bounding box.center)},thick]
\draw[latSymTFTcolor, line width=3pt] (-0.5, 1) -- (10.5,1);
\draw[latSymTFTcolor, line width=3pt] (-0.5, 0) -- (10.5,0);
\draw[latSymTFTcolor, line width=3pt] (-0.5, -1) -- (10.5,-1);
\draw[latSymTFTcolor, line width=3pt] (0, 2) -- (0,-3);
\draw[latSymTFTcolor, line width=3pt] (1, 2) -- (1,-3);
\draw[latSymTFTcolor, line width=3pt] (2, 2) -- (2,-3);
\draw[latSymTFTcolor, line width=3pt] (3, 2) -- (3,-3);
\draw[latSymTFTcolor, line width=3pt] (4, 2) -- (4,-3);
\draw[latSymTFTcolor, line width=3pt] (5, 2) -- (5,-3);
\draw[latSymTFTcolor, line width=3pt] (6, 2) -- (6,-3);
\draw[latSymTFTcolor, line width=3pt] (7, 2) -- (7,-3);
\draw[latSymTFTcolor, line width=3pt] (8, 2) -- (8,-3);
\draw[latSymTFTcolor, line width=3pt] (9, 2) -- (9,-3);
\draw[latSymTFTcolor, line width=3pt] (10, 2) -- (10,-3);
\draw[black, line width=3pt] (-0.5, 2) -- (10.5,2);
\draw[black, line width=3pt] (0, -2) -- (0,-3);
\draw[black, line width=3pt] (1, -2) -- (1,-3);
\draw[black, line width=3pt] (2, -2) -- (2,-3);
\draw[black, line width=3pt] (3, -2) -- (3,-3);
\draw[black, line width=3pt] (4, -2) -- (4,-3);
\draw[black, line width=3pt] (5, -2) -- (5,-3);
\draw[black, line width=3pt] (6, -2) -- (6,-3);
\draw[black, line width=3pt] (7, -2) -- (7,-3);
\draw[black, line width=3pt] (8, -2) -- (8,-3);
\draw[black, line width=3pt] (9, -2) -- (9,-3);
\draw[black, line width=3pt] (10, -2) -- (10,-3);
\draw[latSymTFTcolor, line width=3pt] (-0.5, -2) -- (10.5,-2);
\node[black, left] at (-.5, 2) {\Large $\cdots$};
\node[latSymTFTcolor, left] at (-.5, 1) {\Large $\cdots$};
\node[latSymTFTcolor, left] at (-.5, 0) {\Large $\cdots$};
\node[latSymTFTcolor, left] at (-.5, -1) {\Large $\cdots$};
\node[latSymTFTcolor, left] at (-.5, -2) {\Large $\cdots$};
\node[black, right] at (10.5, 2) {\Large $\cdots$};
\node[latSymTFTcolor, right] at (10.5, 1) {\Large $\cdots$};
\node[latSymTFTcolor, right] at (10.5, 0) {\Large $\cdots$};
\node[latSymTFTcolor, right] at (10.5, -1) {\Large $\cdots$};
\node[latSymTFTcolor, right] at (10.5, -2) {\Large $\cdots$};
\node[black, left] at (-1.5, 2) {\Large $\mathfrak{B}^{\text{sym}}_{\cS}$};
\node[black, left] at (-1.5, 0) {\Large $\mathfrak{Z}(\cS)$};
\node[black, left] at (-1.5, -2.5) {\Large $\mathfrak{B}^{\text{phys}}_{\mathfrak{T}^\cS}$};
\end{tikzpicture}
\caption{
Shows how, in the quantum code description of the SymTFT, the cells of the spatial lattice for a ${2+1}$D SymTFT are organized into the SymTFT sandwich ${(\mathfrak{B}^{\text{sym}}_\cS, \mathfrak{Z}(\cS), \mathfrak{B}^{\text{phys}}_{\mathfrak{T}^\cS})}$. The stabilizers defining the SymTFT ${\mathfrak{Z}(\cS)}$ act only on purple edges in the bulk, and the boundary stabilizers defining the symmetry boundary act on at least one top boundary edges belonging to ${\mathfrak{B}^{\text{sym}}_\cS}$. Stabilizers or terms in the Hamiltonian acting on the bottom boundary edges ${\mathfrak{B}^{\text{phys}}_{\mathfrak{T}^\cS}}$ encode the dynamics of the ${1+1}$D theory $\mathfrak{T}^\cS$. After the interval compactification, only the black edges remain, and they become the edges and sites, respectively, of the ${1+1}$D spatial lattice for $\mathfrak{T}^\cS$.
}
\label{fig:LatticeTopHolo}
\end{figure}
In the Euclidean field theory perspective, the SymTFT is generally a foliated field theory. This foliated field theory does not depend on a background Riemannian metric but can depend on a background foliation structure of spacetime. It is topological in at least the direction in which the interval compactification is performed. Furthermore, it becomes a topological field theory when the foliation structure is turned off. Again, in most cases, the SymTFT will be topological in all directions and be a topological field theory without any foliation backgrounds turned on.

Whether it be from the quantum code or Euclidean field theory perspectives, the SymTFT has numerous powerful applications. Here we review two: discrete gauging and classifying gapped and gapless phases.

Discrete gauging of $\cS$ is performed in the SymTFT formalism by changing the gapped boundary $\mathfrak{B}^{\text{sym}}_\cS$. This is often visualized using the ``quiche'' construction
\begin{equation}\label{quicheSymTFTRev}
\begin{tikzpicture}[scale=.75, baseline={([yshift=-.5ex]current bounding box.center)},thick]
\fill[SymTFTcolor] (0,0) rectangle (4,2);
\draw[black, line width=4pt] (0, 2) -- (4,2);
\node[black, left] at (0, 2) {\Large $\mathfrak{B}^{\text{sym}}_{\cS}$};
\node[black] at (2, 1) {\Large $\mathfrak{Z}(\cS)$};
\end{tikzpicture}.
\end{equation}
The top boundary $\mathfrak{B}^{\text{sym}}_{\cS}$ is gapped, and its topological defects form the symmetry $\cS$. Discrete gauging $\cS$ is performed by replacing the gapped boundary $\mathfrak{B}^{\text{sym}}_{\cS}$ with another gapped boundary $\mathfrak{B}^{\text{sym}}_{\cS^\vee}$. The topological defects on $\mathfrak{B}^{\text{sym}}_{\cS^\vee}$ form the dual symmetry $\cS^\vee$ that arises after gauging. Therefore, two symmetries related to one another by discrete gauging have the same SymTFT ${\mathfrak{Z}(\cS)}$, and ${\mathfrak{Z}(\cS)}$ is a label for the gauging web of a symmetry.

The SymTFT can also be used for classifying gapped and gapless phases of $\mathfrak{T}^\cS$ characterized by $\cS$. Let us now specialize our discussion to ${d+1 = 1+1}$ dimensional theories where $\cS$ is a finite 0-form symmetry described by a fusion category, which we denote by $\cS$. Phases characterized by $\cS$ are classified by condensable algebras $\cA$ of $\cZ(\cS)$~\cite{CW220506244, BBP231217322, BPS240300905}. In the SymTFT formalism, this classification is based on the ``club quiche'' picture shown in Fig.~\ref{fig:TopHoloClass}, where ${\mathfrak{Z}(\cS)/\cA}$ denotes the theory resulting from condensing $\cA$ in $\mathfrak{Z}(\cS)$ and $\mathcal{I}_\cA$ is the topological interface between $\mathfrak{Z}(\cS)$ and ${\mathfrak{Z}(\cS)/\cA}$. 
The topological defects of ${\mathfrak{Z}(\cS)/\cA}$ are described by the Drinfeld center $\cZ(\cS')$ of the fusion category $\cS'$. The interface $\mathcal{I}_\cA$ is a physical realization of the functor ${\cS \to \cS'}$. Physically, $\cS'$ describes the symmetries acting on the gapless degrees of freedom in the phase corresponding to $\cA$.

When $\cZ(\cS')$ is trivial, the theory ${\mathfrak{Z}(\cS)/\cA}$ has no remaining non-trivial topological defects from $\cZ(\cS)$ and $\cA$ is a Lagrangian algebra of $\cZ(\cS)$. When $\cA$ is Lagrangian, it classifies gapped phases. In fact, the gapped boundary $\mathfrak{B}^{\text{sym}}_\cS$ realizing $\cS$ is characterized by a Lagrangian algebra we denote by $\cL_\cS$. When ${\cA = \cL_\cS}$, it corresponds to the phase where the entire $\cS$ symmetry is spontaneously broken. In general, the intersection ${\cA\cap\cL_\cS}$ of $\cA$ and $\cL_\cS$ describes the number of ground states in the phase classified by $\cA$ and SSB pattern. When ${\cA\cap\cL_\cS = 1}$, none of $\cS$ is spontaneously broken, and $\cA$ describes a symmetry protected topological (SPT) phase. Furthermore, when the condensable algebra $\cA$ is not Lagrangian, it classifies a gapless phase which can be a gapless SSB or SPT phase by the same criteria from the intersection ${\cA\cap\cL_\cS}$.

\section{Introduction to group extensions}\label{grpExtApp}

In this Appendix, we review the basics of group extensions and their classification.\footnote{The reader may also refer to~\cite[Chapter IV]{Brown1982} for an alternative introduction to the subject.} In particular, we consider a group $G$ described by a group extension involving a group $Q$ and a $Q$-module $A^{\rho}$. Recall that a $Q$-module $A^\rho$ is an Abelian group $A$ along with a group action ${\rho\colon Q\times A \to A}$ satisfying ${\rho_{q_{1}}(\rho_{q_{2}}(a)) = \rho_{q_{1} q_{2}}(a)}$ and ${\rho_q(a_1 + a_2) = \rho_q(a_1) + \rho_q(a_2)}$ for all ${q_\bullet\in Q}$ and ${a_\bullet\in A}$. Denoting by $\Aut(A)$ the automorphism group of $A$, the group action $\rho$ is equivalently described as the group homomorphism 
\begin{equation}\label{groupActionEq}
\rho\colon Q \to \Aut(A).
\end{equation}
When $G$ is described by this group extension, it is said to be ``an extension of $Q$ by $A$.''

A defining property of a group extension is that the groups $G$, $Q$, and $A$ satisfy the short exact sequence
\begin{equation}\label{SESapp}
\bm{1} \to A \xrightarrow{\iota} G \xrightarrow{\pi} Q \to \bm{1},
\end{equation}
where $\bm{1}$ denotes the order one group. As a short exact sequence, the inclusion and projection homomorphisms $\iota$ and $\pi$, respectively, satisfy ${\mathrm{im}(\iota) = \ker(\pi)}$. Therefore, given a group element ${a\in A}$, its image ${\iota(a)\in G}$ under $\iota$ satisfies ${\pi(\iota(a)) = 1}$, where $1$ is the identity element of $Q$. This implies that $Q$ is isomorphic to the quotient group ${G/\iota(A)}$ and $\iota(A)$ is always a normal subgroup of $G$. Clearly, $G$ can always be described by this group extension if ${G\cong A\times Q}$, in which case the extension is called a trivial extension. However, there are generally many inequivalent ways to extend $Q$ by $A$, each of which describes a different group $G$. Two extensions of $Q$ by $A$ are equivalent if there exists a group isomorphism ${T\colon G_1\to G_2}$ making the commutative diagram
\begin{equation}
\begin{tikzpicture}[ baseline = {([yshift=-.5ex]current bounding box.center)}, scale=2]
\node (0L) at (-.85, 1) {\normalsize $\bm{1}$};
\node (K) at (0, 1) {\normalsize $A$};
\node (Q) at (1, 1.7) {\normalsize $G_1$};
\node (H) at (2, 1) {\normalsize $Q$};
\node (0R) at (2.85, 1) {\normalsize $\bm{1}$};
\node (Q') at (1, 0.3) {\normalsize $G_2$};

\draw[-{>[scale=1.2]}] (0L) -- (K);
\draw[-{>[scale=1.2]}] (K) -- node[above left] {\small $\iota_1$} (Q);
\draw[-{>[scale=1.2]}] (Q) -- node[above right] {\small $\pi_1$} (H);
\draw[-{>[scale=1.2]}] (H) -- (0R);
\draw[-{>[scale=1.2]}] (K) -- node[below left] {\small $\iota_2$} (Q');
\draw[-{>[scale=1.2]}] (Q) -- node[right] {\small $T$} (Q');
\draw[-{>[scale=1.2]}] (Q') -- node[below right] {\small $\pi_2$} (H);
\end{tikzpicture},
\end{equation}
i.e., ${\iota_2 = T\circ \iota_1}$ and ${\pi_1 = \pi_2\circ T}$.

Before continuing further, let us pause here to overview two simple examples:
\begin{enumerate}
\item The group ${G= \Z_2\times \Z_2}$ can be described as an extension of ${Q = \Z_2}$ by ${A = \Z_2}$ with trivial ${\rho}$. In other words, there is the short exact sequence of groups
\begin{equation}
\bm{1} \to \Z_2 \xrightarrow{\iota} \Z_2\times \Z_2 \xrightarrow{\pi} \Z_2 \to \bm{1}.
\end{equation}
Indeed, consider the presentations ${\Z_2\times \Z_2 \cong \< a,b \mid a^2 = b^2 = 1\>}$ and ${\Z_2 \cong \< c \mid c^2 = 1\>}$. Then, we can choose the inclusion homomorphism to satisfy ${\iota(c) = a}$, which is the generator of the normal subgroup ${\iota(\Z_2) = \< a \mid a^2 = 1\>}$ of ${\Z_2\times \Z_2}$. Furthermore, choosing the projection homomorphism $\pi$ such that ${\pi(b) = \pi(ab) = c}$, we find that ${\pi(\iota(c)) = 1}$ and so ${\mathrm{im(\iota) = \ker(\pi)}}$. 
\item The group ${G = \Z_4}$ can also be described as an extension of ${Q = \Z_2}$ by ${A = \Z_2}$ with trivial ${\rho}$. Since $\Z_4$ is not isomorphic to ${\Z_2\times \Z_2}$, this is a different extension than in the first example. It is described by the short exact sequence 
\begin{equation}
\bm{1} \to \Z_2 \xrightarrow{\iota} \Z_4 \xrightarrow{\pi} \Z_2 \to \bm{1}.
\end{equation}
Indeed, let us use the presentations ${\Z_4 \cong \< a \mid a^4 = 1\>}$ and ${\Z_2 \cong \< c \mid c^2 = 1\>}$ such that ${\iota(c) = a^2}$. With this choice of the inclusion homomorphism, ${\iota(\Z_2) = \< a^2 \mid a^4 = 1\>}$ is the normal $\Z_2$ subgroup of $\Z_4$. The projection homomorphism $\pi$ satisfies ${\pi(a) = \pi(a^3) = c}$, which is the quotient map ${\pi\colon \Z_4\to \Z_4/\Z_2}$.
\item The Dihedral group of order 8 ${D_8}$ can be described as an extension of ${Q=\Z_2}$ by ${A=\Z_4}$ with non-trivial $\rho$. The corresponding short exact sequence is
\begin{equation}
\bm{1} \to \Z_4 \xrightarrow{\iota} D_8 \xrightarrow{\pi} \Z_2 \to \bm{1}.
\end{equation}
Indeed, in the presentation ${D_8\cong\left\langle r, s \mid r^2=s^4=1, r s r^{-1}=s^3\right\rangle}$, the image of the inclusion homomorphism ${\iota(\Z_4) = \left\langle s \mid s^4=1\right\rangle}$. Further presenting ${Q = \Z_2 \cong \<c\mid c^2=1\>}$, the projection homomorphism $\pi$ satisfies ${\pi(s^i) = 1}$ and ${\pi(rs^i) = c}$, where ${i=0,1,2,3}$. The group action $\rho$ is non-trivial because ${r s r^{-1} = \rho_r(s) = s^3}$, making ${\Z_4}$ a non-trivial $\Z_2$-module.
\end{enumerate}

The $Q$-module structure on $A$, particularly the action $\rho$ of $Q$ on $A$, can be naturally formulated using the short exact sequence~\eqref{SESapp}. To do so, we introduce a lift ${s\colon Q \to G}$ such that ${\pi(s(q)) = q}$ for all ${q\in Q}$. The group action $\rho$ can then be specified as
\begin{equation}\label{rhodefliftApp}
\rho_q(a) = s(q)\, \iota(a)\, s(q)^{\,-1}.
\end{equation}
Indeed, this expression satisfies the homomorphism condition of~\eqref{groupActionEq} because $\iota$ is a homomorphism. Furthermore, $\rho_q(a)$ does not depend on the choice of lift $s$. Indeed, the lift can be changed by 
\begin{equation}\label{changeLift}
s(q) \to f(q)\, s(q)
\end{equation}
for any ${f(q)\in \iota(A)}$ because ${\mathrm{im}(\iota) = \ker(\pi)}$. However, this change does not affect $\rho_q(a)$ since $A$ is Abelian and $s(q)$ acts on $\iota(a)$ by conjugation. Relatedly, we also note that the product $s(q_1)s(q_2)$ is not necessarily $s(q_1q_2)$. The most general product rule compatible with the homomorphism condition ${\pi(s(q)) = q}$ is ${s(q_1) s(q_2) = c(q_1,q_2)\, s(q_1q_2)}$ for some function ${c\colon Q\times Q \to
\iota(A)}$. This function must satisfy ${c(q,1) = c(1,q) = s(1)}$ for all ${q\in Q}$.

It is convenient to denote the group elements of $G$ by elements ${(a,q)}$ of the set ${A\times Q}$. Using this notation, the inclusion and projection homomorphisms are
\begin{equation}
\iota(a) = (a,1),\hspace{20pt} \pi((a,q)) = q.
\end{equation}
This notation is also useful for writing down the group multiplication of $G$. Indeed, recalling~\eqref{rhodefliftApp} and ${s(q_1) s(q_2) = c(q_1,q_2)\, s(q_1q_2)}$, the group elements of $G$ must satisfy
\begin{align}
(a_1,q)\cdot (a_2,1)  &= (\rho_q(a_2),1)\cdot (a_1,q),
\\
(0,q_1)\cdot (0,q_2) &= (c(q_1,q_2),q_1q_2).
\end{align}
It can be readily checked that the group multiplication rule satisfying these conditions is
\begin{equation}\label{multiplication}
(a_1,q_1)\cdot(a_2, q_2) = (a_1 + \rho_{q_1}(a_2) + c(q_1,q_2), q_1q_2).
\end{equation}
From the group multiplication, $G$ is a trivial extension of $Q$ by $A$ (i.e., ${G\cong A\times Q}$) if $\rho$ and $c$ are trivial. non-trivial extensions arise from non-trivial group actions $\rho$ and functions $c$.

Let us contextualize this notation in the two previously considered examples where ${A = Q = \Z_2}$. We will represent the group elements of $A$ by ${\{0, 1\}}$ with modulo 2 addition and those of $Q$ by ${\{1,-1\}}$ with multiplication. Therefore, the elements of both ${\Z_2\times\Z_2}$ and $\Z_4$ will be labeled by ${\{(0,1),(1,1),(0,-1),(1,-1)\}}$.
\begin{enumerate}
\item The group ${\Z_2\times\Z_2}$ is a trivial extension of ${Q=\Z_2}$ by ${A=\Z_2}$, and therefore both $\rho$ and $c$ are trivial. Its group multiplication indeed is of the form~\eqref{multiplication} with trivial $\rho$ and $c$. 
\item The group $\Z_4$, on the other hand, must be a non-trivial extension. Indeed, the $\Z_4$ group multiplication takes the form of~\eqref{multiplication} with trivial $\rho$ and \begin{equation}
c(q_1,q_2) = 
\begin{cases}
1 \quad & q_1=q_2=-1,\\
0 \quad & \text{else}.
\end{cases}
\end{equation}
This presentation is related to ${\Z_4\cong\<a\mid a^4 = 1\>}$ by ${(0,1)\to 1}$, ${(1,1)\to a^2}$, ${(0,-1)\to a}$, and ${(1,-1)\to a^3}$.
\end{enumerate}

Having discussed how a group $G$ can be described as an extension of $Q$ by $A$, we now move to classifying such group extensions. Isomorphism classes of extensions are related to inequivalent choices of the group action $\rho$ and function ${c(q_1,q_2)}$. As discussed, different group actions correspond to different homomorphisms ${\rho\colon Q\to\Aut(A)}$. So what remains is to deduce the classes of ${c(q_1,q_2)}$ leading to inequivalent extensions. Firstly, we note that for the group multiplication rule~\eqref{multiplication} to be associative, we require
\begin{equation}
\rho_{q_1}\left(c\left(q_2, q_3\right)\right)-c\left(q_1 q_2, q_3\right)+c\left(q_1, q_2 q_3\right)-c\left(q_1, q_2\right)=0.
\end{equation}
This implies that $c$ must be a 2-cocycle ${c\in Z^2(BQ,A^\rho)}$. However, these 2-cocycles do not each lead to inequivalent extensions. Indeed, recall that the change of lift~\eqref{changeLift}, which in our current notation is ${(a,q) \to (a+f(q),q)}$, implements a group isomorphism of $G$ that does not change the extension class. The group multiplication rule~\eqref{multiplication} implies that the 2-cocycle $c(q_1,q_2)$ transforms under this group isomorphism as
\begin{equation}
c(q_1,q_2) \to c(q_1,q_2) + f(q_1) + \rho_{q_1}(f(q_2)) - f(q_1q_2).
\end{equation}
In fact, this is same as shifting $c(q_1,q_2)$ by the 2-coboundary ${\dd f\in B^2(BQ,A^\rho)}$. Therefore, the extension classes depend on the 2-cocycle $c$ only through its cohomology class ${[c]\in H^2(BQ, A^\rho)}$. Thus, in summary, the group extension class of $G$ is specified by the data
\begin{equation}\label{grpExtData}
G = (Q, A, \rho, [c]).
\end{equation}

Before concluding this Appendix, let us discuss two commonly encountered specialized cases of~\eqref{grpExtData}. The first is group extensions for which the group action $\rho$ is trivial. Then, by~\eqref{rhodefliftApp}, the normal subgroup $\iota(A)$ of $Q$ is a subgroup of the center $Z(Q)$ of $Q$. In this case, the group extension is called a central extension. The other commonly encountered scenario is that the cohomology class ${[c]}$ is trivial. Such group extensions are called split extensions. They differ from trivial extensions by the group action $\rho$, making $G$ the semi-direct product group ${Q \ltimes_{\rho} A}$.

\section{Background foliation fields}\label{FoliationApp}

A QFT defined on a ${(d+1)}$-dimensional spacetime $M$ can have various background fields. Perhaps the most familiar is when $M$ is equipped with a Riemannian metric $g_{\mu\nu}$. Then, a QFT on $M$ can depend explicitly on $g_{\mu\nu}$. Another commonly encountered background field for a QFT is a background gauge field $A_\mu$ for a global symmetry $G$. This arises as the $G$-connection of a principal $G$-bundle on $M$.  Lastly, another typical background field is a spin structure. If $M$ is a spin manifold, it can be equipped with a background spin structure $\rho$. This is necessary for QFTs with spinor fields.

A smooth, orientable manifold can also be equipped with a foliation structure. Mathematically, a codimension $n$ foliation is a particular decomposition of a manifold formed by a collection of pairwise-disjoint, connected, immersed codimension $n$ submanifolds called leaves. In particular, each point in spacetime must have a neighborhood whose first $n$ local coordinates in each leaf are constant. Informally, a manifold with a codimension $n$ foliation can be regarded as an infinite number of layered/stacked codimension $n$ manifolds infinitesimally close to one another.  Codimension $n$ foliations are described by an $n$-form foliation field which is never zero. A QFT can couple to this background foliation field, and those that do are called a foliated QFT. A manifold can be simultaneously equipped with multiple foliation structures, and a foliated QFT can depend on more than one foliation field.

Let us focus on a codimension $1$ foliation of spacetime described by the background $1$-form foliation field ${e_{\mu}}$. A codimension $1$ foliation is defined as the kernel of $e_\mu$. This means that at each point $x_\mu$ in spacetime, the leaf containing $x_\mu$ is one whose tangent vector $v_\mu$ satisfies ${ e_\mu v^{\mu} = 0}$. Therefore, $e_\mu$ is orthogonal to each leaf. However, not all 1-forms define a foliation structure. As already mentioned, ${e_{\mu}}$ must be non-zero everywhere. Furthermore, by Frobenius' theorem, the foliation field $e$ must also 
satisfy the integrability condition\footnote{When~\eqref{ede=0} is not satisfied, the leaves defined by $e_\mu$ fail to be ``nice'' codimension $1$ submanifolds and instead densely fill a codimension 0 region of spacetime. For instance, 1-forms $\al$ for which ${\al\wdg\dd\al >0}$ define contact structures of 3-manifolds. The contact structure is given by the kernel of the contact 1-form $\al$, which is the subject of contact geometry. }
\begin{equation}\label{ede=0}
e\wdg \dd e = 0.
\end{equation}
One of the simplest foliations of a manifold are flat foliations where ${\dd e = 0}$.  

The integrability condition implies that $e$  satisfies
\begin{equation}\label{etobt}
\dd e = e \wdg \bt,
\end{equation}
for some 1-form $\bt$. This 1-form $\bt$ is not unique. It is ambiguous up to a shift by a scalar field $c$, which transforms it as  $\bt\rightarrow{\bt + c\, e}$.
In addition to this ambiguity, there is also a gauge redundancy in the foliation field under rescaling by a scalar field $f$: ${e \rightarrow \exp[-f]\, e}$ and ${\bt \rightarrow \bt + \dd f}$. This rescaling does not affect the leaves and the foliation structure. Using this redundancy, we can always make the foliation field closed $\dd e=0$ locally.

There is however an obstruction to making the foliation field closed globally. It is characterized by the de Rham cohomology class  ${[\bt\wdg \dd \bt]}$, known as the Godbillon-Vey invariant, which is a cobordism invariant of foliations~\cite{godbillonVeyInvariant}. The 3-form ${\bt\wdg \dd \bt}$ is closed. By acting the exterior derivative $\dd$ on~\eqref{etobt}, we find ${e \wdg \dd \bt = 0}$, which implies that $\dd\bt = e\wdg \ga$ for some 1-form $\ga$ and, thus, ${\dd(\bt\wdg\dd\bt) = 0}$. The Godbillon-Vey invariant is free from the redundancy in $\beta$. Under $\beta\rightarrow \beta+df$, it transforms as ${\bt\wdg\dd\bt \sim \bt\wdg\dd\bt + \dd (f\dd\bt)}$, which does not change the de Rham cohomology class. Furthermore, under ${\bt\to \bt + c\, e}$, it transforms as ${\bt\wdg\dd\bt \to \bt\wdg\dd\bt + \dd (c\,\dd e)}$, which again does not change the de Rham cohomology class.

\section{\texorpdfstring{$\Z_N$}{ZN} dipole SymTFT as a foliated field theory}\label{ZNdipSymTFTFoliApp}

In Section~\ref{ZNdipEx}, we found that the SymTFT for a $\Z_N$ dipole symmetry in ${1+1}$D is the ${2+1}$D $\Z_N$ symmetric tensor gauge theory
\begin{equation}\label{tensorZnContThy}
\mathscr{L} = \frac{\ii\, N}{2\pi \La}\, (
\,
a_{y}\pp_t b_{xx} - a_{xx}\pp_t b_{y} + b_{t} ( \pp^2_x a_{y} - \pp_y a_{xx} ) - a_{t} ( \pp_y b_{xx} + \pp^2_x b_{y} )
\,).
\end{equation}
It has the $\Z_N$ dipole gauge redundancy
\begin{equation}\label{tensorDipBFGaugeRed}
\begin{aligned}
a_{t} &\sim a_{t} + \pp_t \al,\hspace{30pt} a_{xx}\sim a_{xx} + \pp_x^2 \al,\hspace{30pt}a_{y} \sim a_{y} + \pp_y \al ,\\
b_{t} &\sim b_{t} + \pp_t \bt,\hspace{32pt} b_{xx}\sim b_{xx} - \pp_x^2 \bt,\hspace{33pt}b_{y} \sim b_{y} + \pp_y \bt.
\end{aligned}
\end{equation}
In this Appendix, following \Rf{EHN240110677}, we derive the duality from this $\Z_N$ tensor gauge theory to the foliated field theory~\eqref{FFT}.

We first decompose the tensor gauge fields $a_{xx}$ and $b_{xx}$ in terms of the fields ${a_x}$, ${\t{a}_x}$, ${b_x}$, and ${\t{b}_x}$ as
\begin{align}
a_{xx} = \pp_x a_x - \La\,\t{a}_x,\hspace{40pt} b_{xx} = -(\pp_x b_x - \La\, \t{b}_x).
\end{align}
This introduces a gauge redundancy
\ie\label{eq:more_redundancy}
&a_x\sim a_x+\Lambda\t{\alpha}~,\quad\qquad \t{b}_x\sim \t{b}_x+\Lambda\t{\beta}~,
\\
&\t{a}_x\sim \t{a}_x+\partial_x\t{\alpha}~,\ \,\,\qquad \t{b}_x\sim \t{b}_x+\partial_x\t{\beta}~,
\fe
Using this field redefinition, the Lagrangian~\eqref{tensorZnContThy} can be written as
\begin{equation}\label{tensorZnContThy2}
\begin{aligned}
\mathscr{L} &= -\frac{\ii\,N}{2\pi} \bigg(
\t{a}_x (\pp_y b_{t} - \pp_t b_{y} )
-
\t{b}_x ( \pp_y  a_{t} - \pp_t a_{y} ) 
- \La^{-1}
( \pp_x b_{y} - \pp_y b_x ) (\pp_x a_{t} - \pp_t a_x)
\\
&  
\hspace{230pt}+ \La^{-1} 
(\pp_x a_{y} - \pp_y a_{x}) (\pp_x b_{t} - \pp_t  b_x )
\bigg).
\end{aligned}
\end{equation}
We next introduce the fields $\t{a}_y$ and $\t{b}_y$ related to ${(a_x,a_{y})}$ and ${(b_x,b_{y})}$ by
\begin{align}
\pp_x a_{y} = \pp_y a_x - \La \, \t{a}_y,\hspace{40pt}
\pp_x b_{y} = \pp_y b_x - \La \, \t{b}_y.
\end{align}
Enforcing these constraints using the Lagrange multiplier fields $\t{a}_t$ and $\t{b}_t$, we can write the Lagrangian~\eqref{tensorZnContThy2} as
\begin{equation}\label{tensorZnContThy3}
\begin{aligned}
\mathscr{L} &= -\frac{\ii\,N}{2\pi} \bigg(
\t{a}_x (\pp_y b_{t} - \pp_t b_{y} )
- \t{a}_y (\pp_x b_{t} - \pp_t  b_x )
-
\t{b}_x ( \pp_y  a_{t} - \pp_t a_{y} ) 
+ \t{b}_y (\pp_x a_{t} - \pp_t a_x)
\\
&  
\hspace{50pt} 
+ \t{a}_t (\pp_x b_{y} - \pp_y b_x + \La \, \t{b}_y )
-
\t{b}_t (\pp_x a_{y} - \pp_y a_x + \La \, \t{a}_y)
\bigg).
\end{aligned}
\end{equation}

The presentation~\eqref{tensorZnContThy3} of the $\Z_N$ tensor gauge theory has a manifest foliation structure. Indeed, the action can be  written using differential forms as
\begin{equation}\label{FFTapp}
S[e] = -\frac{\ii\,N}{2\pi} \int \left(\t{a} \wdg \dd b -\t{b}\wdg\dd a - \t{a}\wdg \t{b} \wdg e\right),
\end{equation}
where ${e = \La \dd x}$ is a background foliation 1-form field describing a flat foliation whose leaves are the $(y,t)$ planes of spacetime. In this presentation, the gauge redundancy~\eqref{tensorDipBFGaugeRed} and~\eqref{eq:more_redundancy} is
\begin{equation}
\begin{aligned}
a &\sim a + \dd \al + \t{\al}\, e,
\hspace{40pt} 
b \sim b + \dd \bt + \t{\bt}\, e,\\
\t{a} &\sim \t{a} + \dd \t{\al},
\hspace{72pt} 
\t{b} \sim \t{b} + \dd \t{\bt},
\end{aligned}
\end{equation}
depending explicitly on the foliation field.

\section{\texorpdfstring{$\Z_2$}{Z2} symmetry fractionalization 
in \texorpdfstring{$\Z_2$}{Z2} topological order}
\label{Z2 enriched TC App}

In this Appendix, we will explore an exactly solvable lattice model realizing $\Z_2$ topological order enriched by a $\Z_2$ 0-form symmetry that fractionalizes. It is particularly instructive to compare this example with the spacetime symmetry enrichment discussed in Sections~\ref{sec reflection SymTFT} and~\ref{sec TRS SymTFT}.

To construct such a model, we start by considering $\Z_4$ qudits on sites $\bm{r}$ of the square lattice, which are acted on by the clock and shift operators
\begin{equation}
\cZ^4 = \cX^4 = 1,\hspace{30pt} \cZ\cX = \ii \cX\cZ,
\end{equation}
with periodic boundary conditions. The lattice Hamiltonian is the paramagnet
\begin{equation}\label{eq:Z4 paramagnet appendix}
H_{\text{SPT}}= -\sum_{\bm{r}} (\cX_{\bm{r}} + \cX^\dag_{\bm{r}}),
\end{equation}
which commutes with the $\Z_4$ 0-form symmetry operator $U = \prod_{\bm{r}} \cX_{\bm{r}}$. Gauging the entire $\Z_4$ symmetry maps the model~\eqref{eq:Z4 paramagnet appendix} to the $\Z_4$ Toric Code (i.e., $\Z_4$ gauge theory). To construct the SET model, we will instead gauge the $\Z_2$ sub-symmetry of $\Z_4$ generated by $U^2$. Since $\Z_4$ is a non-trivial extension of $\Z_2$ by $\Z_2$, then from Section~\ref{extensionsSubSec}, this should lead to a $\Z_2$ topological order enriched by a $\Z_2$ symmetry that fractionalizes.\footnote{The field theory description of this SET follows from~\eqref{SymTFTsec2.1}. The action is ${S = \ii\pi \int_X\,b\cup (\del a-\mathrm{Bock}(\cA))}$, where $a$ and $b$ are dynamical $\Z_2$-valued cochains, $\cA$ is a background $\Z_2$ gauge field described by a $\Z_2$-valued cochain, and ${\mathrm{Bock}\colon H^1(X,\Z_2)\to H^2(X,\Z_2)}$ is the Bockstein homomorphism. Embedding these cochains into U(1) gauge fields, we can write this action as ${S = \frac{2\ii}{2\pi} \int_X\,b \wedge (\dd a-\frac12\dd\cA)}$.
}

To gauge the ${\Z_2\subset \Z_4}$ sub-symmetry,  we first introduce qubits onto the edges of the square lattice, whose Pauli operators we denote by $\si^x$ and $\si^z$. The gauging procedure is specified by the Gauss operator
\begin{equation}
G_{\bm{r}} = \cX_{\bm{r}}^2 \, A_{\bm{r}}, \hspace{20pt}\text{where} \hspace{20pt}A_{\bm{r}} = \begin{tikzpicture}[scale = 0.5, baseline = {([yshift=-.5ex]current bounding box.center)}]
\draw[line width=0.015in, gray] (7.5,0) -- (13.5, 0);
\draw[line width=0.015in, gray] (10.5,-3) -- (10.5, 3);
\node[above] at (9, 0) {\normalsize $\textcolor[HTML]{cb181d}{\si^x}$};
\node[above] at (12,0) {\normalsize $\textcolor[HTML]{cb181d}{\si^x}$};
\node at (10.5, 1.5) {\normalsize $\textcolor[HTML]{cb181d}{\si^x}$};
\node at (10.5, -1.5) {\normalsize $\textcolor[HTML]{cb181d}{\si^x}$};
\fill[black] (10.5,0) circle (6pt);
\node at (10, -0.5) {\normalsize $\bm{r}$};
\end{tikzpicture}~.
\end{equation}
That is, the physical Hilbert space is the subspace of ${\bigotimes_{\bm{r}}(\C_{\text{site}}^4\otimes \C^2_{x\text{-link}}\otimes \C^2_{y\text{-link}})\cong \C^{8^{L_xL_y}}}$, spanned by states $\ket{\psi}$ satisfying ${G_{\bm{r}}\ket{\psi} = \ket{\psi}}$. We also enforce the flatness conditions
\begin{equation}
B_{\bm{r}}\equiv \begin{tikzpicture}[scale = 0.5, baseline={([yshift=-.5ex]current bounding box.center)}]
\draw[line width=0.015in, gray] (-3, 0) -- (-3, 3) -- (0, 3) -- (0, 0) -- cycle;
\node[below] at (-1.5, 0) {\normalsize $\textcolor[HTML]{2171b5}{\si^z}$};
\node[above] at (-1.5, 3) {\normalsize $\textcolor[HTML]{2171b5}{\si^z}$};
\node at (-3.5, 1.5) {\normalsize $\textcolor[HTML]{2171b5}{\si^z}$};
\node at (0.6, 1.5) {\normalsize $\textcolor[HTML]{2171b5}{\si^z}$};
\fill[black] (-3,0) circle (6pt);
\node at (-3.5, -0.5) {\normalsize $\bm{r}$};
\end{tikzpicture}
= 1.
\end{equation}
However, we will enforce this energetically instead of kinematically. This causes the Hamiltonian~\eqref{eq:Z4 paramagnet appendix} to become
\begin{equation}\label{HamAfterGaingZ2}
H_{\mathrm{SET}}
=
-\sum_{\bm{r}} (\cX_{\bm{r}} + \cX_{\bm{r}}^\dag + B_{\bm{r}} ).
\end{equation}
Indeed, $H_{\mathrm{SET}}$ commutes with the Gauss operator $G_{\bm{r}}$ and the low-energy subspace satisfies ${B_{\bm{r}} = 1}$.

The Hamiltonian $H_{\mathrm{SET}}$ has a ${\Z_2\times\Z_2}$ 1-form symmetry, whose symmetry operators are generated by ${W(\ga) = \prod_{e\subset \ga} \si^z_e}$ and 
$V(\ga^\vee) = \prod_{e\subset \ga^\vee} \si^x_e$. Indeed, $W$ and $V$ commute with $H_{\mathrm{SET}}$ for all 1-cycles $\ga$ and $\ga^\vee$ of the square lattice and its dual lattice, respectively. This 1-form symmetry is anomalous and spontaneously broken in the model $H_{\mathrm{SET}}$, which gives rise to $\Z_2$ topological order. The Hamiltonian also commutes with the operator ${U = \prod_{\bm{r}}\cX_{\bm{r}}}$. Due to the Gauss law, this operator satisfies ${U^2 = 1}$ and generates a $\Z_2$ 0-form symmetry operator. Furthermore, this operator becomes the identity operator in the ground state subspace. Therefore, the Hamiltonian $H_{\mathrm{SET}}$ is in an SET phase with $\Z_2$ topological order enriched by a $\Z_2$ 0-form symmetry.

As mentioned, we expect this $\Z_2$ 0-form symmetry to be fractionalized. We can diagnose this symmetry fractionalization using the disorder operators of the $\Z_2$ 0-form symmetry $U$ and the $\Z_2$ 1-form symmetry $W$. Here, the disorder operators are the symmetry operators truncated to act on their respective subspaces but with boundaries. They are the movement operators for their respective symmetry defects. The $\Z_2$ 0-form symmetry disorder operator is 
\begin{equation}
U(\Si) = \prod_{\bm{r}\subset\Si}\cX_{\bm{r}},
\end{equation}
which is the truncation of the symmetry operator $U$ to a connected patch $\Si$ of the lattice. The $\Z_2$ 1-form disorder operator for an oriented path $P$ from site $\bm{r}_1$ to $\bm{r_2}$ is
\begin{equation}
W(P) = \cZ_{\bm{r}_1}\left(\prod_{e\subset P} \si^z_e\right)\cZ_{\bm{r}_2}^\da.
\end{equation}
The endpoints of $W(P)$ are dressed by $\cZ$ operator such that $W(P)$ commutes with $G_{\bm{r}}$ and $U$.\footnote{The $\Z_2$ 1-form symmetry operator ${W(\ga) = \prod_{e\subset \ga} \si^z_e}$ is better written as ${W(\ga) = \prod_{(\bm{r},\mu)\subset \ga} \cZ_{\bm{r}}\si^z_{\bm{r},\mu}\cZ^\da_{\bm{r}+\hat{\mu}}}$ such that it is a product of local operators obeying the same algebra with $U$ and $G_{\bm{r}}$ as $W(\ga)$ does. When written this way, the disorder operator $W(P)$ is a straightforward truncation of $W(\ga)$.} Acting $W(P)$ on the ground states creates a pair of $e$ anyon excitations. 

The symmetry fractionalization is manifest in the projective algebra
\begin{equation}
U(\Si) W(P) = \ii^{\mathrm{link}(\pp\Si, \pp P)}\, W(P)U(\Si),
\end{equation}
which shows that a single $e$ anyon carries fractional $\Z_2$ 0-form symmetry charge. We can relate this to the symmetry defects perspective of symmetry fractionalization. Using that $\cX_{\bm{r}}^2 = A_{\bm{r}}$ in the gauge-invariant subspace, the square of the $\Z_2$ 0-form symmetry disorder operator is
\begin{equation}
D(\Si)\times D(\Si) = V(\pp\Si).
\end{equation}
Therefore, fusing two $U$ symmetry defects gives the trivial symmetry operator with a $V$ symmetry defect dressing the fusion junction:
\begin{equation}
\begin{tikzpicture}[baseline={([yshift=-.5ex]current bounding box.center)},>=Triangle, thick]
\draw[Qdef, line width=2pt] (11.5, -1) -- (12.5,0);
\draw[Qdef, ->, line width=2pt,] (11.5, -1) -- (12.1, -.4);
\node[Qdef, below] at (11.5, -1) {$U$};
%
\draw[Qdef, line width=2pt] (13.5, -1) -- (12.5,0);
\draw[Qdef, ->, line width=2pt,] (13.5, -1) -- (12.9, -.4);
\node[Qdef, below] at (13.5, -1) {$U$};
%
\draw[Qdef, line width=2pt] (12.5, 0) -- (12.5,1.41);
\draw[Qdef, ->, line width=2pt,] (12.5, 0) -- (12.5, 1.01);
\node[Qdef, above] at (12.5, 1.41) {$1$};
\fill[Adef] (12.5, 0) circle (4pt);
\node[Adef, above right] at (12.5, 0) {$V$};
\end{tikzpicture}~.
\end{equation}

From the general theory of anyon condensation in SET, condensing the \( e \) anyons is expected to spontaneously break the \(\mathbb{Z}_2\) 0-form symmetry (see~\cite[Example 5.2]{BJL181100434}). This condensation can be implemented by adding the corresponding string operators \( W_{\bm{r},\mu}=\mathcal{Z}_{\bm{r}} \sigma^z_{\bm{r},\mu} \mathcal{Z}^\dagger_{\bm{r}+\hat{\mu}} \) to the Hamiltonian and increasing the coupling \( \lambda \). The resulting Hamiltonian is:
\begin{equation}
H_\lambda = H_{\mathrm{SET}} - \lambda \sum_{(\bm{r},\mu)}\left(\frac{1}{4} \sum_{n=1}^4 W_{\bm{r},\mu}^n \right).
\end{equation}
This Hamiltonian is gauge invariant since it commutes with the Gauss law operator \( G_{\bm{r}} \). It also preserves both the \(\mathbb{Z}_2\) 0-form symmetry \( U \) and the \(\mathbb{Z}_2\) 1-form symmetry \( W(\gamma) \). In the \( \lambda \to \infty \) limit, the Hamiltonian becomes exactly solvable, with ground states stabilized by the mutually commuting constraints
\[
G_{\bm{r}} = W_{\bm{r},\mu} =B_{\bm{r}}= 1.
\]
In practice, it suffices to impose the first two constraints, since the third follows from the second. 

Let us count the number of ground states in this limit. The full Hilbert space (including gauge-non-invariant states) has dimension $4^{L_x L_y} \cdot 2^{2L_x L_y}$. There are $2^{L_x L_y}$ independent Gauss law constraints ${G_{\bm{r}} = 1}$, one at each site. Naively, there are $4^{2L_x L_y}$ constraints of the form ${W_{\bm{r},\mu} = 1}$, one on each of the \( 2L_x L_y \) links.
However, the ${W_{\bm{r},\mu} = 1}$ constraints are overspecified. First, the product of ${W_{\bm{r},\mu}^2 = \mathcal{Z}_{\bm{r}}^2 \mathcal{Z}^{\dagger 2}_{\bm{r}+\hat{\mu}}}$ around each plaquette is automatically 1. Among the \( L_x L_y \) plaquettes, only ${L_x L_y - 1}$ such relations are independent, which reduces the number of independent \( W_{\bm{r},\mu} \) constraints by a factor of \( 2^{L_x L_y - 1} \).
In addition, there is further redundancy from nonlocal relations: the product of \( W_{\bm{r},\mu}^2 \) along the entire \( x \)-direction or \( y \)-direction is also automatically 1. These two additional redundancies contribute an overcounting by a factor of \( 2^2 \). Altogether, the number of \emph{independent constraints} is:
\[
\frac{2^{L_x L_y} \cdot 4^{2L_x L_y}}{2^{L_x L_y - 1} \cdot 2^2} = 2^{-1} \cdot 4^{2L_x L_y}.
\]
Therefore, the dimension of the ground state subspace is:
\begin{equation}
\text{GSD} = \frac{4^{L_x L_y} \cdot 2^{2L_x L_y}}{2^{-1} \cdot 4^{2L_x L_y}} = 2.
\end{equation}
A convenient basis for this two-dimensional ground state subspace is
\begin{equation}
\ket{\mathrm{GS}_1} = \prod_{\bm{r}} \left(\frac{1+G_{\bm{r}}}{2}\right)\ket{1, 1},
\hspace{40pt}
\ket{\mathrm{GS}_2} = \prod_{\bm{r}} \left(\frac{1+G_{\bm{r}}}{2}\right)\ket{\ii,1},
\end{equation}
where we denote by $\ket{a,b}$ the product state satisfying ${\cZ_{\bm{r}}\ket{a,  b} = a\,\ket{a, b}}$ and ${\si^z_{\bm{r},\mu}\ket{ a, b} = b\ket{ a, b}}$. These two ground states satisfy ${U\ket{\mathrm{GS}_1} = \ket{\mathrm{GS}_2}}$.
Therefore, the $\Z_2$ symmetry generated by $U$ is spontaneously broken.

Another way to understand the spontaneous \(\mathbb{Z}_2\) symmetry breaking is as follows. If we momentarily ignore the Gauss law constraint ${G_{\bm{r}} = 1}$, the symmetry operator \( U \) generates a \(\mathbb{Z}_4\) symmetry. Further neglecting the link degrees of freedom simplifies the operator \( W_{\bm{r},\mu} \) to \( \mathcal{Z}_{\bm{r}} \mathcal{Z}^\dagger_{\bm{r}+\hat{\mu}} \), and the ground states stabilized by this operator realize a spontaneously broken phase of the \(\mathbb{Z}_4\) symmetry generated by \( U \).
After restoring the link degrees of freedom and reimposing the Gauss law constraint ${G_{\bm{r}} = 1}$, this amounts to gauging the ${\mathbb{Z}_2\subset \mathbb{Z}_4}$ subgroup generated by \( U^2 \). As a result, the spontaneously broken \(\mathbb{Z}_4\) phase is mapped to a spontaneously broken \(\mathbb{Z}_2\) phase.

Since condensing $e$ anyons throughout space spontaneously breaks the $\Z_2$ 0-form symmetry, the SET's gapped boundary that condenses $e$ anyons also leads to spontaneous symmetry breaking. Indeed, we can condense the $e$ anyons on the spatial boundary ${r_y = L_y}$ by enforcing ${\cZ_{\bm{s}}\si^z_{\bm{s},x}\cZ^\dag_{\bm{s}+\hat{x}} =1}$, where $\bm{s} = (r_x,L_y)$. After a unitary transformation, this polarizes the qubits on the top link to the ${\si^z = 1}$ state and causes the $B_{\bm{r}}$ operators involving the spatial boundary to become
\begin{equation} 
B^{ \text{rough}}_{\bm{s}-\hat{y}}=\begin{tikzpicture}[scale = 0.5, baseline={([yshift=-.5ex]current bounding box.center)}]
\draw[line width=0.015in, gray] (-3,0) -- (0, 0);
\draw[line width=0.015in, gray] (-3,0) -- (-3, 3);
\draw[line width=0.015in, gray] (0,0) -- (0, 3);
\node[above] at (-1.5, 0) {\normalsize $\textcolor[HTML]{2171b5}{\si^z}$};
\node[left] at (-3, 1.5) {\normalsize $\textcolor[HTML]{2171b5}{\si^z}$};
\node[right] at (0, 1.5) {\normalsize $\textcolor[HTML]{2171b5}{\si^z}$};
\node[above] at (-3, 3) {\normalsize $\textcolor[HTML]{2171b5}{\cZ^\dag}$};
\node[above] at (0, 3) {\normalsize $\textcolor[HTML]{2171b5}{\cZ}$};
\fill[black] (-3,0) circle (6pt);
\node[below] at (-3, 0) {\normalsize $\bm{s}-\hat{y}$};
\end{tikzpicture}.
\end{equation}
This is the rough boundary of the SET. In the low-energy subspace where ${B^{ \text{rough}}_{\bm{r}} = 1}$, the $\Z_4$ qudits on the boundary sites $\bm{s}$ will obey $(B^{ \text{rough}}_{\bm{s}-\hat{y}})^2 = \cZ_{\bm{s}}^2\cZ_{\bm{s}+\hat{x}}^2 = 1$ for all $\bm{s}$. Therefore, in the thermodynamic limit, $\cZ_{\bm{s}}^2$ acquires an expectation value and the symmetry generated by ${U=\prod_{\bm{r}}\cX_{\bm{r}}}$ is spontaneously broken on the boundary. In the presence of the rough boundary, however, $U$ is no longer a $\Z_2$ symmetry operator. It is now an order four operator, satisfying
\begin{equation}
    U^2 = \prod_{\bm{r}} A_{\bm{r}} = \prod_{r_x} \sigma^x_{\bm{s}-\hat{y},y} \cX^2_{\bm{s}}.
\end{equation}
Therefore, with this boundary condition, the spontaneous symmetry breaking pattern is ${\Z_4\ssb \Z_2}$.

The above analysis considered a basis in which the physical Hilbert space did not have a tensor product factorization. Before concluding this Appendix, let us solve the Gauss law by entering a unitary frame where the physical Hilbert space has a tensor product factorization. In particular, we use the onsite unitary operator
\begin{equation}
W = \prod_{\bm{r}}\left(\frac{1+\cZ_{\bm{r}} + A_{\bm{r}}(1-\cZ_{\bm{r}})}2\right)
\end{equation}
to perform the basis transformation 
\begin{equation}
\begin{aligned}
\cX_{\bm{r}} &\to \cX_{\bm{r}}\ee^{\ii\frac{\pi}{4}(1-A_{\bm{r}})},\hspace{40pt} \cZ_{\bm{r}} \to \cZ_{\bm{r}},\\
\si^x_{\bm{r},\mu} &\to \si^x_{\bm{r},\mu}, \hspace{75pt} \si^z_{\bm{r},\mu} \to \si^z_{\bm{r},\mu} \cO_{\bm{r}} \cO_{\bm{r}+\hat{\mu}} .
\end{aligned}
\end{equation}
where ${2\cO_{\bm{r}} = A_{\bm{r}}(\cZ_{\bm{r}}^\da - \cZ_{\bm{r}} ) - \cZ_{\bm{r}}^\da - \cZ_{\bm{r}}}$. In this basis, the Gauss law is ${\cX^2_{\bm{r}} = 1}$. Therefore, each $\Z_4$ qudit becomes a qubit after gauging whose Pauli operators are $Z_{\bm{r}}\equiv \cZ^2_{\bm{r}}$ and $\cX_{\bm{r}}\equiv X_{\bm{r}} \equiv \cX^\dag_{\bm{r}}$. The SET Hamiltonian then becomes
\begin{equation}
W H_{\mathrm{SET}}W^\da = -\sum_{\bm{r}}\left(X_{\bm{r}} + X_{\bm{r}}A_{\bm{r}})+ B_{\bm{r}}\right) + \mathrm{H.c.},
\end{equation}
and the $\Z_2$ symmetry operator $U$ in this frame is
\begin{equation}\label{UinWframe}
WUW^\dag = \prod_{\bm{r}}X_{\bm{r}}\ee^{\ii\frac{\pi}{4}(1-A_{\bm{r}})}.
\end{equation}
The $\Z_2$ 1-form symmetry operators are still $W(\ga) = \prod_{e\subset \ga} \si^z_e$ and $V(\ga^\vee) = \prod_{e\subset \ga^\vee} \si^x_e$. However, their associated string operators are now ${\si^z_{\bm{r},\mu} \t\cO^+_{\bm{r}} \t\cO^-_{\bm{r}+\hat{\mu}}}$ and $\si^x_{\bm{r}}$ where ${2\t\cO_{\bm{r}}^\pm = \pm A_{\bm{r}}(1 - Z_{\bm{r}} ) - 1 - Z_{\bm{r}}}$. The $e$ string operator is quite complicated, which is required for it to be symmetric under~\eqref{UinWframe}.

\addcontentsline{toc}{section}{References}

\bibliographystyle{ytphys}
\baselineskip=0.85\baselineskip
\bibliography{local.bib} 

\end{document}